\documentclass[fleqn,usenatbib]{mnras}

\usepackage{newtxtext,newtxmath}

\usepackage[T1]{fontenc}
\usepackage{ae,aecompl}

\usepackage{graphicx}	
\usepackage{amsmath}	
\usepackage{amssymb}	
\usepackage{color}
\usepackage{multicol}

\title[Reevaluating Old Stellar Populations]{Reevaluating Old Stellar Populations}

\author[Stanway \& Eldridge]{
E. R. Stanway$^{1}$\thanks{E-mail: e.r.stanway@warwick.ac.uk}
and J. J. Eldridge$^{2}\thanks{E-mail: J.Eldridge@auckland.ac.nz}$
\\
$^{1}$Department of Physics, University of Warwick, Gibbet Hill Road, Coventry, CV4 7AL, United Kingdom\\
$^{2}$Department of Physics, Private Bag 92019, University of Auckland, New Zealand
}

\date{Accepted 2018 May 17. Received 2018 May 14; in original form 2018 March 27}

\pubyear{2018}

\begin{document}
\label{firstpage}
\pagerange{\pageref{firstpage}--\pageref{lastpage}}
\maketitle

\begin{abstract}
Determining the properties of old stellar populations (those with age $>$1\,Gyr) has long involved the comparison of their integrated light, either in the form of photometry or spectroscopic indexes, with empirical or synthetic templates. Here we reevaluate the properties of old stellar populations using a new set of stellar population synthesis models, designed to incorporate the effects of binary stellar evolution pathways as a function of stellar mass and age. We find that single-aged stellar population models incorporating binary stars, as well as new stellar evolution and atmosphere models, can reproduce the colours and spectral indices observed in both globular clusters and quiescent galaxies. The best fitting model populations are often younger than those derived from older spectral synthesis models, and may also lie at slightly higher metallicities.
\end{abstract}

\begin{keywords}
methods: numerical -- binaries: general -- galaxies: stellar content -- globular clusters: general
\end{keywords}



\section{Introduction}\label{sec:intro}

The most massive galaxies in the local Universe are mature, quiescent systems in which there is little or no ongoing star formation, and in which the most massive stars have lived and died, leaving their lower mass siblings to dominate the galaxy's integrated light. Similar old stellar populations can be found in other environments, including in the halos of spiral galaxies like the Milky Way, and in the globular clusters found in galaxies of all types. These latter objects are typically the result of a single starburst, their stars forming a coeval mass sequence or isochrone.

In the case of local globular clusters, i.e. those in the Milky Way and elsewhere in the Local Group, it is sometimes possible to spatially resolve the light of individual stars and thus determine the stellar population properties in detail. These have revealed examples of clusters with a broad or bifurcated main sequence \citep[e.g.][]{2004ApJ...605L.125B}, which is not expected from single-aged simple stellar populations, and, in many cases, also blue stragglers - surviving luminous blue stars which appear younger than the rest of the population \citep[e.g.][]{2014ApJ...782...49S}. However in the majority of old stellar populations, such detailed investigation of resolved stellar properties is impossible due to a lack of depth or angular resolution, and the integrated light of the unresolved stars must instead be used to constrain the properties of the population as a whole. In this scenario observable characteristics of the source spectral energy distribution (SED), including photometric colours and spectroscopic emission lines or indices, are compared to those determined for models of known age and composition. Either a best fitting template or a relation calibrated on such templates is then used to characterise the population. As a result, such analyses are strongly dependent on the properties of the template stellar population models.

While early attempts to produce empirical galaxy templates depended on the use of observed galaxy spectra and their combination into `typical' galaxy type composites \citep[e.g.][]{1980ApJS...43..393C},  the use of synthetic stellar population synthesis (SPS) templates is now widespread \citep[see][for a recent review]{2013ARA&A..51..393C}. In these, modelled evolutionary tracks for individual stars are combined to generate a population with the desired age, initial composition and initial mass distribution. The physical parameters of the stellar models are then used to select atmosphere models which are again combined to generate a synthetic spectrum. The evolutionary tracks and atmospheres can themselves be empirically-derived models, purely theoretical or a mixture of the two. 

While this approach has been highly successful, allowing for the efficient characterization of the stellar populations of galaxies in huge survey datasets, the results are still model-dependent, and their interpretation has proven vulnerable to uncertainties in stellar evolution theory. Treatment of stars in different evolutionary stages varies between extant spectral synthesis models. In some cases, notably at metallicities and ages where the post-main sequence Asymptotic Giant Branch (AGB) dominates the integrated light, slight differences in model prescription can lead to substantially different predictions for a population SED, particularly in the near-infrared \citep[e.g.][]{2003MNRAS.344.1000B,2005MNRAS.362..799M,2009MNRAS.394L.107M,2018MNRAS.tmp..499D}. To add to the complication, the most widely-used SPS models make the fundamental assumption that every star evolves independently of any companions, i.e. that all stars are single stars.  For massive stars (M$>>$M$_\odot$) in particular, justification for using single star models is now far from clear, with observations suggesting a multiplicity fraction approaching unity \citep[e.g.][]{2017ApJS..230...15M}. Given that binary interactions can substantially effect a star's evolution through the Hertzsprung-Russell (HR) diagram, neglecting these can also affect conclusions drawn from the integrated light of populations including binaries. While the lower, near- or sub-Solar, mass stars which dominate at late ages have a lower binary fraction than the massive stars which dominate starbursts, there may still be significant differences in the evolution of a population incorporating binaries, due to the effects of mass transfer and stellar interactions which rejuvenate or truncate the lifetimes of stars, changing the apparent age of the population.

A new generation of population and spectral synthesis models are now attempting to probe this hitherto unconsidered parameter space by incorporating new stellar evolution pathways. In some recent work, the use of stellar models which evolve singly but while rotating at a substantial fraction of their break-up speed has been explored \citep[for example, this is the case in the very recent detailed study of ][]{2018arXiv180110185C}. Other stellar population synthesis codes have begun to consider the effects of binary evolution \citep[e.g.][]{2003A&A...400..429B,2015MNRAS.447L..21Z,2017A&A...608A..11G,2018arXiv180203018G}. Amongst these are our Binary Population and Spectral Synthesis \citep[BPASS,][]{2009MNRAS.400.1019E,2012MNRAS.419..479E,2017PASA...34...58E} models, which have proven remarkably successful at matching the properties of young stellar populations ($<$100\,Myr) across a broad range of environments and redshifts \citep[e.g.][]{2016MNRAS.457.4296W,2016MNRAS.460.3170W,2017arXiv171108820S,2018arXiv180105125R}.  However previous versions of BPASS have focused on the treatment of low metallicity, massive stars that dominate the young stellar populations of distant galaxies but are relatively rare in the local Universe. As a result, they have incorporated a relatively simplistic prescription for the low stellar mass pathways which dominate in forming old stellar populations and have been unreliable at stellar population ages $>1$\,Gyr. 

In this paper, we present new BPASS v2.2 models which address these issues, and use them to reevaluate the physical properties of old stellar populations in both globular clusters (GCs) and nearby quiescent galaxies. In section \ref{sec:v2.2} we introduce the population synthesis models, describing the improvements made to our treatment of old stellar populations.  In section \ref{sec:obs} we use these new models to fit the properties of observational samples, discussing the resultant shifts in our view of old stellar populations in section \ref{sec:disc}. Finally, we present our conclusions in section \ref{sec:conc}.


\begin{figure*}
\hspace{-1.5cm}
\begin{center}
\includegraphics[width=0.80\columnwidth]{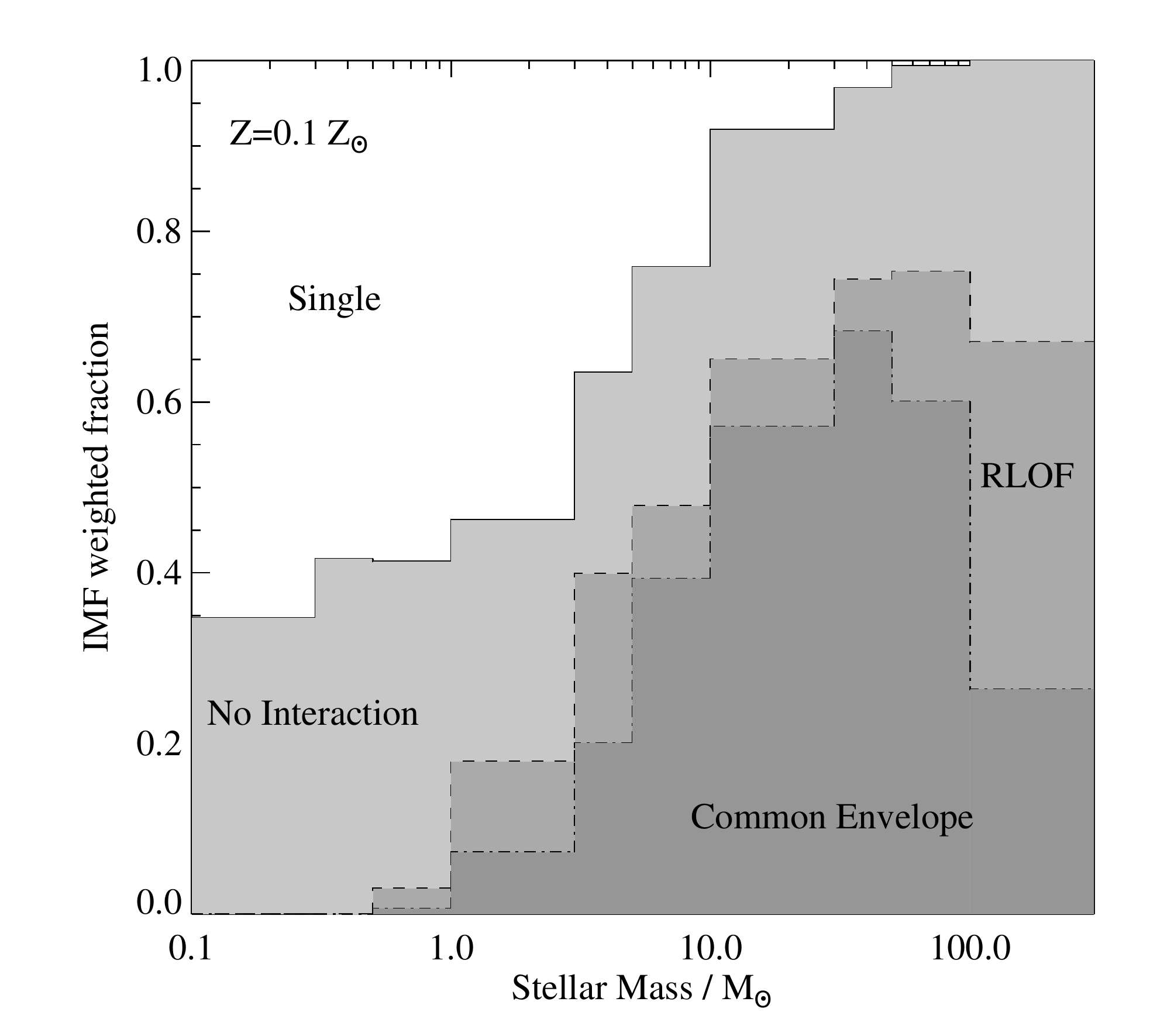}
\includegraphics[width=0.80\columnwidth]{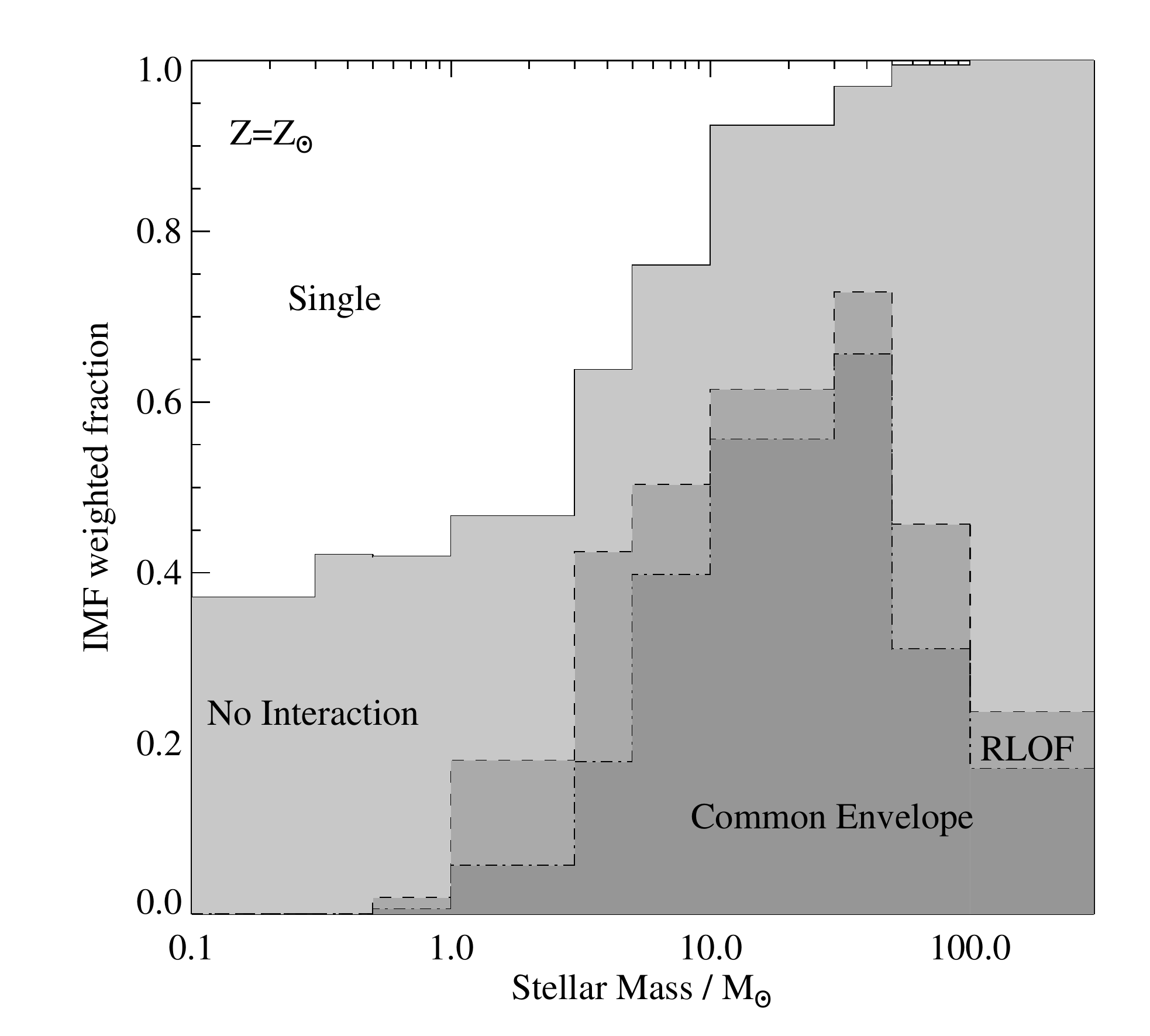}
\caption{The IMF- and binary-parameter weighted fraction of stars which undergo binary interactions within the age of the Universe, as a function of stellar mass, as adopted for the BPASS v2.2 binary models. The binary fraction remains fixed at all metallicities, as do the initial period and mass ratio distributions, with the fraction of sources eventually going through one or more interaction phases (including Roche-lobe overflow, RLOF, and Common Envelope evolution) determined by metallicity-dependent evolutionary processes.}
\label{fig:intfrac}
\end{center}
\end{figure*}

\section{New Synthetic Stellar Population Models}\label{sec:v2.2}

\begin{figure*}
\includegraphics[width=0.99\columnwidth]{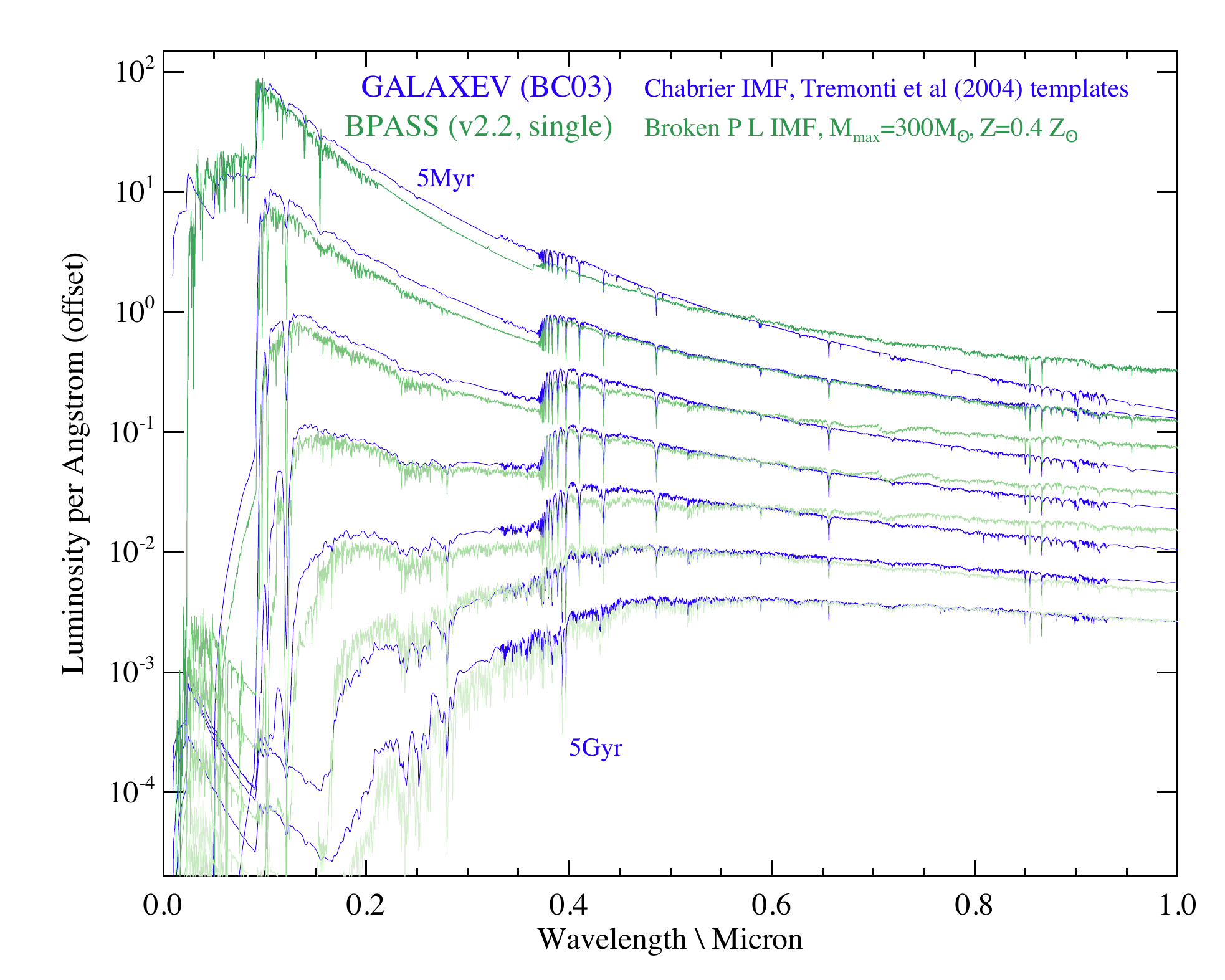}
\includegraphics[width=0.99\columnwidth]{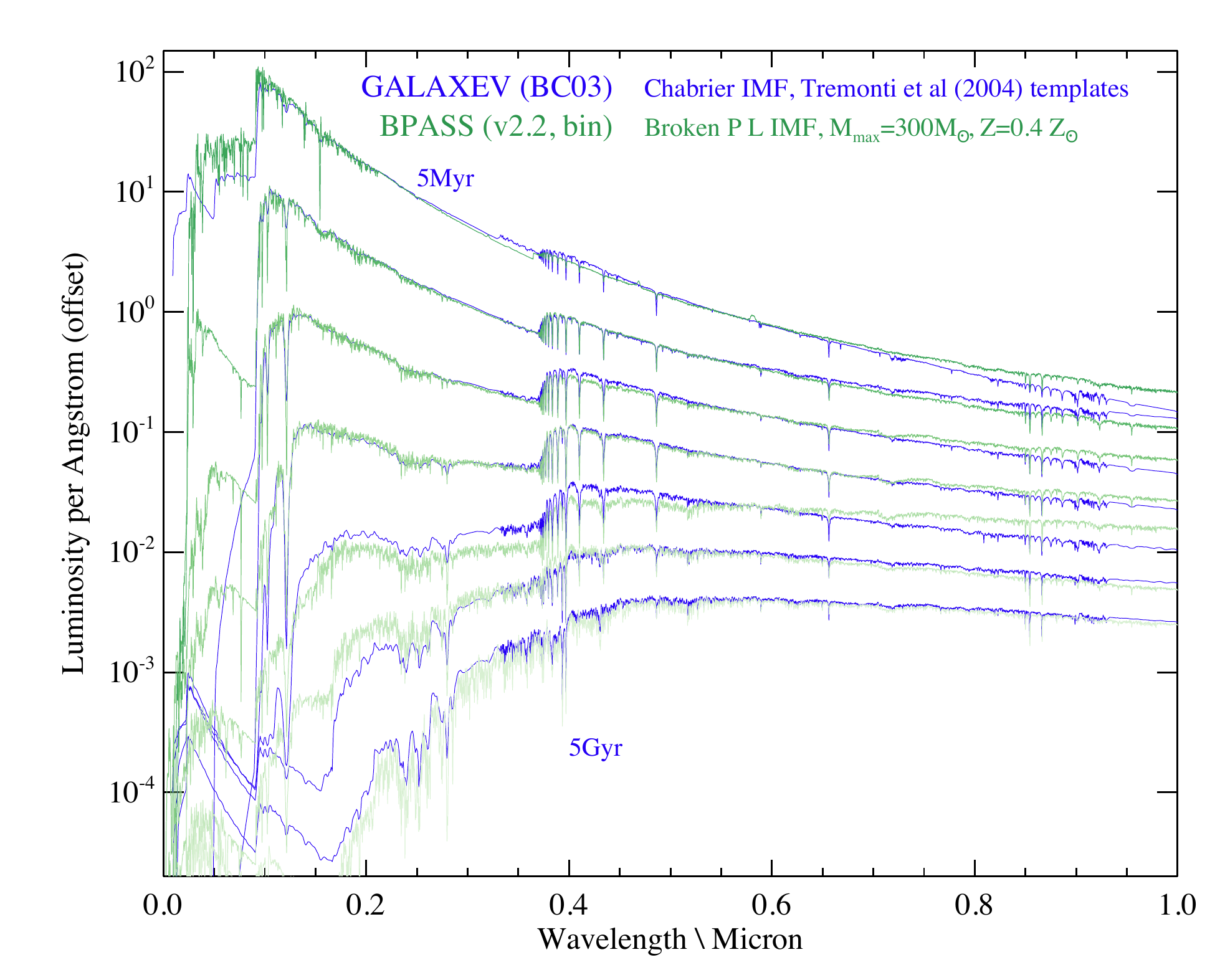}
\includegraphics[width=0.99\columnwidth]{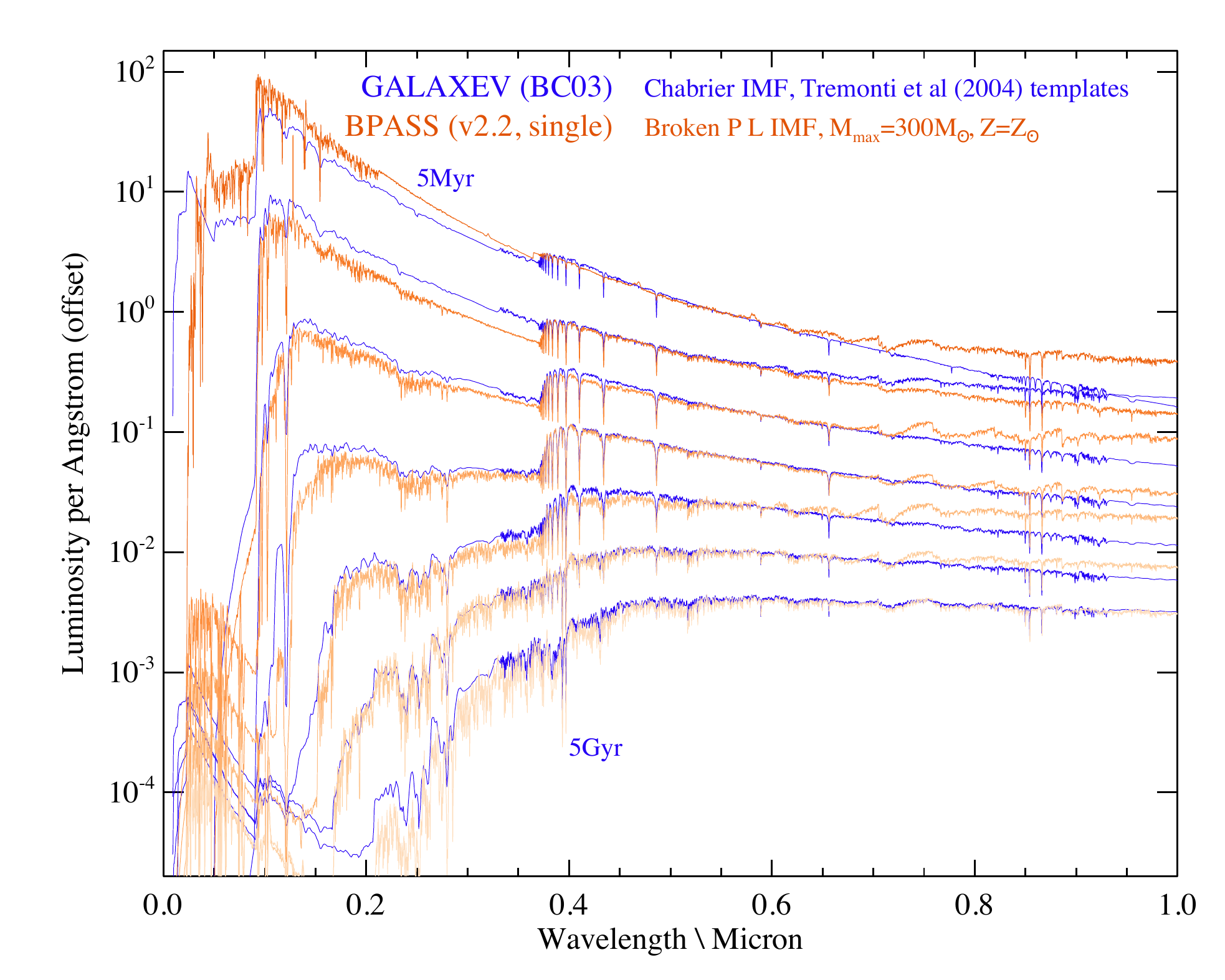}
\includegraphics[width=0.99\columnwidth]{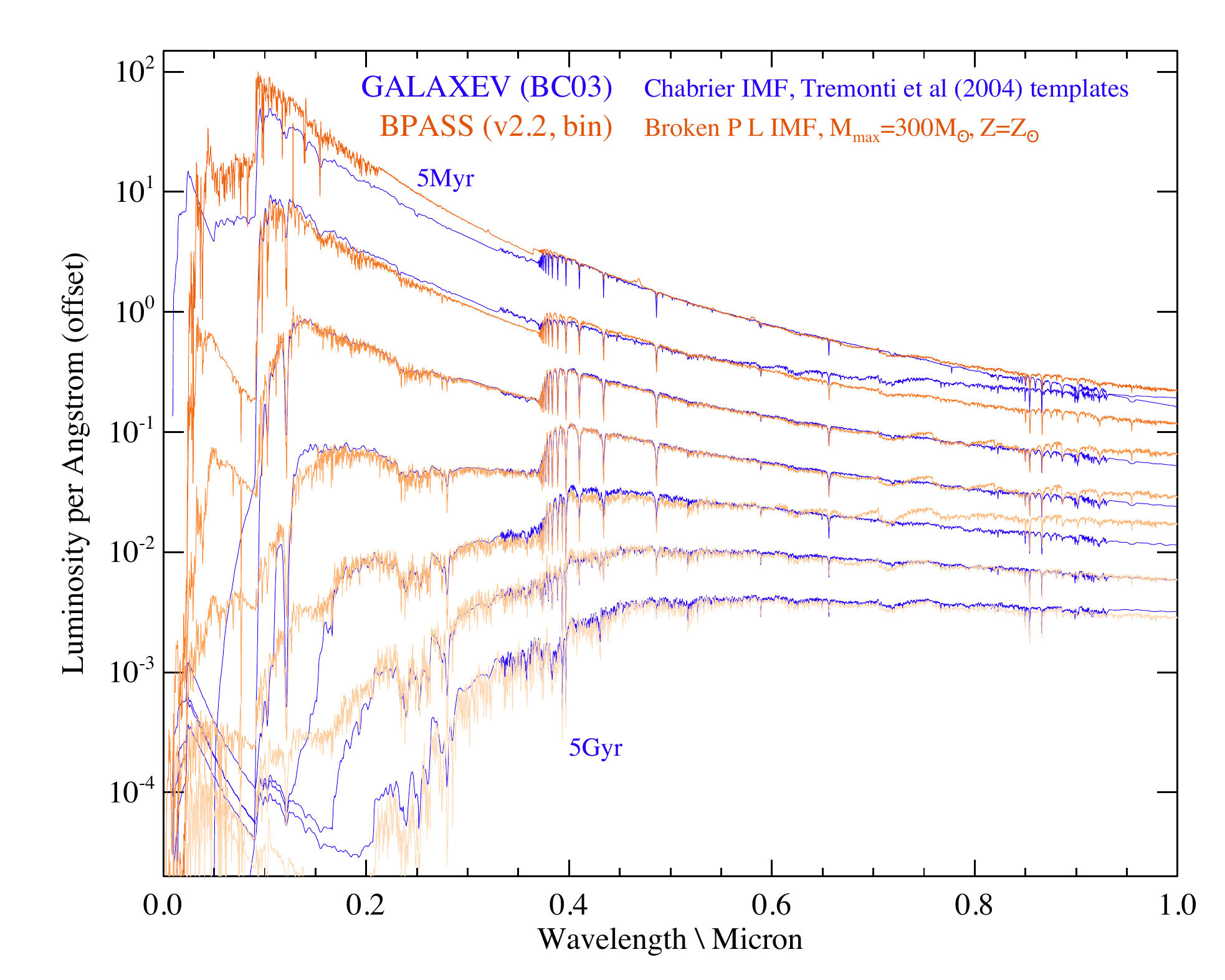}
\caption{BPASS v2.2 spectral energy distributions at 0.4\,Z$_\odot$ and Z$_\odot$ (green and orange respectively), compared to the equivalent SEDs drawn from the GALAXEV models of Bruzual \& Charlot 2003 (shown in blue). Results from our single star models are shown on the left and illustrate the differences in single star stellar evolution and atmospheres between our models and those of BC03. Results from our v2.2 binary models are shown on the right. Spectra are normalised to match between model sets in luminosity per unit wavelength at 5500\AA\ and are offset in luminosity for clarity. SEDs are shown at stellar population ages of 5, 25, 100, 250 and 690\,Myr, 1.4 and 5\,Gyr.}
\label{fig:bc03comp}
\end{figure*}

\begin{figure*}
\begin{center}
\includegraphics[width=01.8\columnwidth]{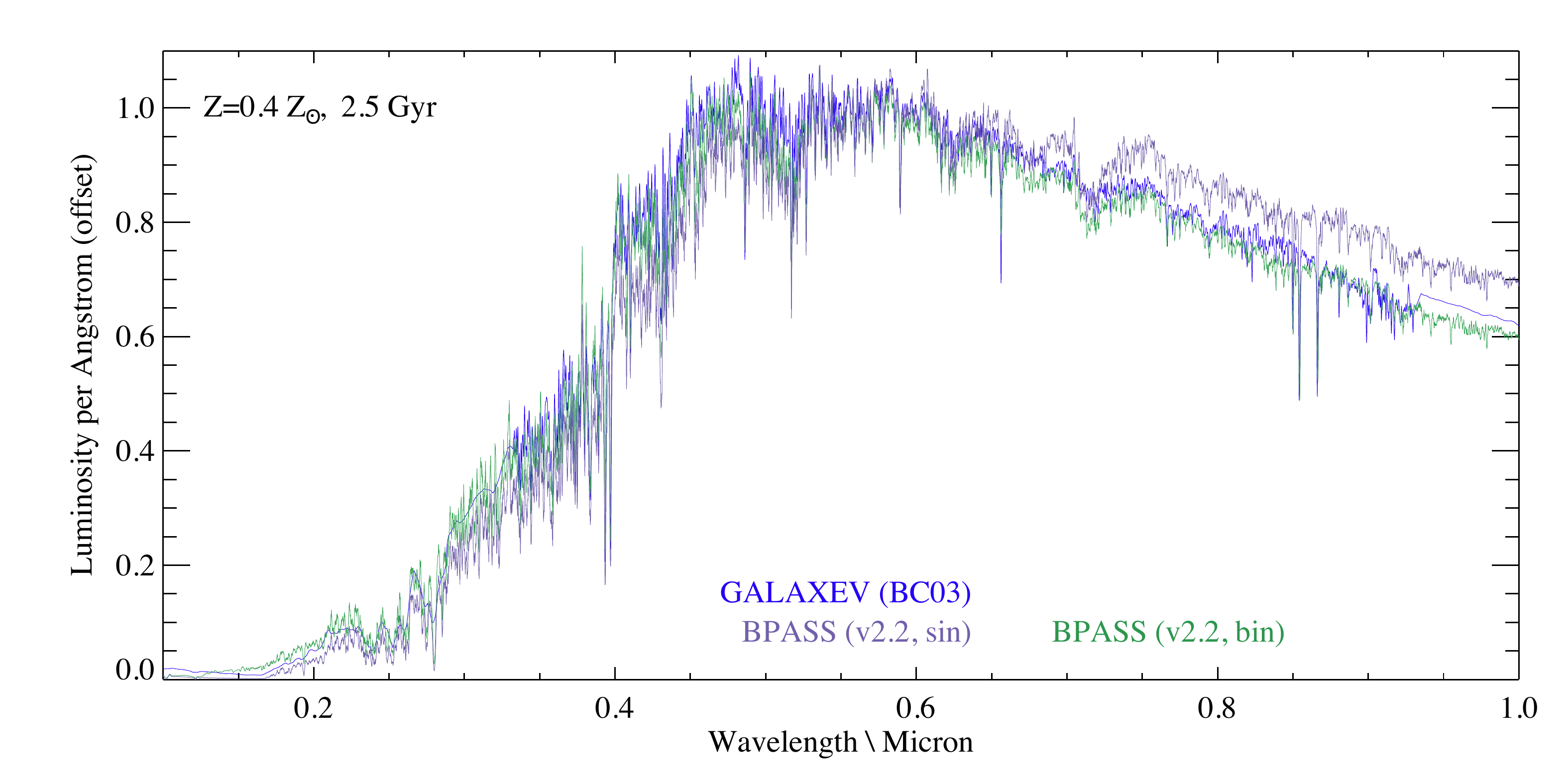}
\caption{BPASS v2.2 spectral energy distributions at 0.4\,Z$_\odot$ for single and binary stellar populations, compared to the equivalent SEDs drawn from the GALAXEV models of Bruzual \& Charlot 2003 (shown in blue) at a single age (2.5 Gyr). The BPASS models have been smoothed for clarity. All models are normalised at 5500\AA.}
\label{fig:bc03old}
\end{center}
\end{figure*}

To evaluate the properties of old stellar populations we introduce a new version 2.2 of the Binary Population and Spectral Synthesis models. BPASS v2.1 and its input physics was described in full detail in Eldridge, Stanway et al. (\citeyear[][hereafter ES17]{2017PASA...34...58E}). The models combine individual detailed stellar evolution models to predict the integrated light of a simple (coeval) stellar population at 51 ages after the initial starburst, ranging from 1\,Myr to 100\,Gyr in increments of log(age/years)=0.1. Models are produced at 13 metallicities, ranging from 0.05 per cent of Solar to twice Solar (where Z$_\odot=0.020$). The input stellar models include both isolated single stars, and treatment of binary star evolution with a range of periods and mass ratios. However BPASS v2.1 was unable to reproduce the properties of integrated stellar populations at ages $>1$\,Gyr due to its focus on the massive end of the stellar initial mass function, which dominates at early times, and the resultant poor sampling of the low mass stellar population. 

Here we describe modifications made in the new v2.2\footnote{From version 2.1 (Kiwi) onwards, BPASS models are associated with version names. Version 2.2, with its focus on old stellar populations, is named ``Tuatara" after the New Zealand-native reptile which has been identified as a {\em living fossil} and which retains characteristics of its origin as a species $\sim$220\,Myr ago.} models which are targeted at refining predictions for old stellar populations.

\subsection{New Stellar Models}\label{sec:stars}

BPASS v2.2 incorporates an extended, more finely-sampled, grid of low mass stellar evolution models, which dominate at late times. We modify the grid described in ES17, adding single star models such that the grid samples: 
\begin{enumerate}
\item  every 0.02\,M$_\odot$ in the range 0.1-2.0\,M$_\odot$,
\item every 0.05\,M$_\odot$ in the range 2.0-3.0\,M$_\odot$, 
\item every 0.1\,M$_\odot$ in the range 3.0-10\,M$_\odot$,
\item every 1\,M$_\odot$ in the range 10-100\,M$_\odot$, and 
\item every 25\,M$_\odot$ in the range 100-300\,M$_\odot$.
\end{enumerate}

Due to computational demands, our binary grid is necessarily coarser. Nonetheless, we also add additional binary evolution models with a refined grid of secondary star masses which samples:
\begin{enumerate}
\item every 0.1\,M$_\odot$ in the range 0.1-2.1\,M$_\odot$,
\item masses 2.3, 2.5, 2.7, 3.0, 3.2, 3.5, 3.7\,M$_\odot$,
\item every 0.5\,M$_\odot$ in the range 4-10\,M$_\odot$,
\item every 1\,M$_\odot$ in the range 10-25\,M$_\odot$, and
\item masses 30, 35, 40, 50, 60, 70, 80, 100, 120, 150, 200, 300, 400, 500\,M$_\odot$.
\end{enumerate}
The upper end of this mass distribution accounts for the fact that post-merger and mass-transfer binaries will sample these models, and that our primary grid extends to 300\,M$_\odot$. We do not model below the hydrogen burning mass limit (i.e. we exclude brown dwarfs). The single-star models below a mass of approximately 2\,M$_{\odot}$ have all been recomputed to improve handling of the helium flash. Typically these models ignite helium in a degenerate core and experience a short-lived episode of runaway core thermonuclear burning known as the helium flash, before core expansion ends the runaway reaction. Our code is not currently able to evolve a stellar model through this rapid evolutionary phase. Therefore we use an initially more massive stellar model, which ignites helium in a less degenerate and lower mass star, to create a substitute model which is burning helium in the core with the same helium core mass and hydrogen envelope mass as the original star, in order to continue the evolution. We match these models at the point of helium ignition. We have not yet calculated similar composite models for the binary evolution model set because the number of such models failing at the helium flash is smaller due to binary interactions, and also because the binary fraction at these masses is relatively small.

Our new models incorporate an improved prescription for the rejuvenation of secondary stars, as a result of mass transfer and subsequent short-lived rotational mixing events \citep[e.g.][]{1998A&A...334...21V,1999NewA....4..173V}, at the population synthesis stage. This now accounts for the evolutionary state of the secondary star at the epoch where the mass transfer occurred. The effects of rejuvenation on the star are reduced if the mass transfer occurs towards the end of hydrogen core burning. 

For a primary model in which mass transfer occurs from a primary star with a hydrogen rich envelope, we determine the time $t_{\dot{M}}$ when mass transfer first occurs. We then use the hydrogen-burning lifetime from our single star models to estimate the hydrogen-burning lifetime of the secondary star with initial mass $M_{\rm 2i}$, which we designate $t_{\rm 2i}$, and for a star with its final, post-mass-transfer mass $M_{\rm 2i}$, which yields lifetime $t_{\rm 2f}$. Limiting $t_{\dot{M}}$ to be not greater than $t_{\rm 2i}$, we  use these parameters to reset the starting age for the secondary model due to rejuvenation as,
\begin{equation}
t_{\rm rej} = t_{\dot{M}} - t_{\rm 2f} \frac{M_{\rm 2i} t_{\dot{M}}}{M_{\rm 2f} t_{\rm 2i}}.
\end{equation}
This now means that secondary stars that undergo accretion do not have their ages set to unrealistically old ages (which was the case for low mass stars in our v2.1 models). 

We have also revisited the implementation of Asymptotic Giant Branch evolution in our models. AGB stars start to substantially influence the integrated light of a stellar population at ages of a few hundred Myr. There are large uncertainties associated with the stellar wind mass loss rates that should be used at this stage, and their effect on the subsequent evolution \citep[see e.g.][]{2014ApJ...790...22R}. We now identify when our models enter the AGB phase, assuming this is when the mass enclosed by hydrogen and helium burning shells move to within 0.03\,M$_{\odot}$ of each other and that the helium core is less than 1.3\,M$_{\odot}$. We chose this value after reviewing our detailed models. 

Rather than following the thermal pulses in detail, which is important for nucleosynthesis by AGB stars, we make several approximations and use results from \citet{hurley2000} to continue the model evolution. In future, this will provide flexibility to vary the assumptions of AGB evolution without having to recalculate a large number of detailed models each time. During the thermal pulsing phase the computational time required to track detailed models is significant. In the approximate approach, we fix the core mass and luminosity of our model at the values when the AGB phase is entered. We then switch to using the mass-loss rates of \citet{2005ApJ...630L..73S,2007A&A...465..593S}. While several other AGB wind prescriptions now exist in the literature, we defer investigation of whether other mass-loss rate schemes produce any differences to future work. We then scale our model time-step adaptively such that 5 per cent of the total stellar mass is removed by the derived mass-loss rate in any given time interval. We repeat the process until the final envelope mass decreases below 0.1\,M$_{\odot}$, leaving only the degenerate core, and then assume the model has become a carbon-oxygen white dwarf. 
This change removes those AGB models that became unphysically over-luminous in BPASS v2.1 due to the previously-implemented mass-loss rates not taking account of the AGB superwind. We note that this will likely not capture the full luminosity evolution of these systems, but provides a reasonable first estimate for their contribution to an integrated stellar population. It will also enable us to implement more complex AGB prescriptions, such as those used by \citet{2007A&A...466..641I}, in future in order to include the complex effects of third dredge-up on the stellar properties.

To model the evolution of white dwarfs formed as the result of thermal pulses, we again employ the method of \citet{hurley2000} to derive the luminosity and radius of the white dwarf on the cooling track. White dwarfs forming as a result of binary interactions are already modelled in full in our detailed code. We terminate the evolution when the total age reaches 100\,Gyrs.

\subsection{New Binary Distribution Parameters}
\label{sec:binary_params} 

BPASS v2.1 and earlier versions used a simple distribution of binary periods and mass ratios that was based on behaviour in high mass stellar binaries \citep{2013A&A...550A.107S}, but applied at all masses. All stars were considered to be in binaries, with their interaction fraction determined by the period distribution and the details of stellar evolution in each case. This led to an unphysically high interacting binary fraction amongst low mass stars at late times, and hence our populations never became dominated by the old, red stars seen in quiescent stellar systems.

In BPASS v2.2 we implement a new set of weighting parameters for binary fraction, period and mass ratio distribution, as a function of the mass of the primary star. We base our new binary parameter distribution on those determined by the detailed study of \citet[][hereafter MS17]{2017ApJS..230...15M} which analysed a large compilation of observational data to determine the parameters of early-type and main sequence binaries.

We adopt the multiplicity statistics presented in table 13 of MS17 with only minor alterations to accommodate our model parameter space: 
\begin{enumerate}
\item The maximum period in our binary models is log(Period/days)=4. We find that a negligible fraction of the binary stars at this limit interact in the lifetime of the Universe. As a result, we treat binaries with a larger period as effectively single in our population synthesis;
\item Our model prescription does not implement precisely equal mass binaries, due to computational limitations. We use the MS17 values for the {\it twin fraction} (i.e. number of near equal-mass binaries) to boost the number of binaries in our highest mass ratio bin ($M_2/M_1$=0.9);
\item We interpolate the derived relations between the mass bin centroids reported by MS17, and extrapolate the relations in the range 0.1-0.8\,M$_\odot$, below the mass limit considered by MS17. The resultant binary fraction is broadly consistent with that determined by \citet{2006AJ....132..663B} for very low mass stars. 
\item We apply the parameters for O-type stars, as derived by MS17, to Very Massive Stars ($>100$\,M$_\odot$). Note that the binary fraction for stars with M$>16$\,M$_\odot$ exceeds 90 per cent, and that for M$>60$\,M$_\odot$ is unity.
\end{enumerate}

To accommodate this set of parameter distributions, our integrated stellar populations now include both binary and single star models, in proportions given by the initial mass function and binary fraction distribution. For low mass stars ($<$2\,M$_\odot$) the population is dominated by single stars (60 per cent) and binaries which do not interact in the age of the Universe as figure \ref{fig:intfrac} demonstrates.  Hence our emphasis on improving the single star model mass grid in this version of BPASS.

\begin{figure*}
 \includegraphics[width=0.66\columnwidth]{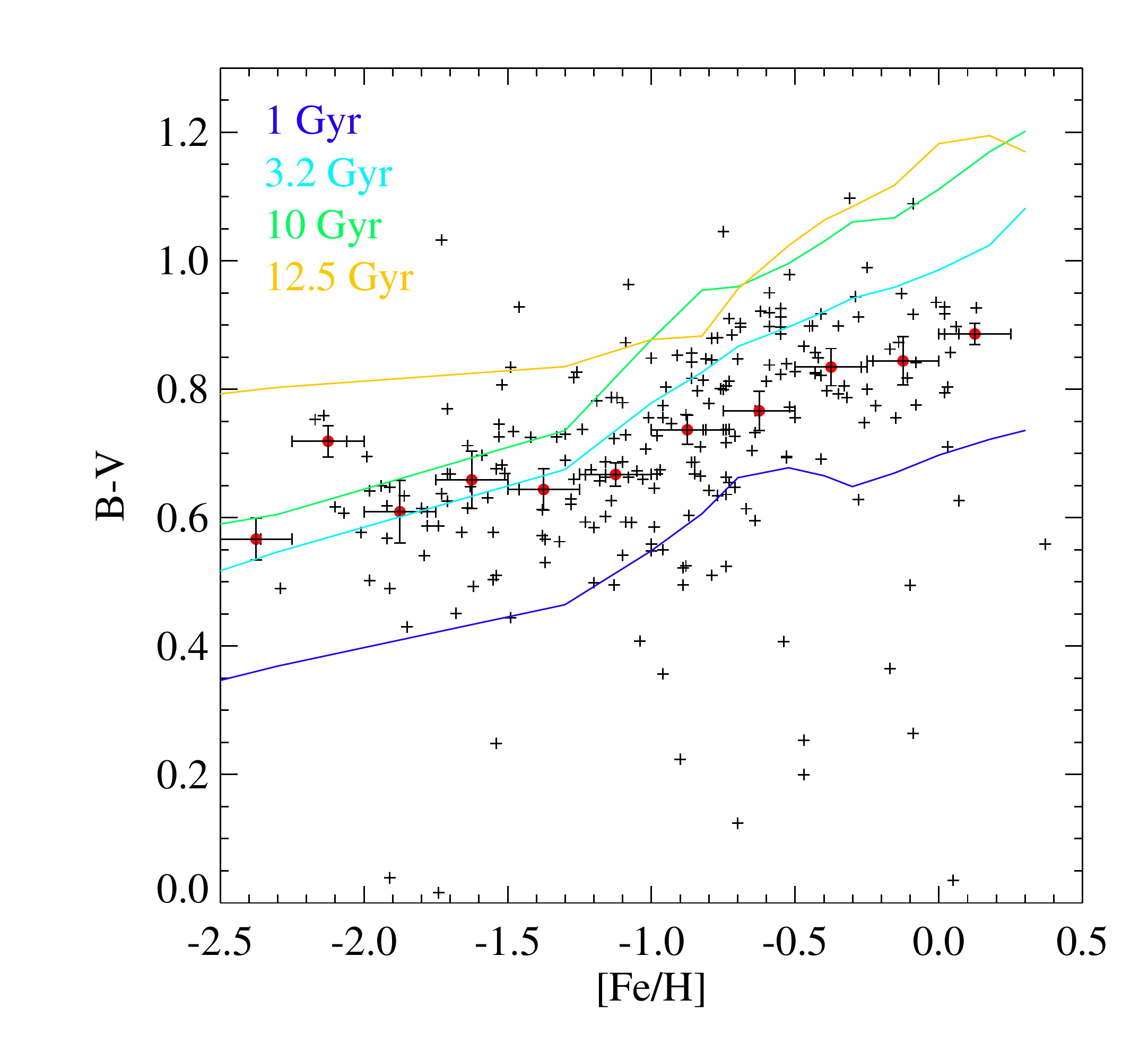}
 \includegraphics[width=0.66\columnwidth]{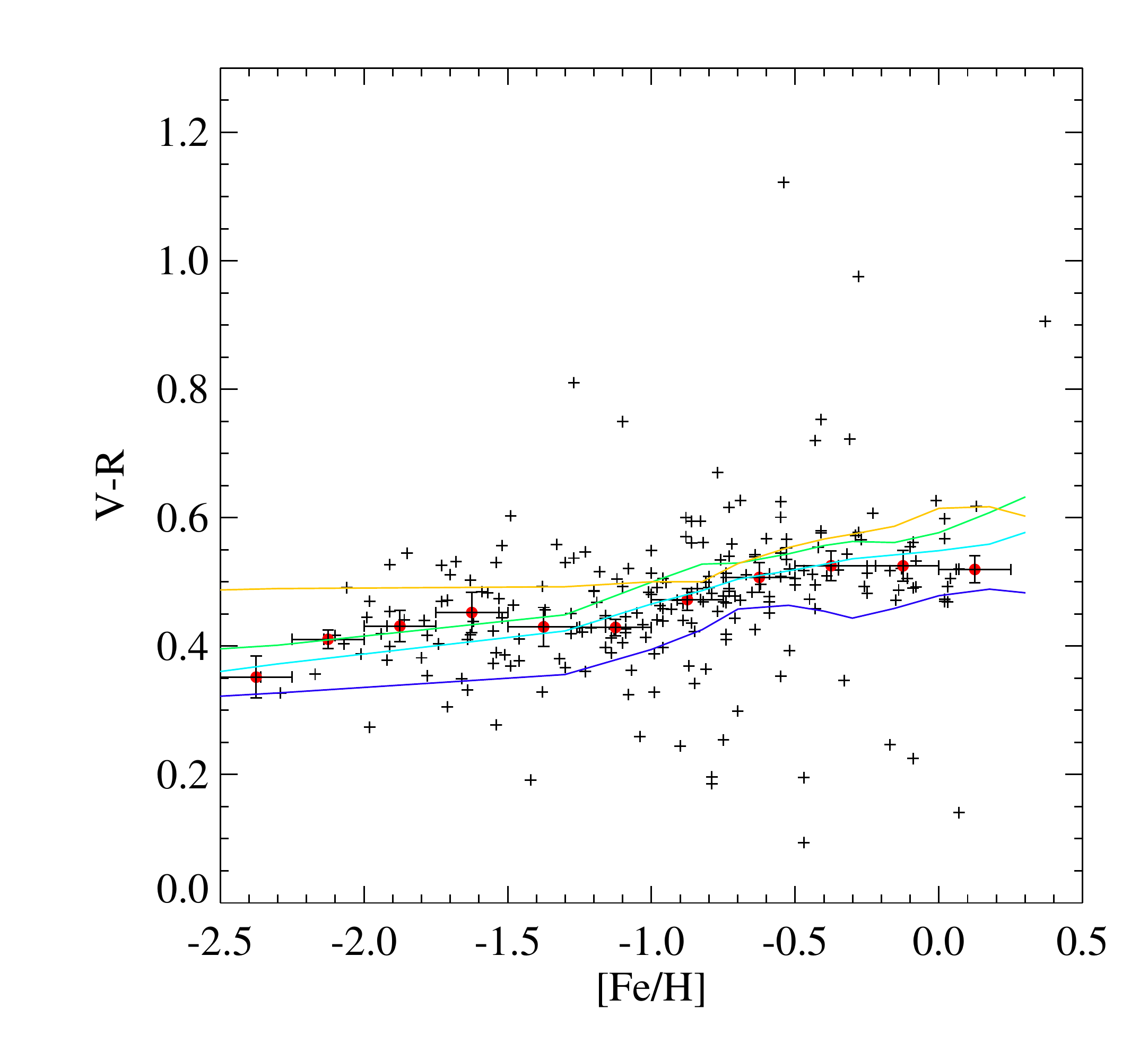}
 \includegraphics[width=0.66\columnwidth]{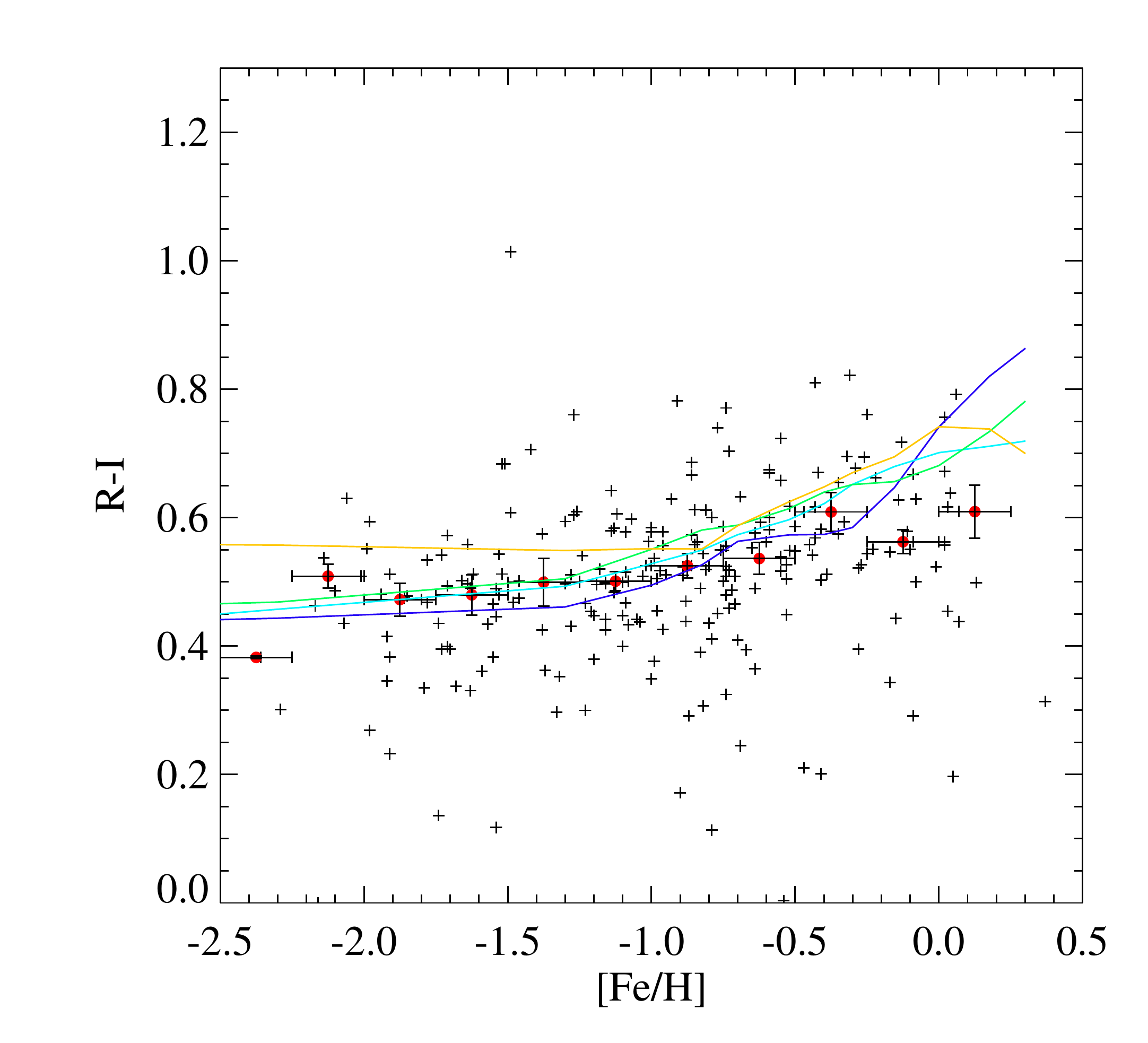}
 \includegraphics[width=0.66\columnwidth]{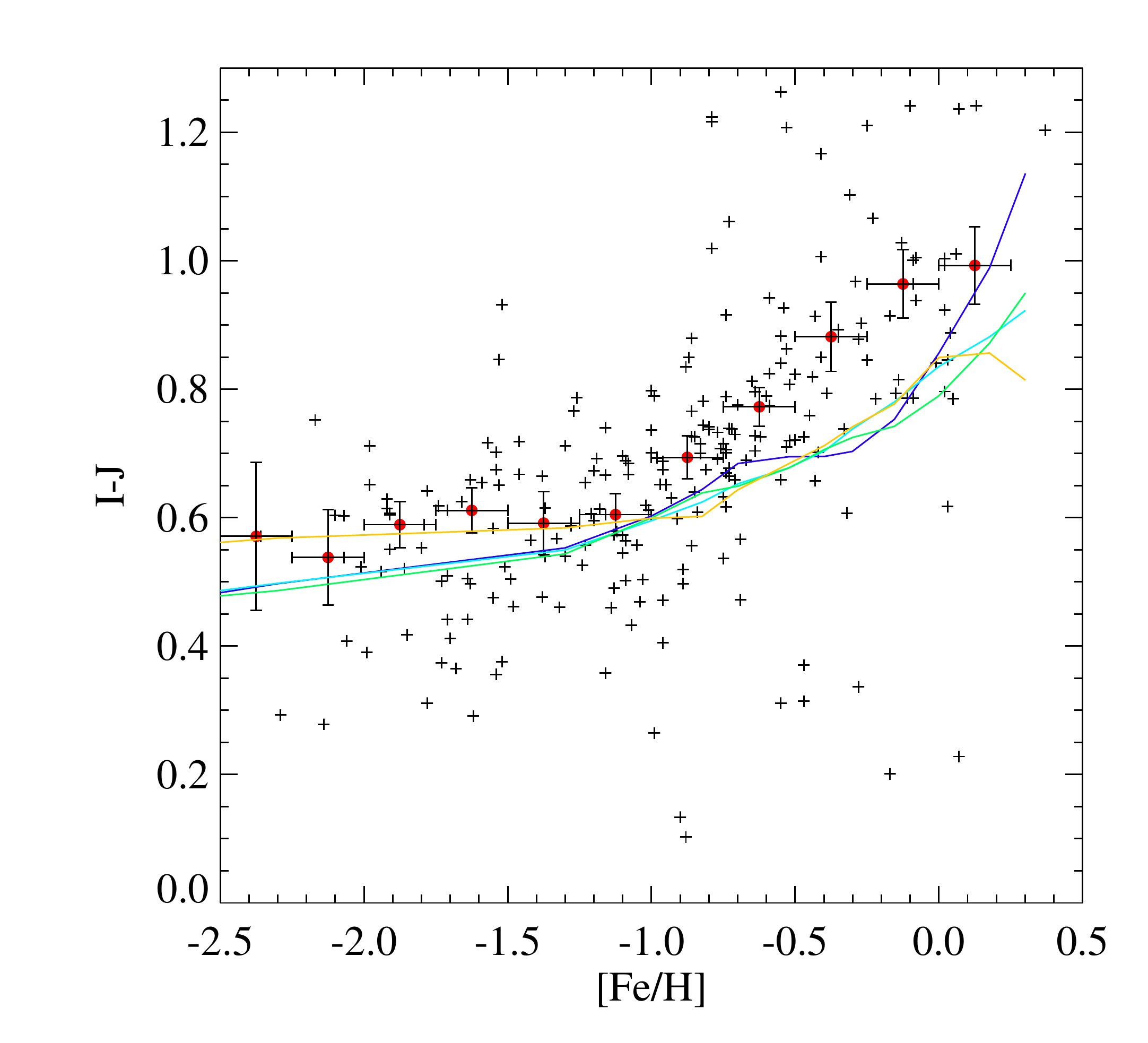}
 \includegraphics[width=0.66\columnwidth]{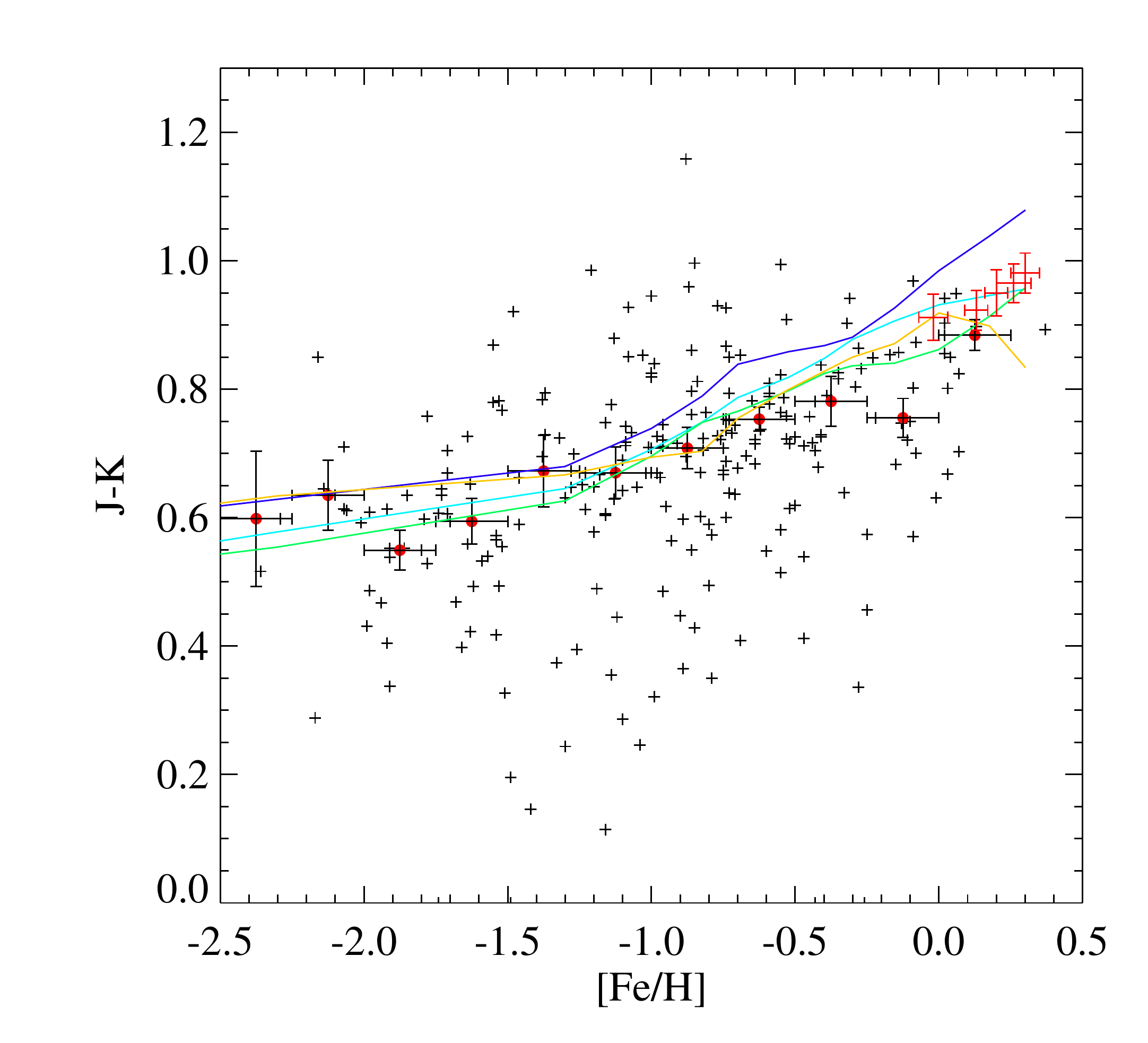}

 \caption{Colours of globular clusters from the sample of \citet{2004A&A...415..123P}. The red points in J-K are the typical colours of eliptical galaxies from \citet{1538-3881-152-6-214}. The GC colours have been corrected for reddening by foreground dust extinction. The colours are given in Vega magnitudes. The mean and standard deviation in metallicity bins of 0.2 dex are overplotted  as filled circles. Solid lines show the predicted colours of BPASS v2.2 models at four stellar population ages. Typical uncertainties on observed metallicity are $\pm$0.3, while photometric and calibration uncertainties are of order 0.1\,magnitudes. \label{fig:gc_cols}}
\end{figure*}

\subsection{New Atmosphere models}\label{sec:atmospheres}

For the purposes of spectral synthesis, we complement our new stellar evolution models and binary parameter distributions by a revised set of stellar atmosphere models. 

We continue to use the Potsdam Wolf-Rayet \citep[PoWR,][]{2015A&A...577A..13S} models for WR stars and our custom set of {\sc WMBASIC} models for massive O stars, as discussed in ES17. However, at metallicities $Z=0.001$ (0.05\,$Z_\odot$) and above, we substitute the BASELv3 atmospheres used for main sequence and giant stars in BPASS v2.1 with a new grid of higher resolution stellar atmosphere models provided by C. K. Conroy \citep[known as the CKC14 models, see also ][]{2014ApJ...780...33C}.  These comprise 750 individual stellar atmospheres per metallicity in the range T=2500-25,000\,K and log($g$)=-1 to 5.5. Where necessary, we interpolate in both temperature and surface gravity between grid points in this atmosphere model set. The initial grids supplied in the set have metallicity mass fractions of $Z=0.0001$, 0.0004, 0.0013, 0.0042, 0.0134 and 0.0424. We have interpolated in $\log(Z)$ between these atmosphere models to create a model set in our own metallicity grid. Interpolation is performed between matching temperature and gravity grid points (i.e. for fixed physical parameters) in adjacent metallicity atmosphere grids. 

We also now incorporate a grid of white dwarf atmospheres in our spectral synthesis (where previously the contribution of these was approximated as a hot black body).  This addition has important consequences for the stellar population at late ages where absorption in the Balmer series of Hydrogen is commonly used as a stellar population diagnostic, since this can have a strong contribution from white dwarfs.

We adopt the synthetic spectral grid of \citet{2017ApJS..231....1L}. The spectral grid spans 17000\,K$<$Teff$<10^5$\,K and 7.0$<$log($g$)$<$9.5, and the spectra cover a wavelength range from 900\,\AA\ to 2.5\,$\mu$m. Outside this wavelength range, we extend the spectra with a simple black body, matched to the white dwarf temperature. Note that this grid only includes DA (i.e. pure hydrogen) white dwarf atmospheres. Given that a few percent of white dwarfs are classed as DB (with strong helium features), we expect to slightly underestimate the strength of helium absorption lines at late times.

\subsection{New Initial Mass Functions}

We have generated BPASS stellar population models for the seven initial mass functions (IMFs) described in ES17. These are broken power law functions, with a lower IMF slope, $\alpha_1$, applied between 0.1\,M$_\odot$ and $M_1$  and an upper slope, $\alpha_2$, applied between M$_1$ and M$_{\rm max}$, i.e.
\[
N(M<M_{\rm max}) \propto \int^{M_1}_{0.1}{\left(\frac{M}{M_\odot}\right)^{\alpha_1}}\,dM 
\]
\begin{equation}
\hspace{2.5cm} + \ M_1^{\alpha_1} \int^{M_{\rm max}}_{M_1}{\left(\frac{M}{M_\odot}\right)^{\alpha_2}}\,dM
\label{eqn:imf}
\end{equation}

For BPASS v2.2 we also calculate models for two additional IMFs which, rather than acting as a broken power law instead introduce an exponential cut-off at masses below 1\,M$_\odot$ and a high mass slope of -2.3, following the prescription of \citet{2003PASP..115..763C}. We calculate this model with upper mass cut-offs of 100 and 300\,M$_\odot$. All of these IMFs are summarised in table \ref{tab:imfs}. 

The results presented here are based on our simple stellar populations (i.e. a single age starburst) and are given for our default broken power-law IMF with slopes of $\alpha_1=-1.3$ for stars with 0.1-0.5\,M$_\odot$ and $\alpha_2=-2.35$ for 0.5-300\,M$_\odot$ (model ``135\_300" in table \ref{tab:imfs}).

BPASS v2.2 output SEDs at a range of ages (1-10\,Gyr) are shown in the appendix for Solar and 0.4 Solar metallicities and all nine IMFs. For old stellar populations, the dominant effect of the IMF is to alter the normalisation (i.e. the total number of low mass stars for a given initial total mass) rather than the spectral shape.

\begin{table}
\begin{tabular}{ccccc}
Model  &  $\alpha_1$             & $\alpha_2$ & M$_1$ & M$_{\rm max}$ \\
       & (0.1\,M$_\odot$ - M$_1$)  & (M$_1$ - M$_{\rm max}$) & &  \\
\hline
100\_100  &  -1.30   &  -2.00 & 0.5 &  100\,M$_\odot$ \\ 
100\_300  &  -1.30   &  -2.00 & 0.5 &  300\,M$_\odot$ \\ 
135\_100  &  -1.30   &  -2.35 & 0.5 &  100\,M$_\odot$ \\ 
135\_300  &  -1.30   &  -2.35 & 0.5 &  300\,M$_\odot$ \\ 
135all\_100  &  -2.35   &  -2.35 & 0.5 &  100\,M$_\odot$ \\ 
170\_100  &  -1.30   &  -2.70 & 0.5 &  100\,M$_\odot$ \\ 
170\_300  &  -1.30   &  -2.70 & 0.5 &  300\,M$_\odot$ \\
chab100  & exp cutoff & -2.3 & 1.0 &  100\,M$_\odot$  \\
chab300  & exp cutoff & -2.3 & 1.0 &  300\,M$_\odot$  \\
\end{tabular}
\caption{The range of stellar initial mass functions (IMFs) calculated in BPASS v2.2. Our default IMF is ``135\_300" and results in this paper are calculated using that IMF unless stated otherwise. ``exp cutoff" indicates an exponential cut-off in the mass distribution following the prescription of \citet{2003PASP..115..763C}.\label{tab:imfs}}
\end{table}

\begin{figure}
\includegraphics[width=0.9\columnwidth]{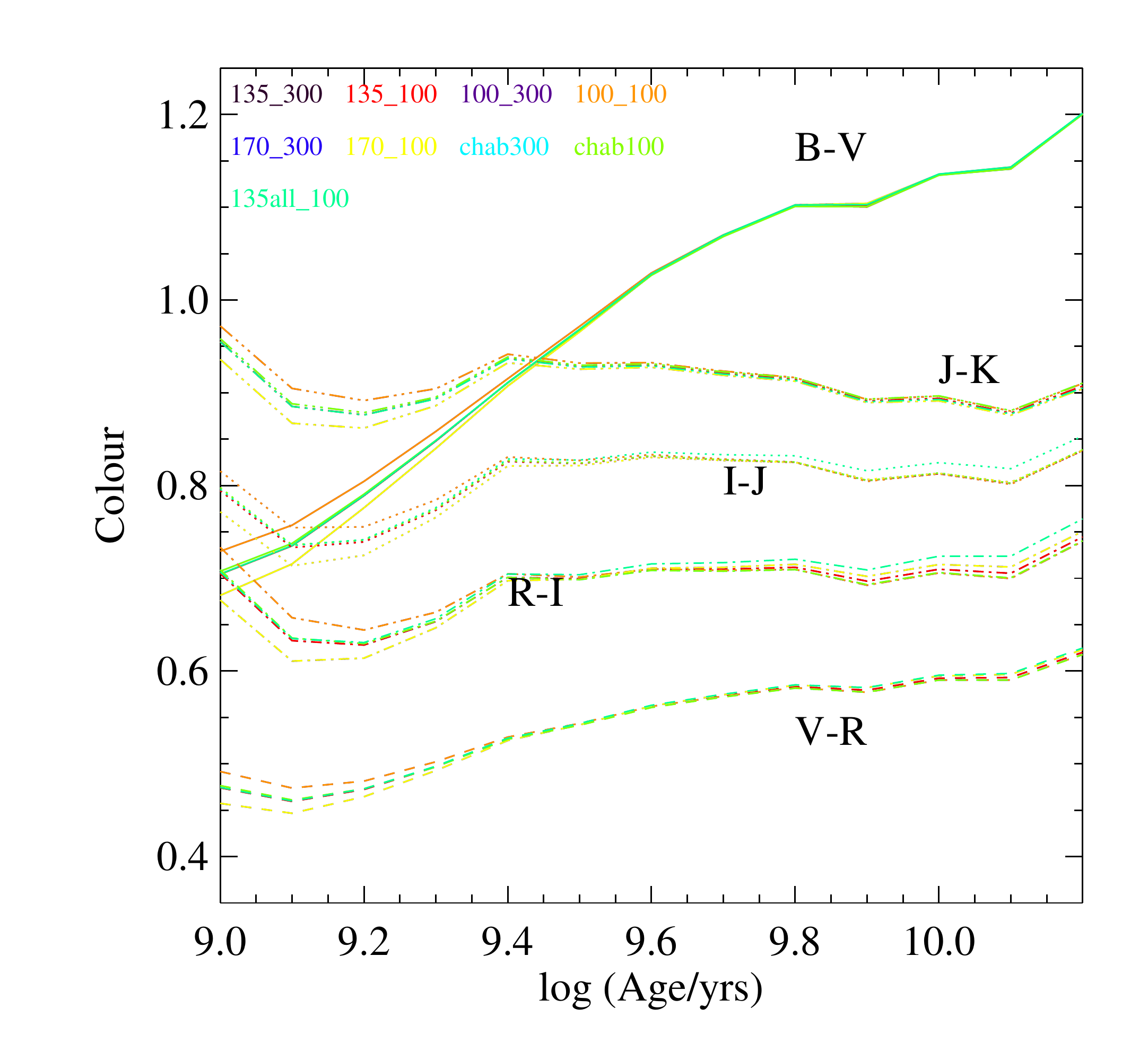}
\caption{The evolution of integrated population photometric colour with age as a function of IMF. Models are shown for all nine IMFs calculated in BPASS v2.2, and at Solar metallicity. The effect of IMF on colour is very small ($<$0.05\,mag) in all cases, but is highest at relatively young ages ($\sim$1\,Gyr).\label{fig:cols_imf}}
\end{figure}

\subsection{Comparison with Other Stellar Population Synthesis Models}\label{sec:comp}

The evolution of the resultant spectral energy distribution for an integrated stellar population is shown in Fig. \ref{fig:bc03comp}, and compared to that of the widely-used GALAXEV models of \citet{2003MNRAS.344.1000B} at two metallicities. For a simple comparison, we use the GALAXEV galaxy template set identified by \citet{2004ApJ...613..898T}. For clarity we present the comparisons separately for our preferred binary parameter distributions (right) and for a population constructed solely from single star evolution pathways (left). The overall evolution of the SED shape is comparable in the two model sets, and at both metallicities, while differences arise from the differences in stellar evolution and atmosphere treatment.

A certain amount of residual jumpiness remains in the models at late ages. Our binary grid remains relatively coarse and slight differences in assumptions regarding the underlying stellar physics with metallicity or in model mass sampling can exchange mature stars between time bins. As a result, our 10\,Gyr and 12.5\,Gyr models, which are adjacent time bins in BPASS models, show some evidence for jumps between adjacent metallicities, due to stars reaching certain evolutionary states in one bin or the other. In fact, this behaviour is unlikely to be physical -- comparison with equivalent models in our far more finely gridded single star evolution set suggests that it a sampling artifact. As a result, we encourage users to smooth our raw models at fixed metallicity and wavelength over adjacent time bins when considering populations with ages beyond 1\,Gyr. The models used in this paper have been smoothed with a rolling boxcar average over two time bins before being plotted.

At ages above 25\,Myr (below which massive binaries dominate the integrated light), the BPASS v2.2 binary and GALAXEV models are very similar from $\sim2000-8000$\AA. The main difference between them arise from the treatment of massive, cool stars (including the AGB), which are visible in the GALAXEV SED at 32\,Myr but do not begin to dominate the integrated light until ages 100-300\,Myr in BPASS v2.2, since binary interactions tend to strip the more massive stars which might otherwise attain this evolutionary state.  At intermediate times, the binary populations redden more rapidly than those of GALAXEV, with the red colours extending into the infrared. This is true to differing extents in both our single star models (demonstrating the differences in our handling of the AGB phase and our adopted atmospheres) and in our binary models.  Unsurprisingly, given that the fraction of very low mass stars interacting in the age of the Universe is very low, the integrated light from our BPASS v2.2 binary population looks much like our matching single star population at very late ages ($>$5\,Gyr), although differences remain, particularly in those regions of the ultraviolet and optical most affected by post-main sequence phases and populated by the products of stripping or accretion processes.

At these later ages, a number of effects compete to produce the final BPASS v2.2 binary spectrum.  To illustrate the differences in the old stellar populations more clearly, we compare both the BPASS v2.2 single and binary models against the equivalent GALAXEV model at Z=0.4\,$Z_\odot$ and an age of 2.5\,Gyr in figure \ref{fig:bc03old}. The differences in the initial mass function (our population is slightly richer in low mass dwarfs at late times than a Chabrier IMF), in adopted handling of the AGB phase and in the stellar atmosphere models used means that our single star population is rather redder than that of \citet{2003MNRAS.344.1000B} at the same metallicity. The effect of binary interactions up to these ages is to suppress some of the luminosity of the AGB population and result in a spectrum which is bluer than that seen in single stars alone, but is in close agreement to the colour and spectral energy distribution of the GALAXEV models through the optical. We stress that this is not the result of fine tuning but rather an unexpected consequence of carefully applying independent aspects of stellar population evolution (atmospheres, IMF, AGB winds, binary interactions) which together produce a result broadly similar to the older models (which are known to perform well in most respects in the local Universe) but which differs in absorption line strengths and in detailed characteristics.

The GALAXEV and BPASS binary models also differ substantially in their predictions for ionizing photon flux (shortward of 912\AA), as extensively discussed in ES17, \citet{2016MNRAS.456..485S} and \citet{2018arXiv180107068X}. At sub-Solar metallicity and late ages BPASS v2.2 predicts redder ultraviolet-optical colours than the GALAXEV models. We note that the stars primarily responsible for this difference are helium stars that are the result of a binary interaction with initial masses between approximately 10 to 20M$_{\odot}$. The impact of these stars on stellar populations has been studied by multiple authors \citep[e.g.][]{2003A&A...400..429B,2015MNRAS.447L..21Z,2018arXiv180203018G}. We defer discussion of the ultraviolet upturn behaviour of BPASS v2.2 models to a future paper (in prep).

Note that the changes made in BPASS v2.2 have not significantly affected the SED or other properties of our integrated stellar populations at ages $<$100\,Myr, while minor changes at ages between 100\,Myr and 1\,Gyr act to make the population redder than our v2.1 models, and also reduce the visible influence of molecular absorption in cool giant star atmospheres on the optical/near-infrared region of the spectrum. The ionizing photon production rate for young stellar populations is entirely dominated by the most massive stars, and its numerical values and metallicity evolution are very similar in this version to those described in ES17 for the v2.1 models.


\begin{figure*}
 \includegraphics[width=0.6\columnwidth]{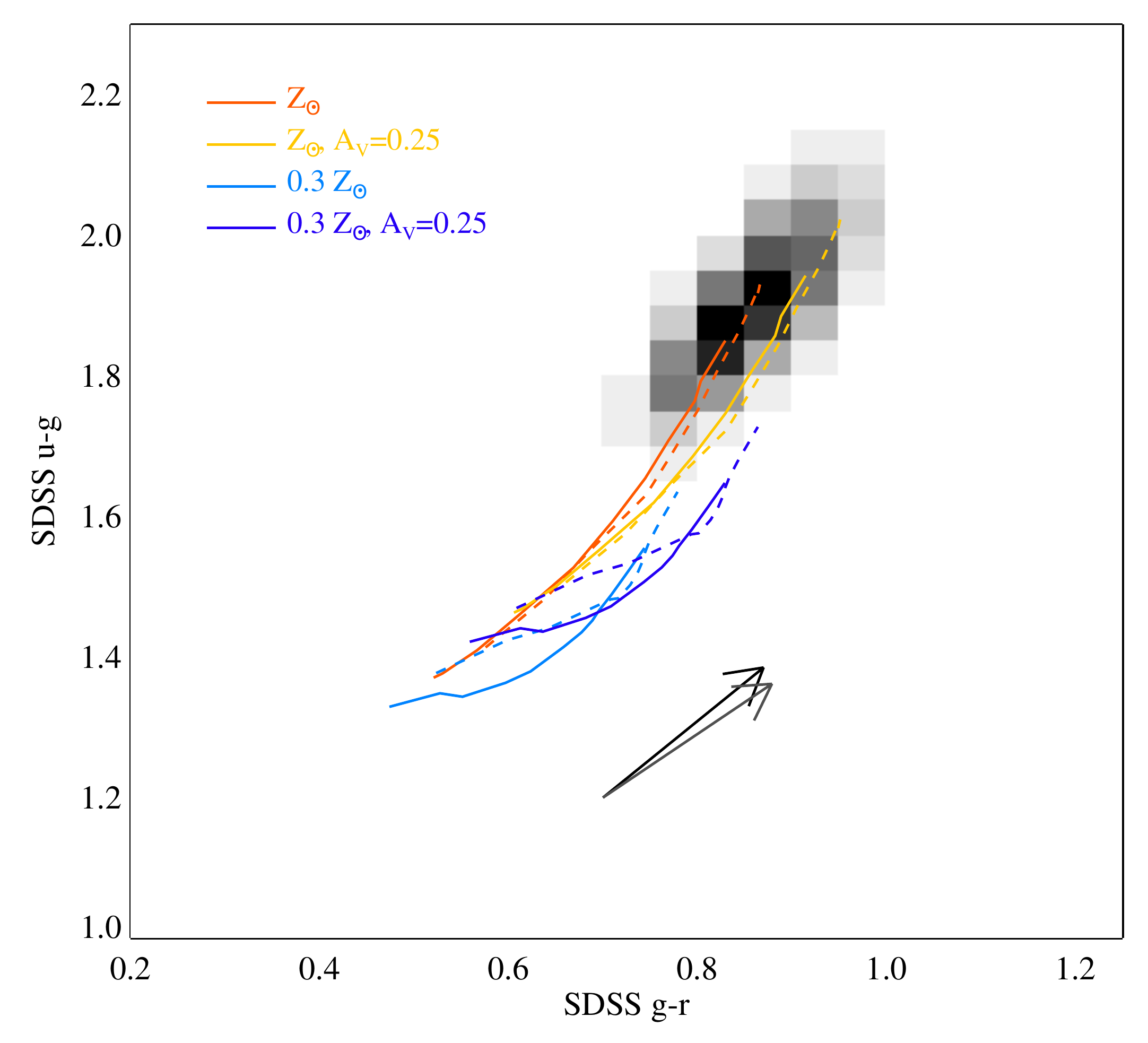}
 \includegraphics[width=0.6\columnwidth]{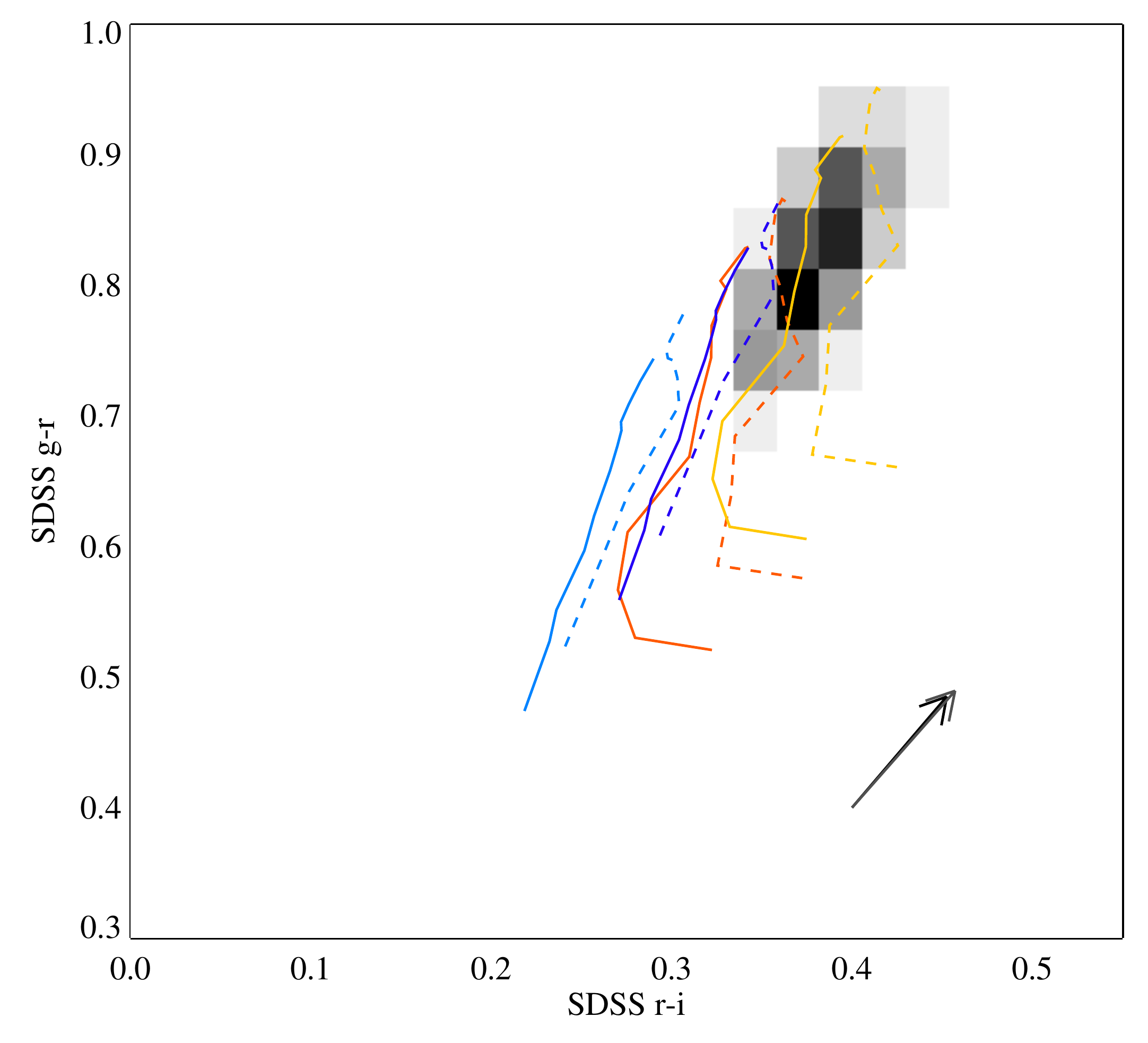}
 \includegraphics[width=0.6\columnwidth]{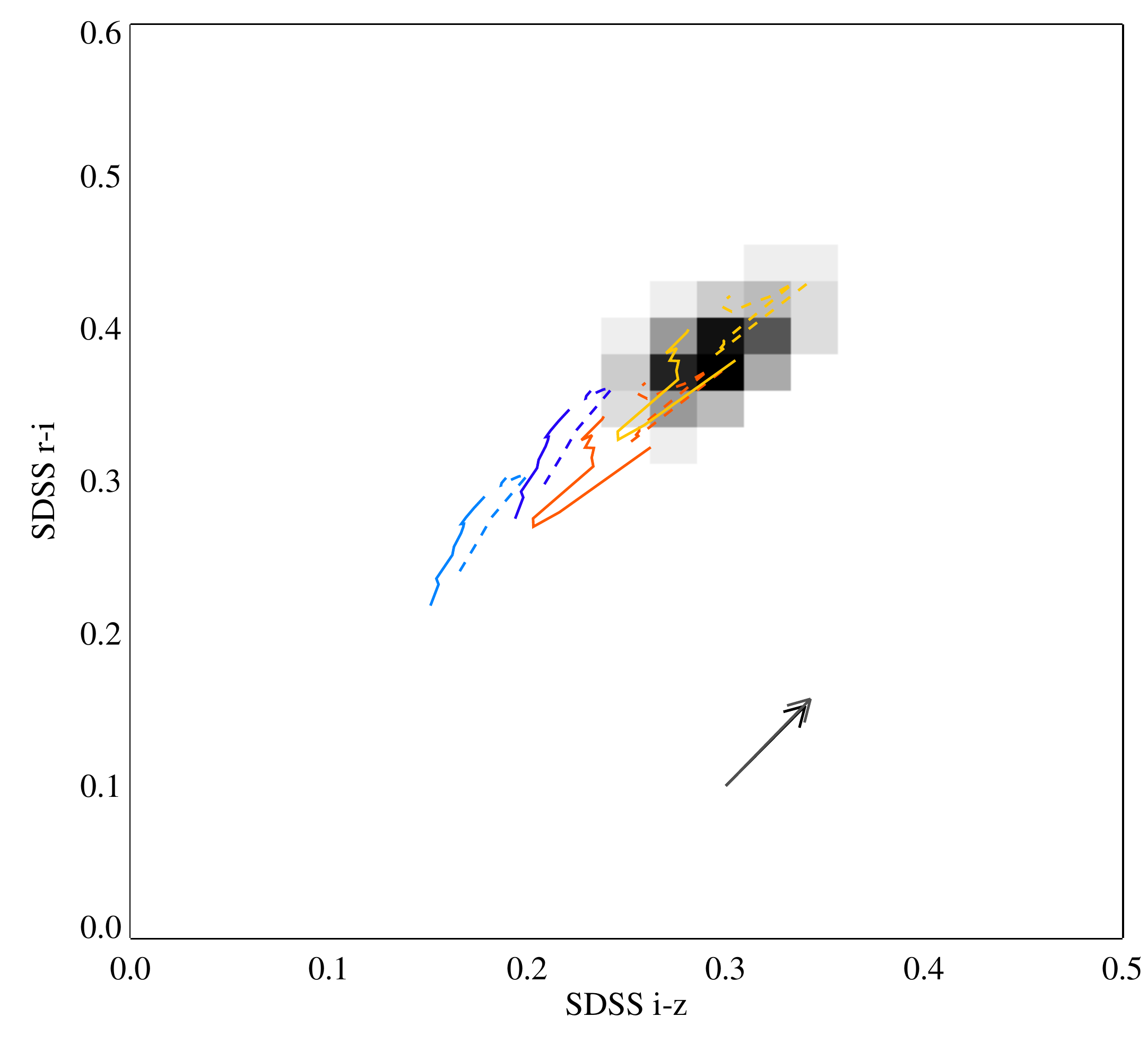}
 \includegraphics[width=0.6\columnwidth]{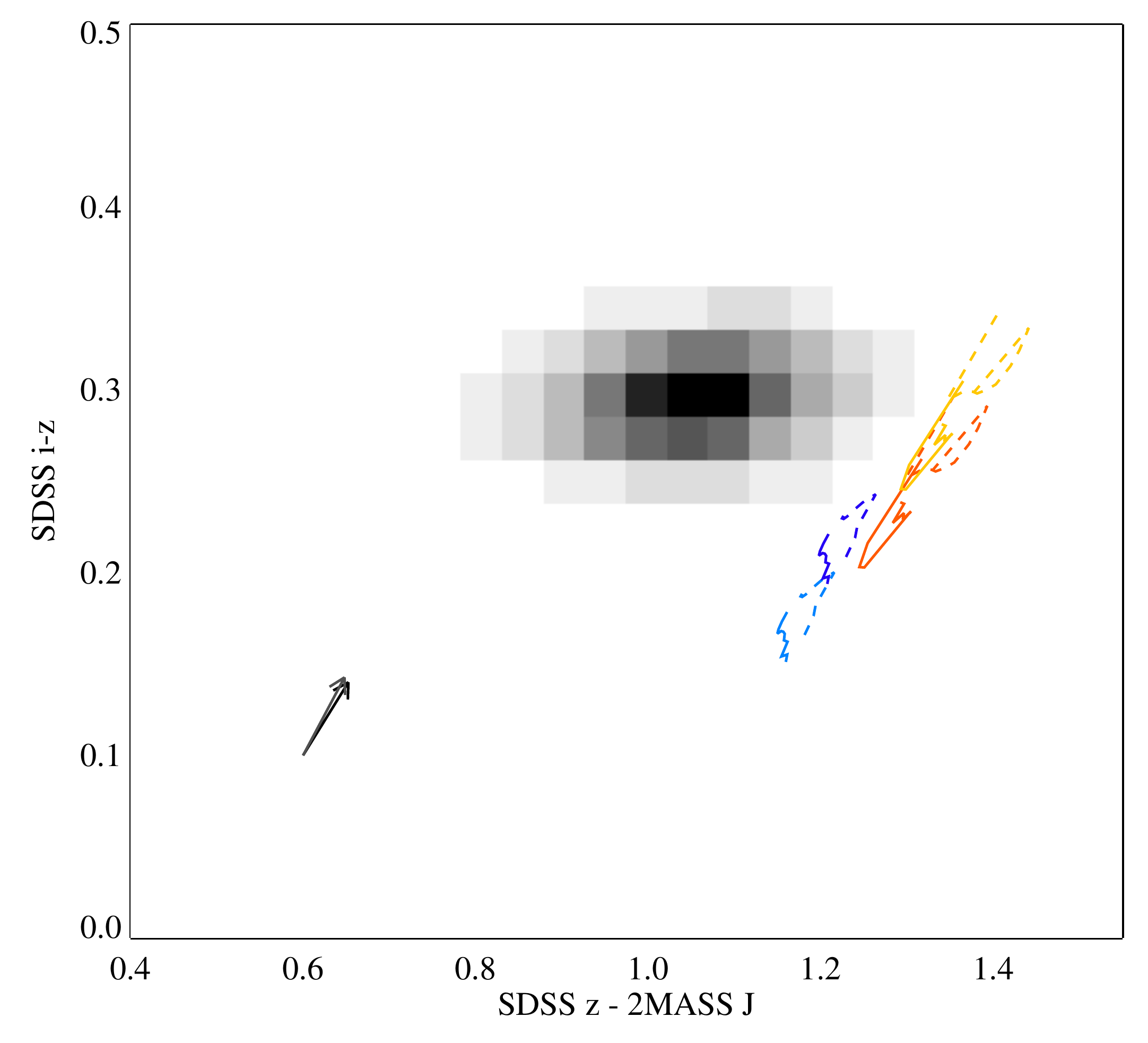}
 \includegraphics[width=0.6\columnwidth]{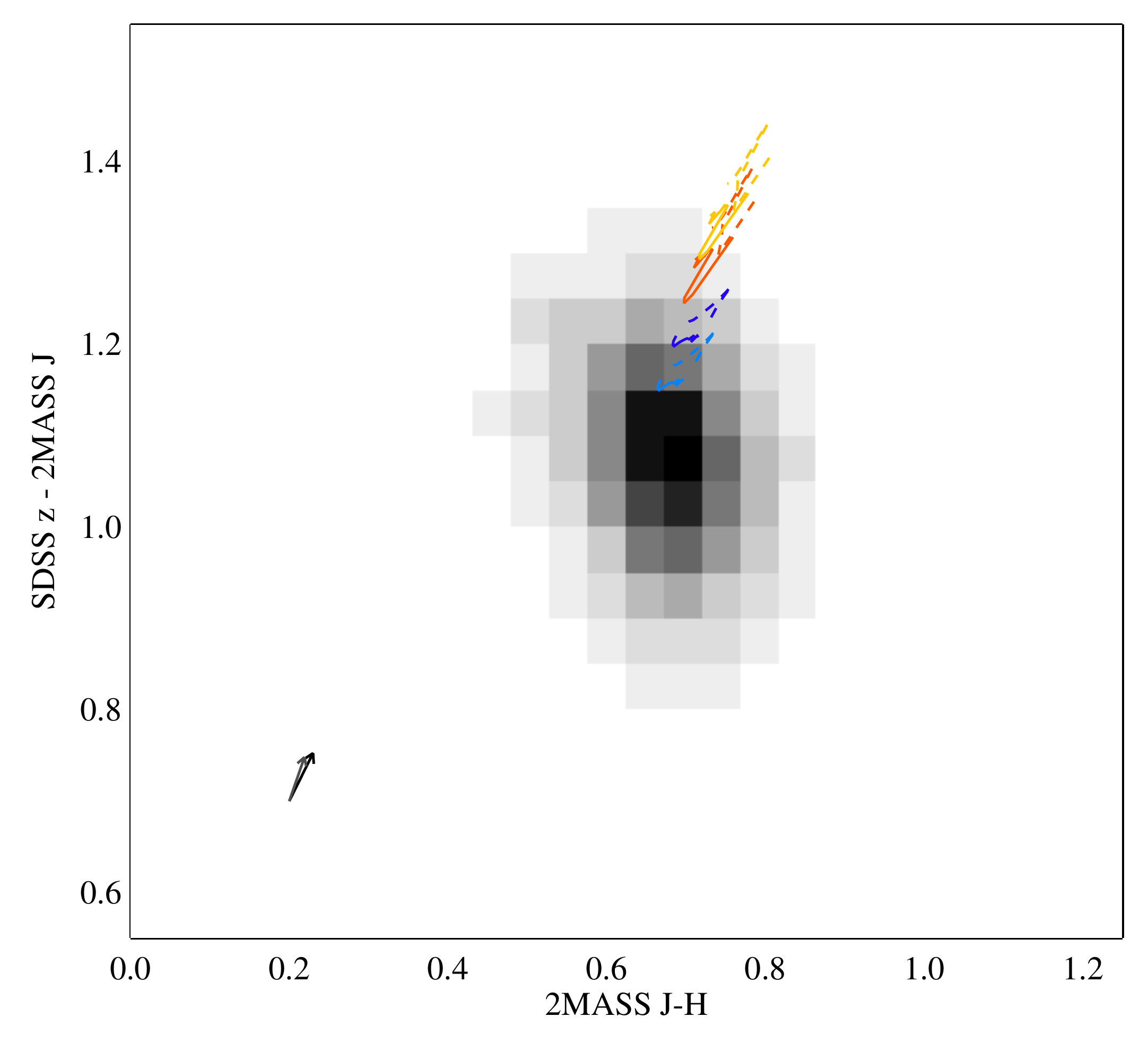}
 \includegraphics[width=0.6\columnwidth]{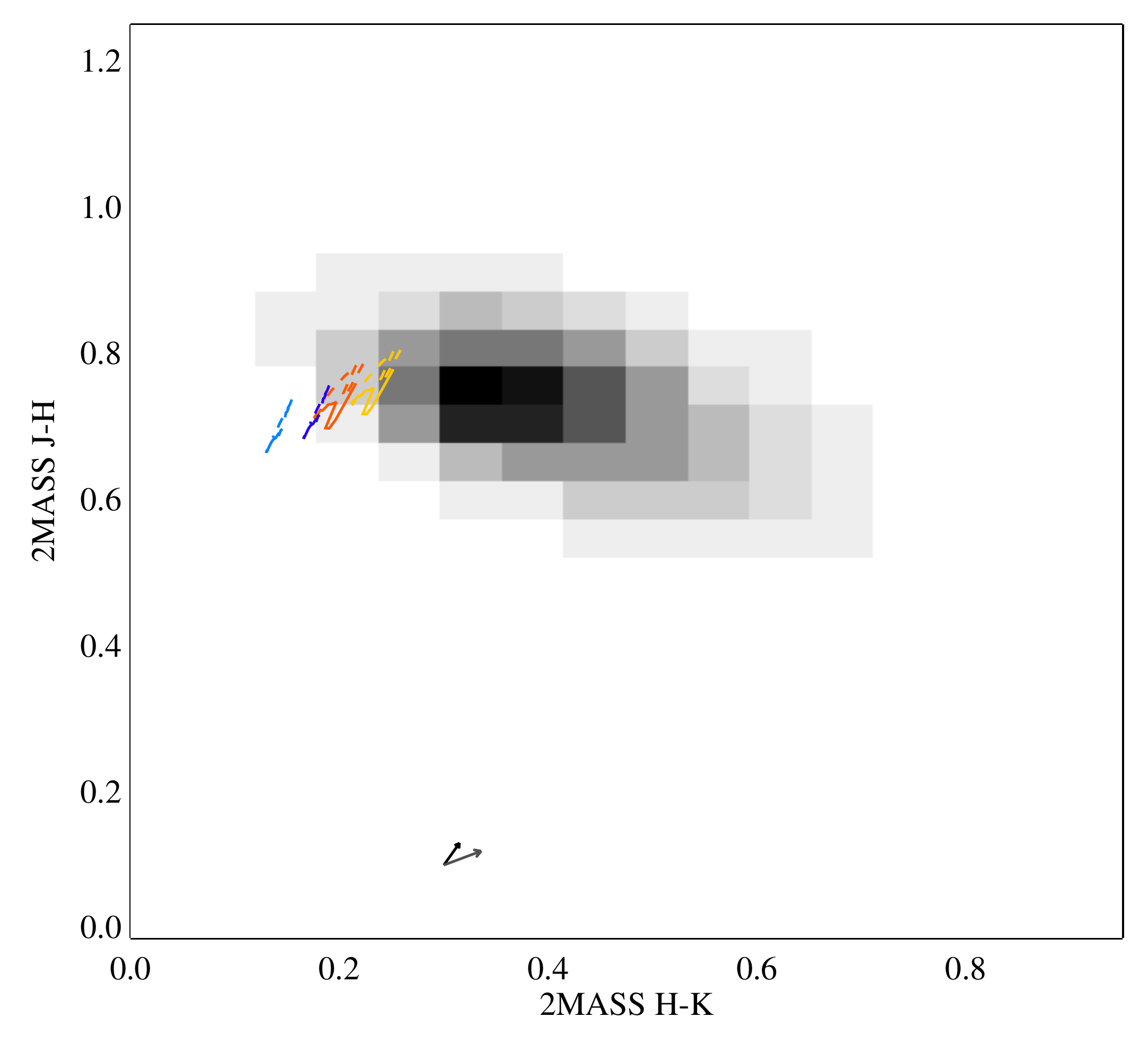}

 \caption{Colours of quiescent galaxies selected from the Sloan Digital Sky Survey. SDSS magnitudes are in the AB system, while source-matched 2MASS catalogue is in Vega magnitudes. Arrows indicate the shift in colour that would result from a reddening $A_V$=0.5\,mag, in black assuming the Fitzpatrick (1999) Milky Way extinction law, extended into the infrared by Indebetouw et al (2005), and in grey assuming the \citet{2003ApJ...594..279G} extinction law derived for the LMC. Overplotted tracks indicate the evolution of BPASS v2.2 models from 1\,Gyr to 12.5\,Gyr in age for single star models (dashed) and our preferred binary models (solid), at metallicities of $Z_\odot$ (red) and 0.3\,$Z_\odot$ (blue). We also show the colours of models assuming an extinction $A_V$=0.25\,mag and the LMC extinction law.\label{fig:sdss_cols}}
\end{figure*}

\section{Reclassifying Old Stellar Pops}\label{sec:obs}

\subsection{Photometry}\label{sec:obs_phot}

The most straightforward observable property of an old stellar population is its photometric colour at optical wavelengths. While this may be somewhat modified by wavelength- and composition-dependent dust extinction, it reflects the underlying shape of the integrated light stellar spectral energy distribution. Here we consider the physical properties suggested by BPASS v2.2 for the old stellar populations in two classes of objects:  globular clusters and elliptical galaxies. We calculate synthetic photometry by convolving a simple stellar population SED model with the appropriate photometric filter\footnote{retrieved from the SVO Filter Profile Service \citep{2012ivoa.rept.1015R}}.  In each case we use the photometry and metallicities given by the original authors as our observational data set, rather than rederiving them, and we take a metallicity mass fraction $Z=0.020$ as our canonical Solar metallicity.

\subsubsection{Globular Clusters}

We first consider the photometric colours of a sample of unresolved globular clusters drawn from seven early type galaxies in order to probe a range of ages and metallicities \citep{2004A&A...415..123P}. 

In Fig. \ref{fig:gc_cols} we compare their photometric properties to those of our models at ages ranging from 1\,Gyr to 12.5\,Gyr (log(age/years)=9-10.1). The data were originally obtained at the ESO Very Large Telescope using the FORS2 and ISAAC instruments. In each case, photometry is given on the Vega system, in the appropriate Bessel and Johnson-Cousins filter sets. We correct the GC photometry for foreground dust extinction using the estimates made by the original authors and a Galactic dust extinction law. We note that any variation in local dust law or extinction estimates will add scatter to the observed distribution, although this is expected to be relatively low in such systems. We also show the mean and standard deviation for the population in bins of 0.2\,dex in metallicity.

As the figure demonstrates, the agreement between BPASS v2.2 old stellar population models and the data sample is good, with the cluster population consistent with ages $>$1\,Gyr in all colours. In the optical bands, there is a good match to the models at ages $\sim5-10$\,Gyr. BPASS models also provide a good match to the observed data in the near-infrared $J-K$ colour, but somewhat underestimates the redness of the population at high metallicities in $I-J$ (by $\le 0.1$\,mag). This suggests that further investigation of AGB evolution (which affects this colour strongly) may be required in future to refine these models.

Globular clusters are relatively small systems and may have very low total masses \citep[$10^4-10^6$\,M$_\odot$, see][]{2006ARA&A..44..193B}. A consequence of this is that the initial mass function may not be fully sampled, and in particular that high mass stars can be either under or over represented relative to their less massive siblings (i.e. where an IMF may predict 0.3 stars at 100\,M$_\odot$, the actual number must necessarily be zero or one). This stochastic variation can, in theory, lead to differences in the stellar populations of globular clusters with the same total mass, age and metallicity \citep[see ][for further discussion]{2012MNRAS.422..794E}. Given that the effects of stochastic IMF sampling are strongest at the short-lived, high mass end of the population (where the overall number of stars is smaller), we expect this to be a larger issue for young globular clusters, observed during their formation at high redshift, than for the old clusters under discussion here. Nonetheless some effects of stochasticity may remain, in the contribution of stars rejuvenated by mass transfer from a companion, or from other long-lived evolutionary states. 

Stochastic sampling of the IMF is not feasible in BPASS v2.2, since a large number of instances would need to be calculated, for each potential GC mass and for each metallicity. However we can make a preliminary estimate of the impact of the IMF by comparing the colours measured in the nine different IMFs which form part of this BPASS release. While each is assumed to be fully sampled, the range of slopes and cut-off mass treatments effectively reproduces systems either deficient or over-rich in stars of different initial masses. In Fig. \ref{fig:cols_imf} we show the effect of IMF variation on photometric colour at Solar metallicity for old stellar populations. Models are shown for all nine IMFs calculated in BPASS v2.2, and at Solar metallicity. The effect of IMF on colour is very small ($<$0.05\,mag) in all cases, but is highest at relatively young ages ($\sim$1\,Gyr), diminishing with increasing age. This suggests that for the old stellar populations typically observed in globular clusters, the effects of stochastic IMF sampling are unlikely to be reflected strongly in the colour of the integrated population.

\subsubsection{SDSS Quiescent galaxies}

A further test of the photometric colour for old stellar populations can be obtained by comparison to elliptical galaxies. These are quiescent galaxies with no evidence for ongoing star formation and typically red colours. Their stellar populations are old, $>$1\,Gyr, but may reflect an extended star formation history before quiescence, and are also subject to dust extinction.  In Fig. \ref{fig:gc_cols}, the globular clusters are overplotted in $J-K$ colour with the metallicity-colour relation derived by \citet{1538-3881-152-6-214} for elliptical galaxies. While the metallicity range probed by this sample is small, there is close agreement with model colours of simple stellar populations at ages of 5-10\,Gyr with the same metal content. To avoid adding the additional free parameters required by an extended star formation history, we continue to treat the galaxies as simple stellar populations from here onwards, in effect establishing the properties of the dominant stellar population, while noting that this will compromise any fit for the youngest or burstiest ellipticals.

In Fig.~\ref{fig:sdss_cols} we compare the colours of BPASS v2.2 models at 0.3\,Z$_\odot$ and Z$_\odot$ and ages 1 to 12.5\,Gyr with the observed colours of galaxies selected from the Sloan Digital Sky Survey (SDSS). We make use of the GALSPEC galaxy processing algorithm catalogue\footnote{Accessed from SDSS CasJobs as tables DR14:dbo.galSpecInfo and galSpecIndx.} associated with SDSS Data Release 14 \citep{2017arXiv170709322A} and based on the original work of \citet{2004MNRAS.351.1151B}. Galaxies are selected from the catalogue with Galactic extinction $E(B-V)<0.2$, stellar masses log(M/M$_\odot$)=10-14 and inferred star formation rates $<0.03$\,M$_\odot$\,yr$^{-1}$ based on GALSPEC fitting. We also limit the galaxies to low redshifts ($z<0.1$) to avoid any significant distortion of the photometry by $k$-correction effects. We use photometry for the resultant sample in the SDSS $u, g, r, i$ and $z$ bands, calibrated on the AB magnitude system and corrected for Galactic foreground extinction, and from the Two Micron All Sky Survey \citep[2MASS,][]{2006AJ....131.1163S} in the $J, H$ and $K_S$ bands, calibrated in Vega magnitudes. We do not attempt to correct individual galaxies for internal extinction, since this is a model-dependent process which would require full spectrophotometric fitting. However, we indicate the expected colour shift associated with an internal dust extinction $A_V=0.5$\,mag given two different extinction laws, by arrows on each panel. The Milky Way extinction law of \citet{1999PASP..111...63F}, modified in the near-infrared by \citet{2005ApJ...619..931I} is shown in black, and that of \citet{2003ApJ...594..279G} for the Large Magellanic Cloud in grey.

In order to evaluate what a reasonable internal dust extinction assumption for these galaxies might be, we also consider the best fitting extinction for old stellar populations determined for the same photometric sample from spectral energy distribution fitting to each galaxy using the FSPS \citep{2009ApJ...699..486C} stellar population templates\footnote{Accessed from the SDSS CasJobs SQL server as table DR14:dbo.stellarMassFSPSGranEarlyDust}.  The median extinction for this sample was $A_V=0.25$\,mag (on old stars) and we overplot tracks with this extinction in Fig.~\ref{fig:sdss_cols}.

As was the case for globular cluster photometry, we find generally good agreement with our models at central frequencies $<8000$\AA. Given near-Solar metallicity and a moderate dust extinction, consistent with that determined using other analyses, our models will overlay the photometry through the whole of the optical. At longer wavelengths, from the SDSS $z$-band into the near-infrared, the picture is somewhat more complicated. The  BPASS v2.2 models appear to overestimate $z-J$ colour by $\sim0.2$\,mag, while remaining broadly consistent with the observed $J-H$ colours.  At the same time, the $H-K$ colour is underestimated by $\sim0.1$\,mag. We note that many extinction laws suggest that these wavelengths are relatively unaffected by dust, and removing dust corrections longwards of the $i$-band may be appropriate. However, this would leave the $\sim$0.1-0.2\,magnitude offsets observed between models and data in both $z-J$ and $H-K$ unresolved. As was the case for globular clusters, we tentatively attribute these differences to uncertainties in our treatment of the cool giant stars which dominate at these wavelengths. 

As both Fig.  \ref{fig:gc_cols} and  \ref{fig:sdss_cols} also make clear, however, colour-colour information is generally unable to provide a precise estimate of age and metallicity for quiescent galaxies. The optical colours of simple stellar populations typically evolve by only a few tenths of a magnitude across an order of magnitude in age and a comparable range across three orders of magnitude in metal enrichment. Since dust extinction, age and iron abundance are all associated with redder colours, there is inevitably a degeneracy between these parameters for any given object, if photometry alone is considered.

\subsection{Mass-to-Light ratios}\label{sec:mlratio}

The mass-to-light ratio of a stellar population, defined as the stellar mass present per unit of emitted luminosity in a given band, and normalised to Solar units, is a crucial parameter that can be derived from SPS models. While the most accurate stellar masses are determined by fitting the full spectral energy distribution of a galaxy, in many cases (e.g. in large extragalactic surveys) this is not possible, and the near-infrared, rest-frame $K$-band magnitude alone (sometimes in combination with one or two photometric colours) is often used as a proxy for stellar mass \citep[e.g.][]{2003MNRAS.346....1K}. The selection of this band is due to its sensitivity to cool, low mass stars which usually dominate the stellar mass at late times, and also due to the relative insensitivity of this band to dust extinction. Given that quiescent galaxies in particular occupy a narrow range of colours at near-Solar metallicities, as Fig. \ref{fig:sdss_cols} demonstrates, the single band may be sufficient to normalise the SED.

\begin{figure}
\includegraphics[width=0.95\columnwidth]{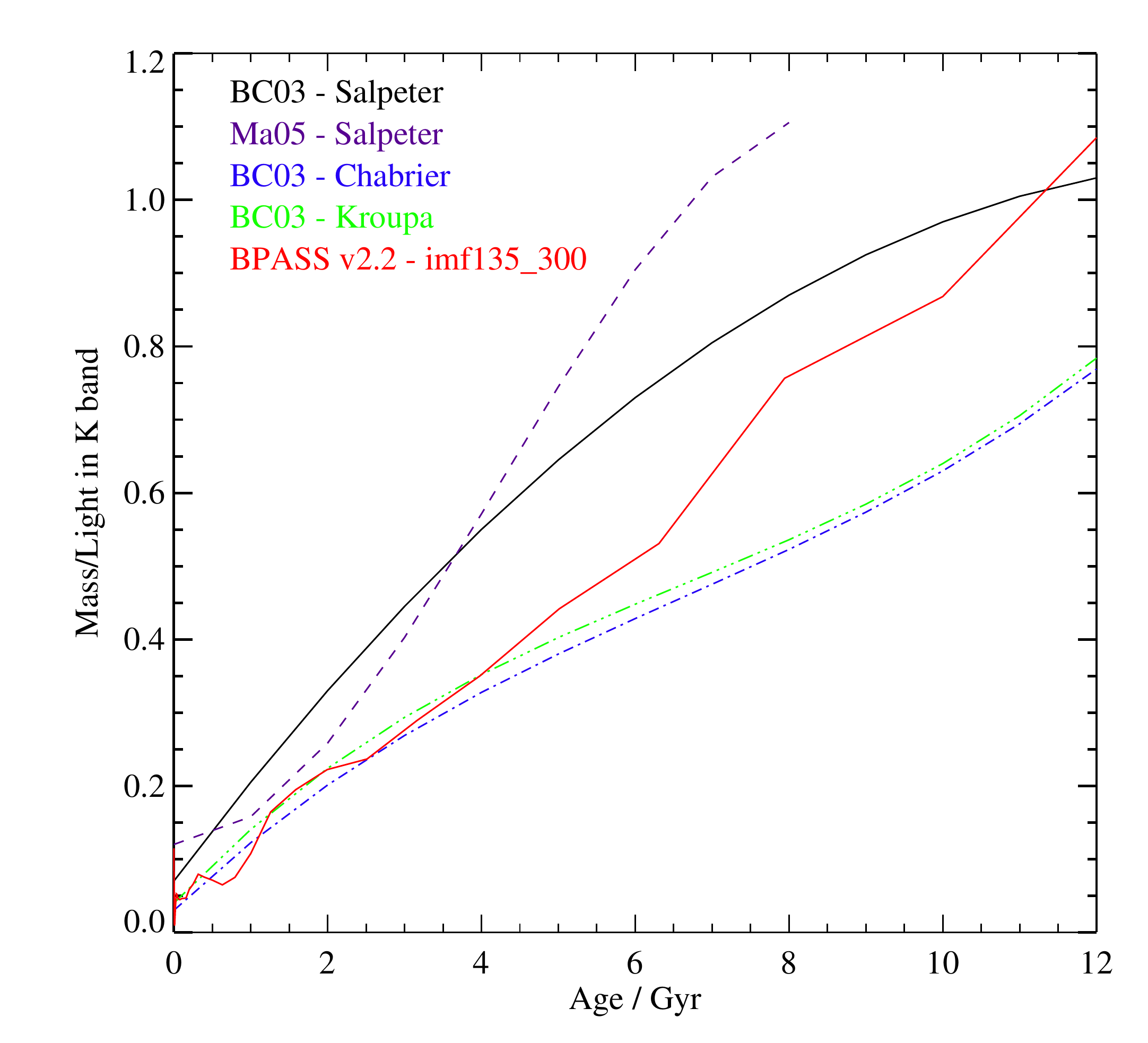}
\caption{The (rest-frame) K-band mass-to-light ratio derived from SPS models. We show the analytic fits to variation of mass-to-light ratio with age derived from a range of stellar population models (up to 8\,Gyr) by \citet{2009MNRAS.394..774L}, as described in section \ref{sec:mlratio}, together with the results of BPASS v2.2, all measured at Solar metallicity.}
\label{fig:mlratio}
\end{figure}

The mass-to-light ratios derived from the GALAXEV models \citep{2003MNRAS.344.1000B} and the models of \citet{2005MNRAS.362..799M} were compared in detail by \citet{2009MNRAS.394..774L}, who derived empirical relations between stellar population age and the inferred ratio, as well as the required $k$-corrections if galaxies were observed at higher redshifts. In Fig. \ref{fig:mlratio} we compare the $K$-band mass-to-light ratio of BPASS v2.2 (with our default, broken power-law, M$_\mathrm{max}=300$\,M$_\odot$ IMF) to the relations derived by \citet{2009MNRAS.394..774L} for the GALAXEV models with three different IMFs based on those of \citet{1955ApJ...121..161S}, \citet{2001MNRAS.322..231K} and \citet{2003PASP..115..763C}. We note that the cubic relations were derived for models at ages $<8$\,Myr, and extrapolation to later ages may not be valid (in the case of the Maraston 2005 models, there is actually a turn over). All models are at Solar metallicity. 
The stellar mass for BPASS models includes main sequence and post-main sequence stars, but not stellar remnants, pre-main sequence or brown dwarf stars.

As might be expected, the BPASS models track the mass-to-light ratios of GALAXEV SPS models assuming a Chabrier or Kroupa IMF (both of which are similar in form to our broken power-law) at stellar population ages up to $\sim6$\,Gyr. Beyond this, the BPASS v2.2 mass-to-light ratio diverges from the GALAXEV model, due to differences in the stellar wind prescriptions used and the effects of mass transfer prolonging the lifetimes of low mass stars. The Salpeter IMF extends the same power-law slope down to very low mass stars, and so the overestimates the number of sub-Solar mass stars. These contribute a large fraction of the integrated mass, but a small fraction of the light, and so models with this IMF produce an overestimate of the mass for a given $K$-band luminosity.

For future reference, we also show how the $K$-band mass-to-light ratio evolves with metallicity in Fig. \ref{fig:mlratio2} (and provide numerical values in the appendix, table \ref{tab:a1}). The ratio is relatively independent of metallicity at ages $<5$\,Gyr. Beyond this age, the mass-to-light ratio evolves less smoothly and becomes more metal dependent, due to the effect of metallicity on determining the luminosity and radius of stars when they enter the giant phases.

\begin{figure}
\includegraphics[width=0.95\columnwidth]{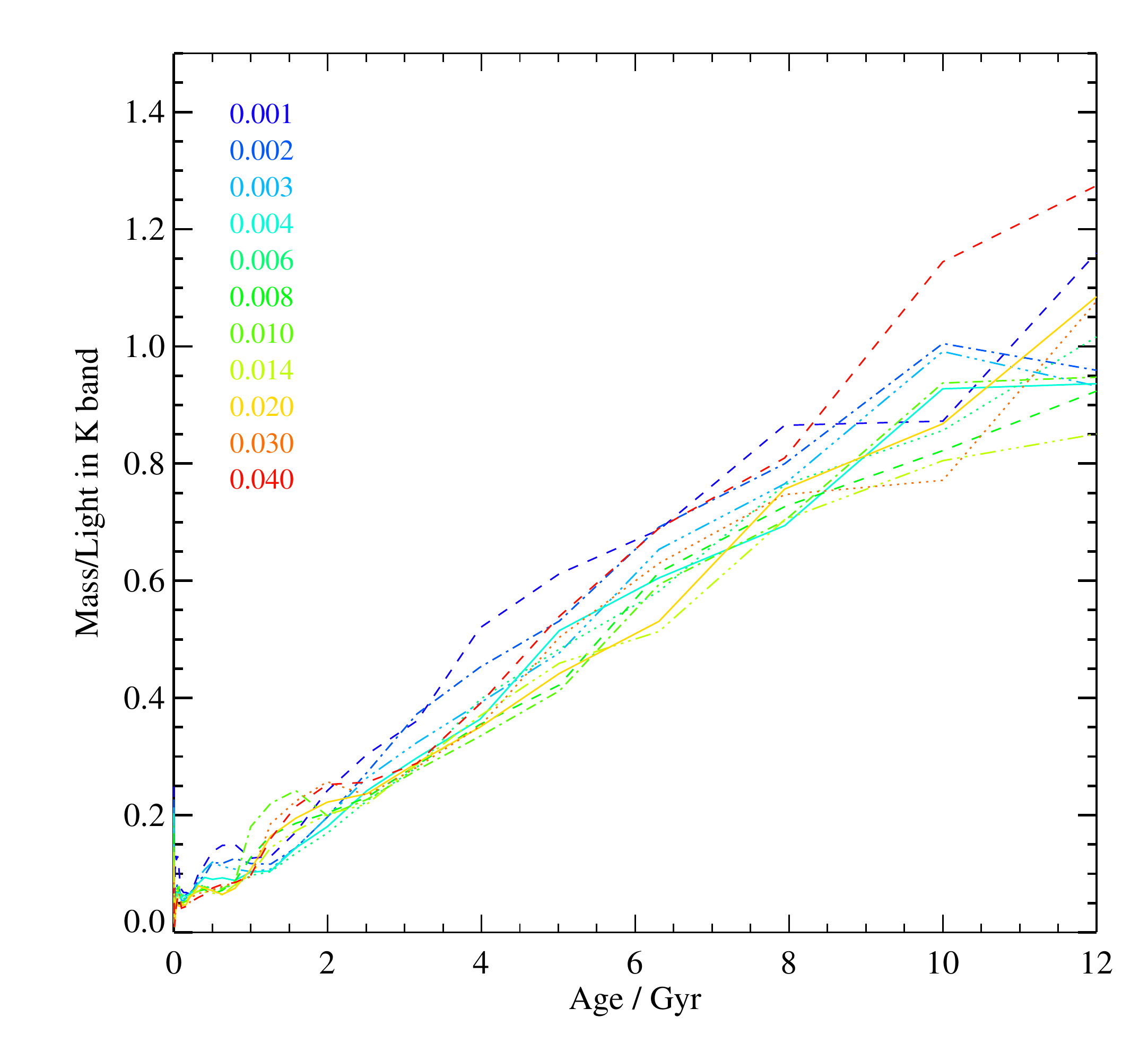}
\caption{The (rest-frame) K-band mass-to-light ratio for BPASS v2.2 as a function of stellar population age and stellar metallicity.}
\label{fig:mlratio2}
\end{figure}

\begin{figure*}
 \includegraphics[width=0.67\columnwidth]{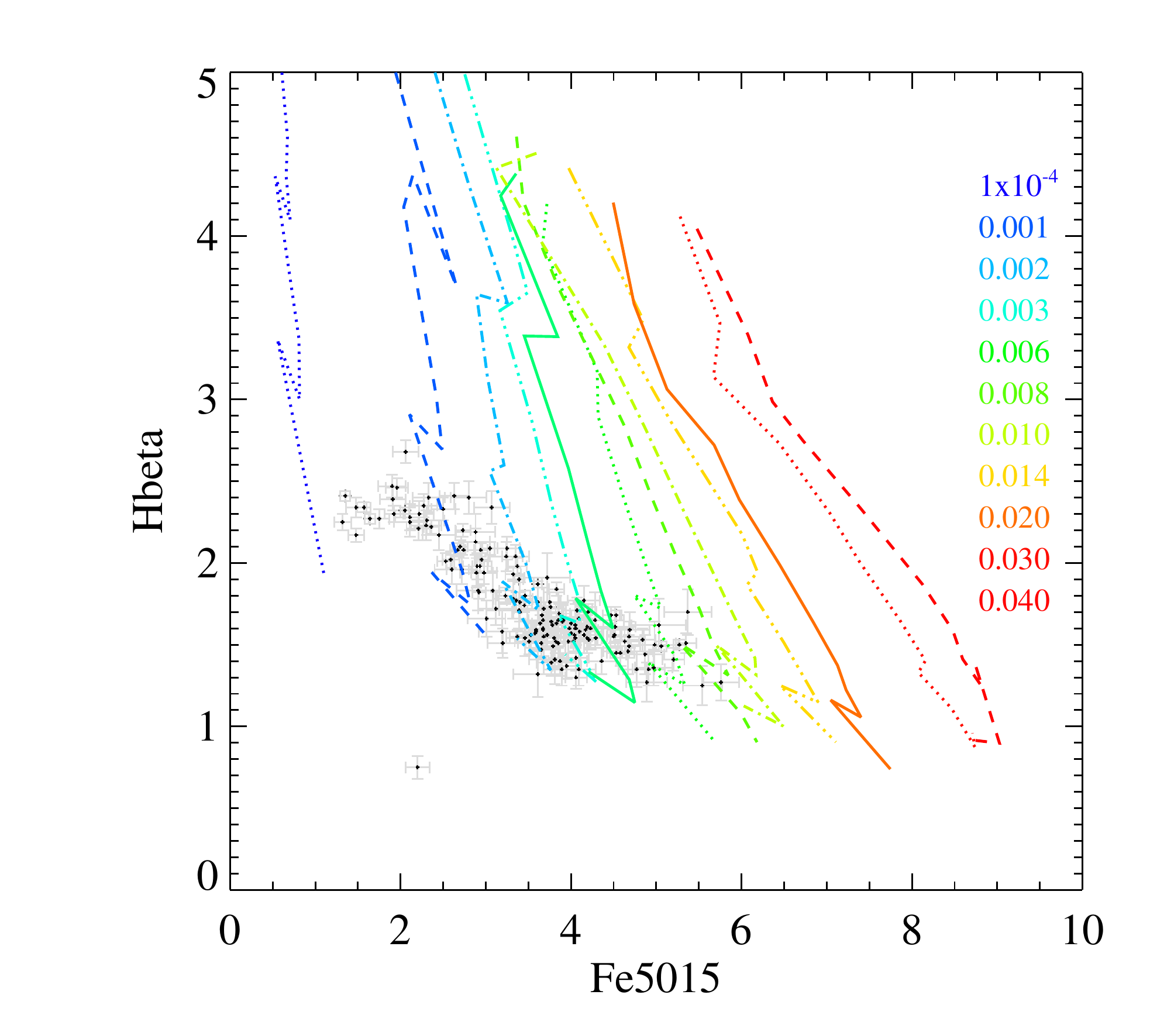}
 \includegraphics[width=0.67\columnwidth]{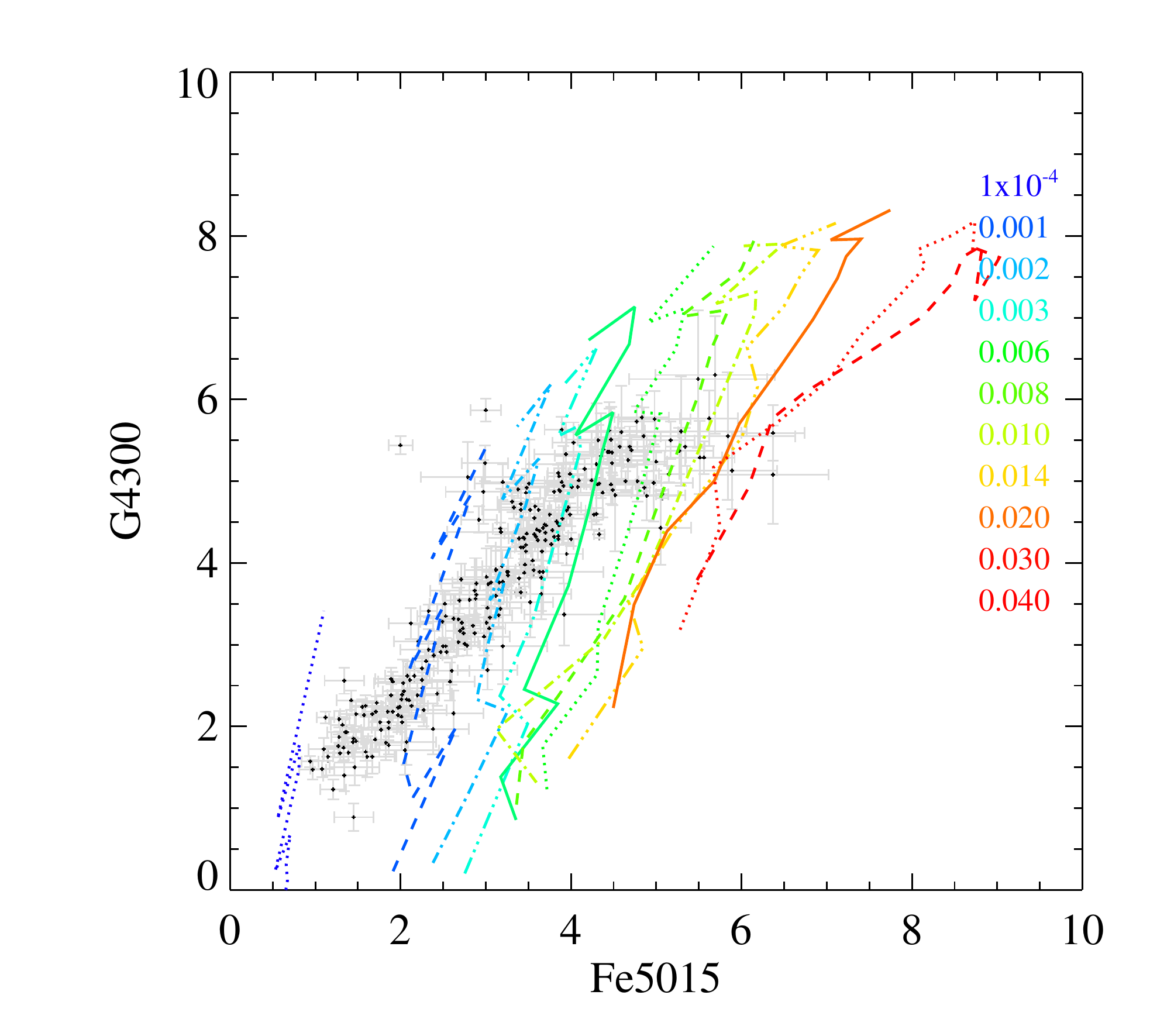}
 \includegraphics[width=0.67\columnwidth]{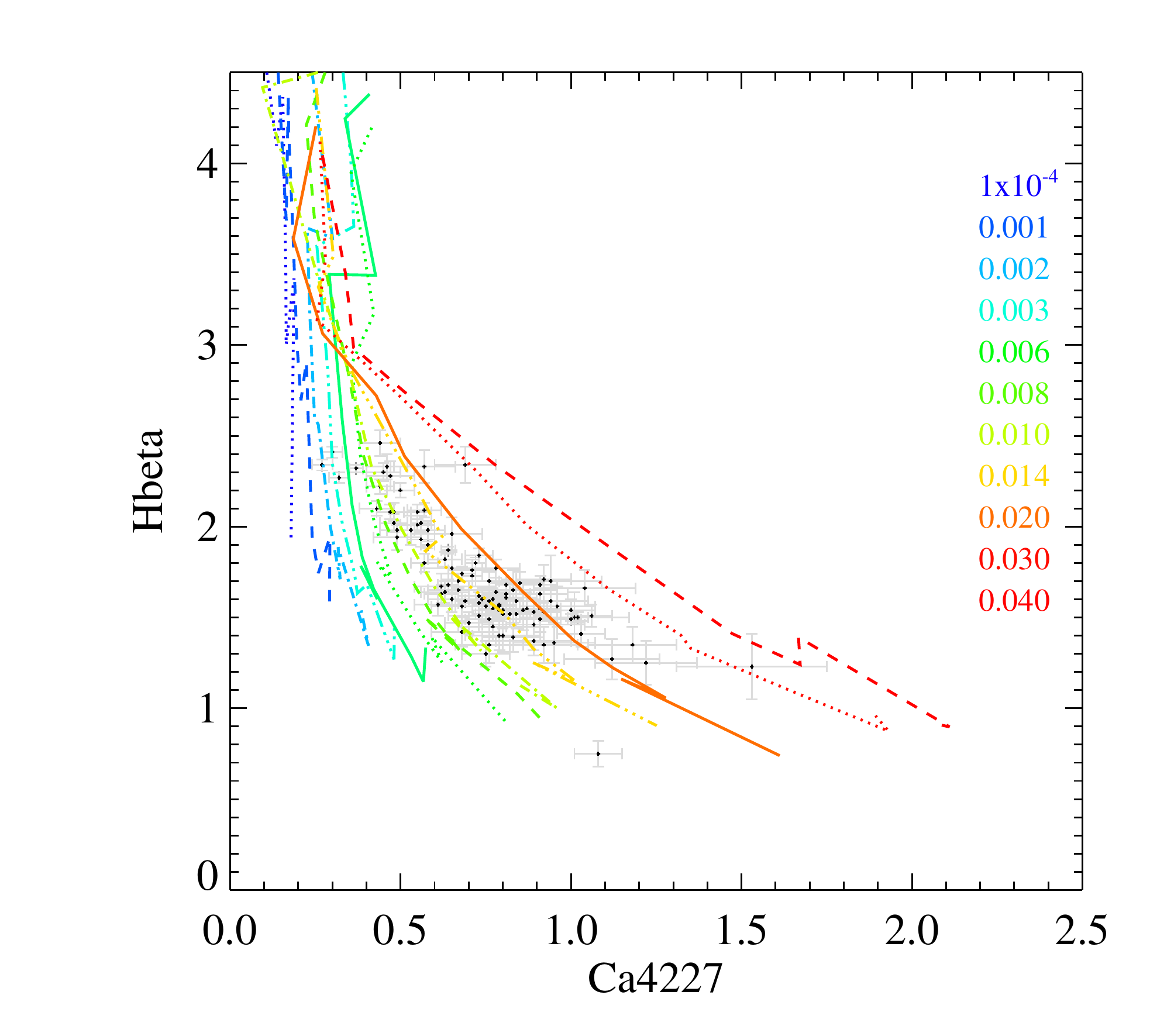}
 \includegraphics[width=0.67\columnwidth]{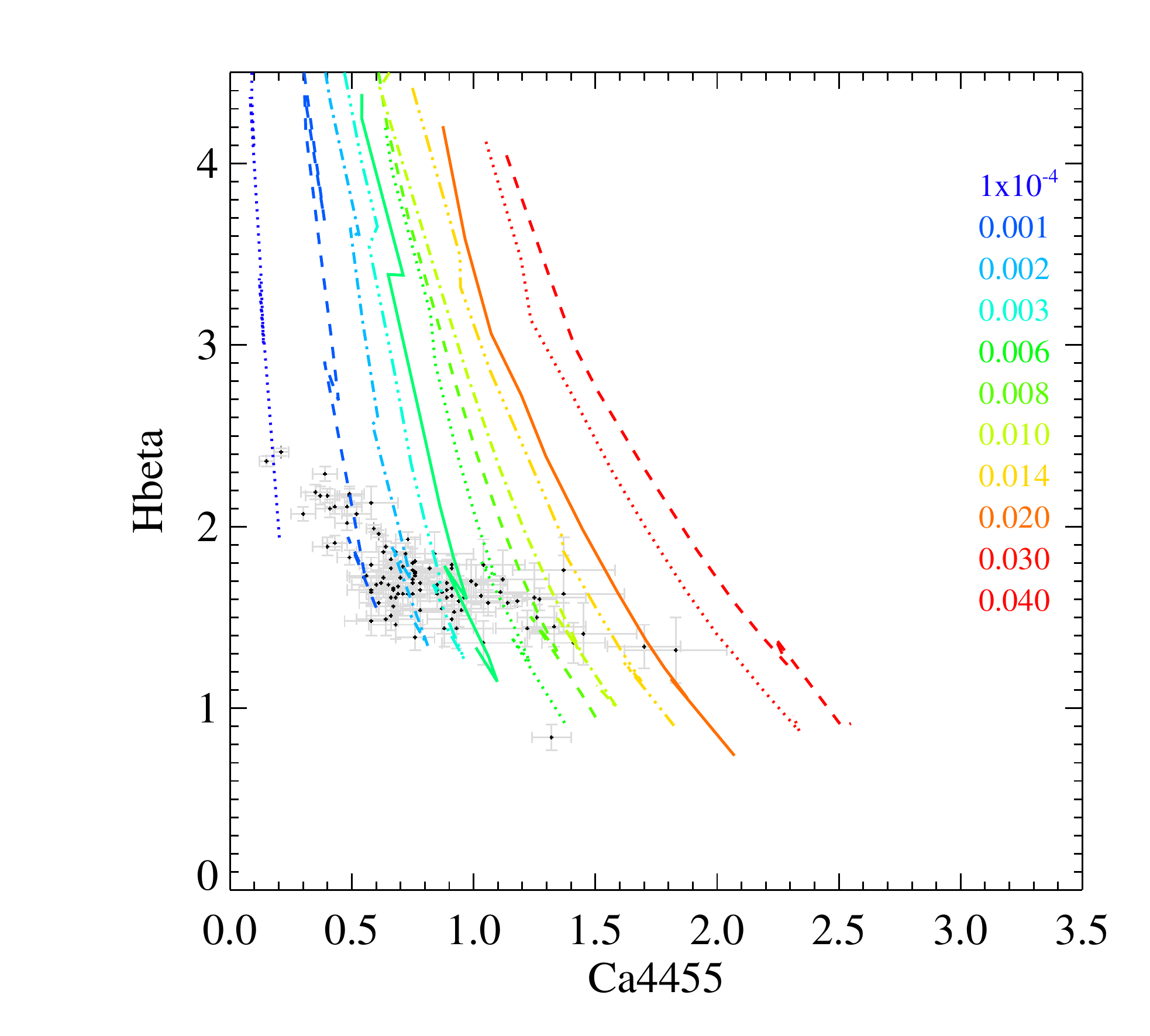}
 \includegraphics[width=0.67\columnwidth]{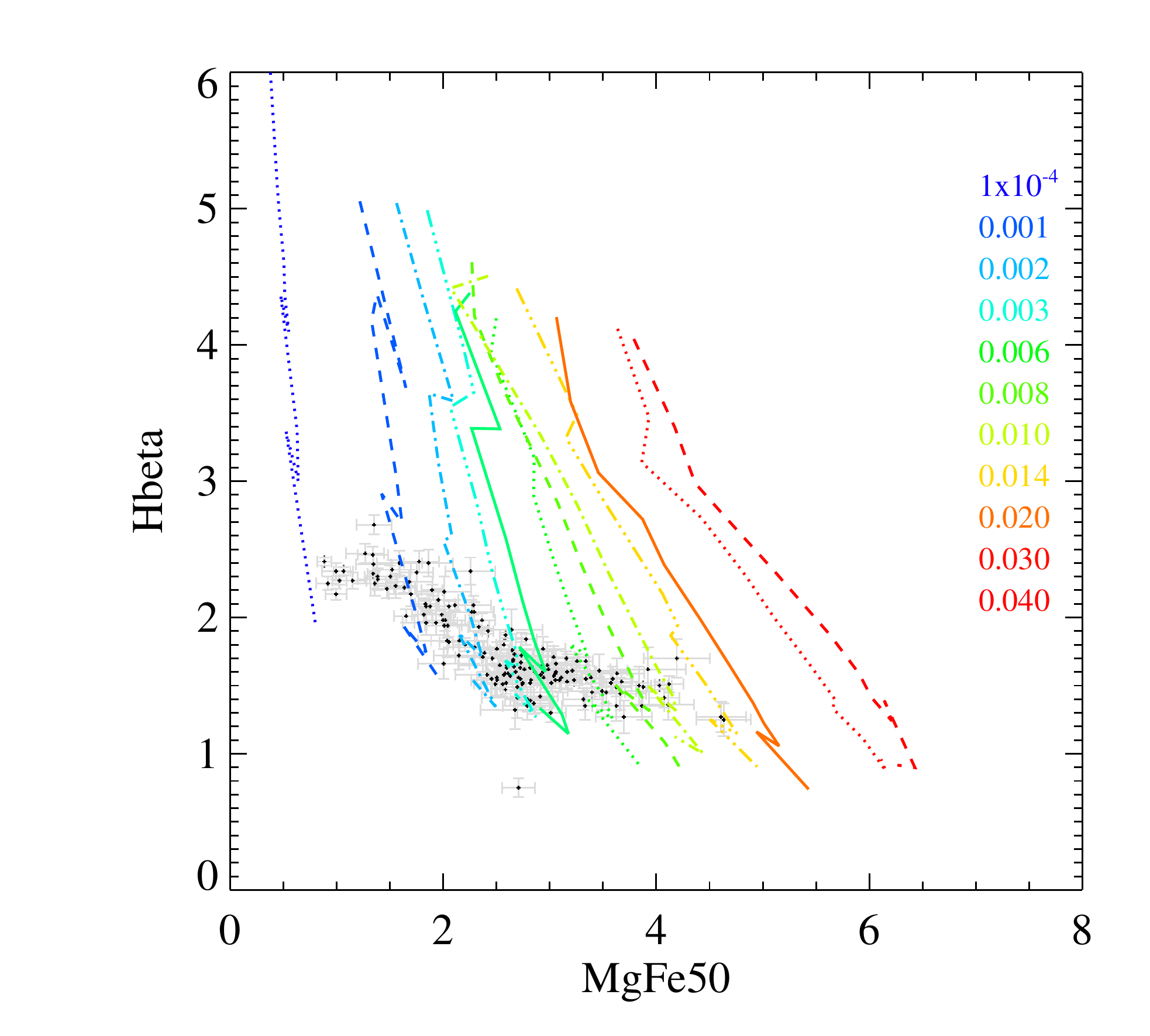}
 \includegraphics[width=0.67\columnwidth]{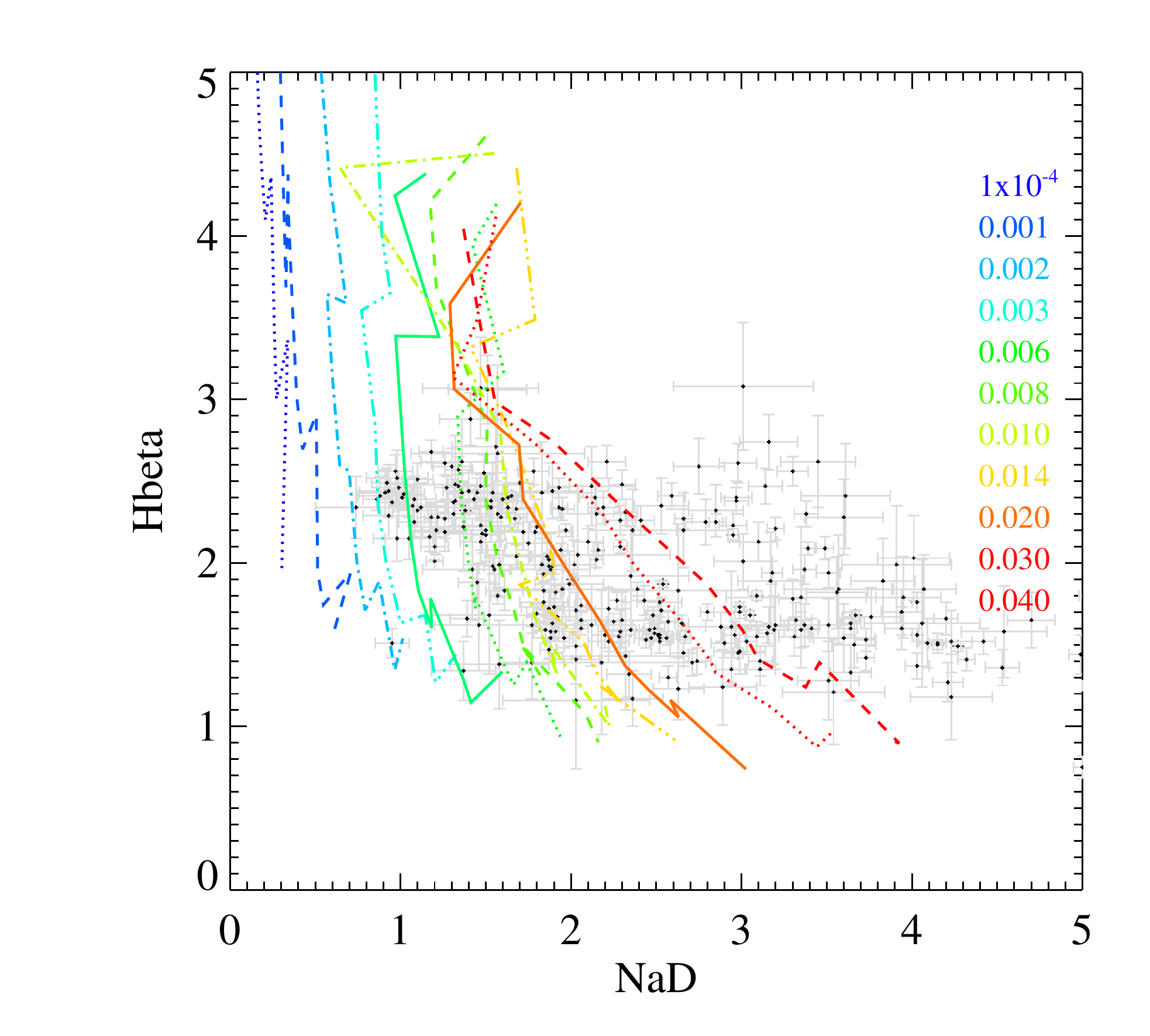}
 \caption{The Lick indices measured for BPASS v2.2, compared to those observed in the globular cluster systems of M31 and the Milky Way \citep{2012AJ....143...14S}. Lines indicate the evolutionary tracks of an aging, single-epoch, integrated starburst at fixed metallicity in the age range log(age/years)=9.0-10.1 (i.e. 3\,Gyr-12.5\,Gyr). \label{fig:lick_obs}}
\end{figure*}

\subsection{Lick Indices}\label{sec:obs_lick}

Where spectroscopy is available, more detailed information on the parameters of a stellar population may be obtained from the relative strength of emission and absorption features associated with atomic, ionic and molecular species. These are usually measured using an equivalent width, i.e. as the line strength relative to the inter-line continuum. A complication arises for populations in which photospheric absorption line blanketing makes identifying a true continuum level difficult. The Lick index system \citep{1994ApJS...94..687W,1997ApJS..111..377W} was established to address this issue. In this system, the mean fluxes in narrow spectral windows (typically $\sim$40\AA\ in width) are used to define a pseudo-continuum, and the flux in similar windows centred on notable absorption features are measured relative to these. Lick indices are frequently used in the analysis of globular clusters and absorption line galaxies, particularly in cases where there is insufficient signal to noise for full pixel-by-pixel spectrophotometric fitting to the data.

Since each spectral window is centred on features relating to a different species, Lick indices can be used as indicators of metallicity, and of the relative abundance of different elements in a stellar population. Using this method, old stellar populations are frequently found to be enhanced in $\alpha$-process elements relative to a Solar composition - the result of early rapid star formation in which the dominant source of metal enrichment was core-collapse supernovae rather than stellar winds. The MgFe50\footnote{[MgFe50]$'=0.5 (0.69$\,Mg\,b + Fe5015$)$, added to the original index definitions by \citet{2010MNRAS.408...97K}.} index 
is one of a number of weighted combinations of the original Lick Mg- and Fe-based indices that have been added to the original set, primarily in order to identify reliable metallicity indicators that are largely independent of $\alpha$-enhancement.

The relative elemental abundances in BPASS are fixed at Solar ratios, and are largely constrained by the stellar opacity tables and atmosphere models used. Thus we expect BPASS v2.2 models to be inconsistent with the data in indices which are affected by $\alpha$-elements, but to be consistent with observations in other indices, particularly those dominated by lines of hydrogen and iron. Hence, BPASS v2.2 models should enable the age and iron abundance of both globular clusters and elliptical galaxies to be estimated, when $\alpha$-independent indices are selected.

We calculate lick indices for the BPASS models following the prescription of \citet{1997ApJS..111..377W}, having first smoothed the integrated light SEDs to a wavelength-dependent resolution that varies from 8.4 to 10.9\,\AA\footnote{We use table 1 of \citet{2007ApJS..171..146S} to define the wavelength dependence of the resolution.} in order to match the resolution of the original Lick calibration spectra.

\subsubsection{Globular Clusters}

To evaluate the performance of BPASS v2.2 in reproducing the observed values of Lick indices in old stellar populations, we compare selected model indices against the Milky Way and M31 globular cluster index measurements of \citet{2012AJ....143...14S}, as a function of age and metallicity. Fig. \ref{fig:lick_obs} demonstrates that BPASS v2.2 probes a comparable parameter space to the observed data points in many of the Lick indices, and that in some cases, such as MgFe50  and the Ca4455 index, there are clear trends in index strength with metallicity at a given stellar population age. The globular clusters appear to be consistent with a relatively narrow range of ages (measured primarily by the H$\beta$ index) but a comparatively broad range of metallicities. 

In some indices, notably the NaD index, there is a significant discrepancy between the models and the data, with much higher metallicities inferred for the globular cluster. The NaD index is problematic in that it is also affected by Na absorption in interstellar material along the line of sight to the target \citep{2003MNRAS.339..897T}. However in other indices, notably those dominated by magnesium (e.g. Mg\,b), any observed discrepancy may well have a physical origin in $\alpha$-element enhancement.

\begin{figure}
 \includegraphics[width=0.99\columnwidth]{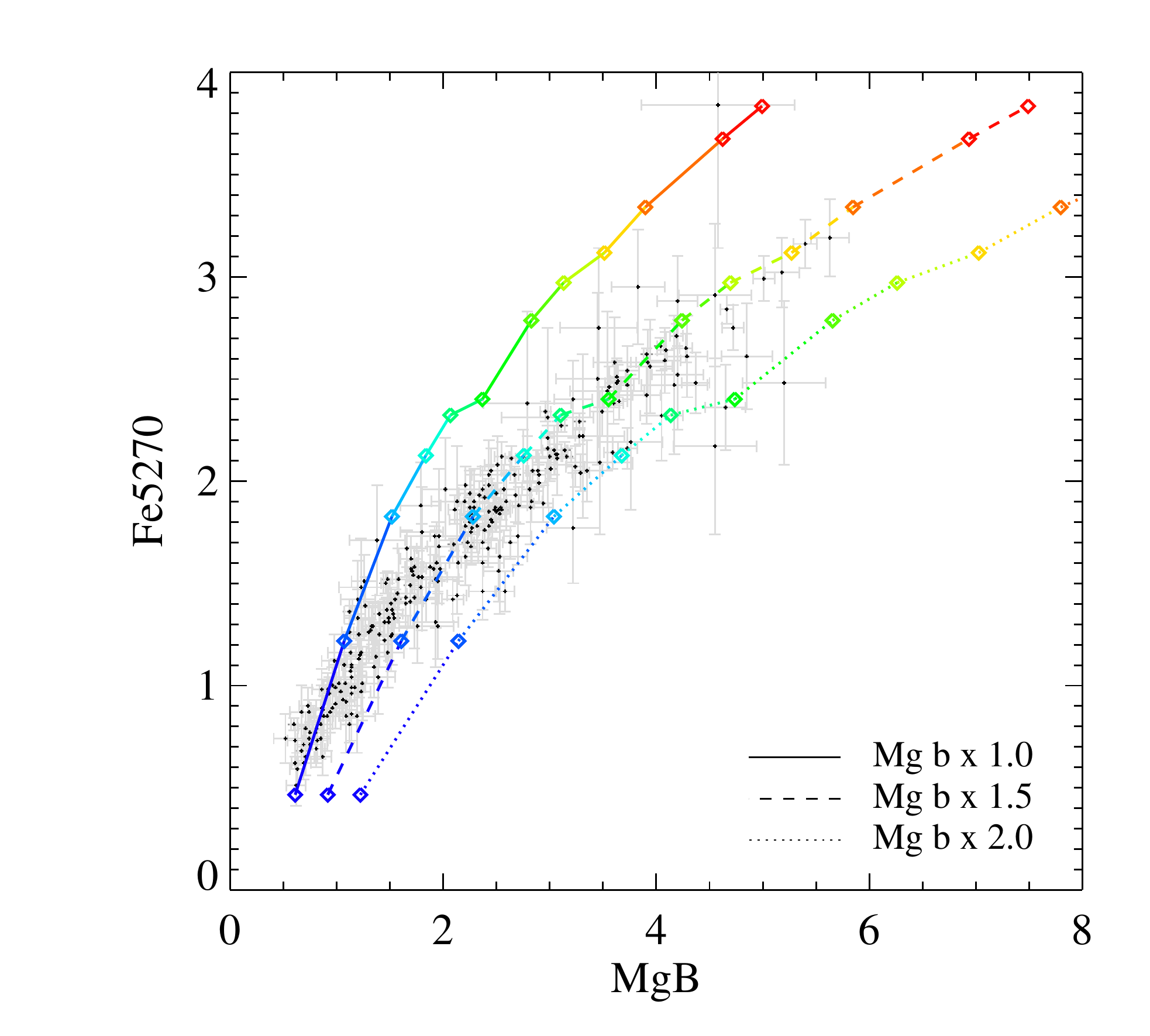}
 \caption{The Lick indices measured for BPASS v2.2, compared to those observed in the globular cluster systems of M31 and the Milky Way \citep{2012AJ....143...14S}, now considering the effects of $\alpha$-element enhancement. Tracks indicate the indices at fixed age, log(age/years)=9.7, and varying metallicity (colour-coded as in Fig. \ref{fig:lick_obs}). We also show indices in which the Mg\,b line index has been artificially enhanced by a factor of 1.5 and 2. \label{fig:lick_alpha}}
\end{figure}

In Fig. \ref{fig:lick_alpha} we show the lick indices in a band known to be strongly dependent on $\alpha$/Fe (Mg\,b), plotted against an index believed to be largely independent of $\alpha$-element enhancement \citep[Fe5270, see][]{2003MNRAS.339..897T}. Models are shown at a single age (log(age/years)=9.7, 5\,Gyr) and at varying metallicity. We also show the metallicity trend when the Mg\,b index is increased by a linear factor of 1.5 or 2.0. The models show a good match to the globular cluster data when Mg\,b is scaled by a factor of 1.5 relative to the Fe-dominated index. Translating from linear scalings to $\alpha$ enhancement is non-trivial since both the index and pseudo-continuum bandpasses can be effected by shifts in abundance ratio, and is usually done through modelling. This is not currently possible with BPASS, but \citet{2003MNRAS.339..897T} demonstrated that an increase in the strength of the Mg\,b index relative to pure iron indices by a factor of 1.5 corresponded to [$\alpha$/Fe]$\sim0.2$ (in logarithmic units) in their models.

For future reference we tabulate key indices in the appendix, and provide full tables of BPASS v2.2 index measurements with age, metallicity and IMF on the main BPASS project website\footnote{bpass.auckland.ac.nz}.

\begin{figure}
\hspace{-0.6cm}
 \includegraphics[width=0.53\columnwidth]{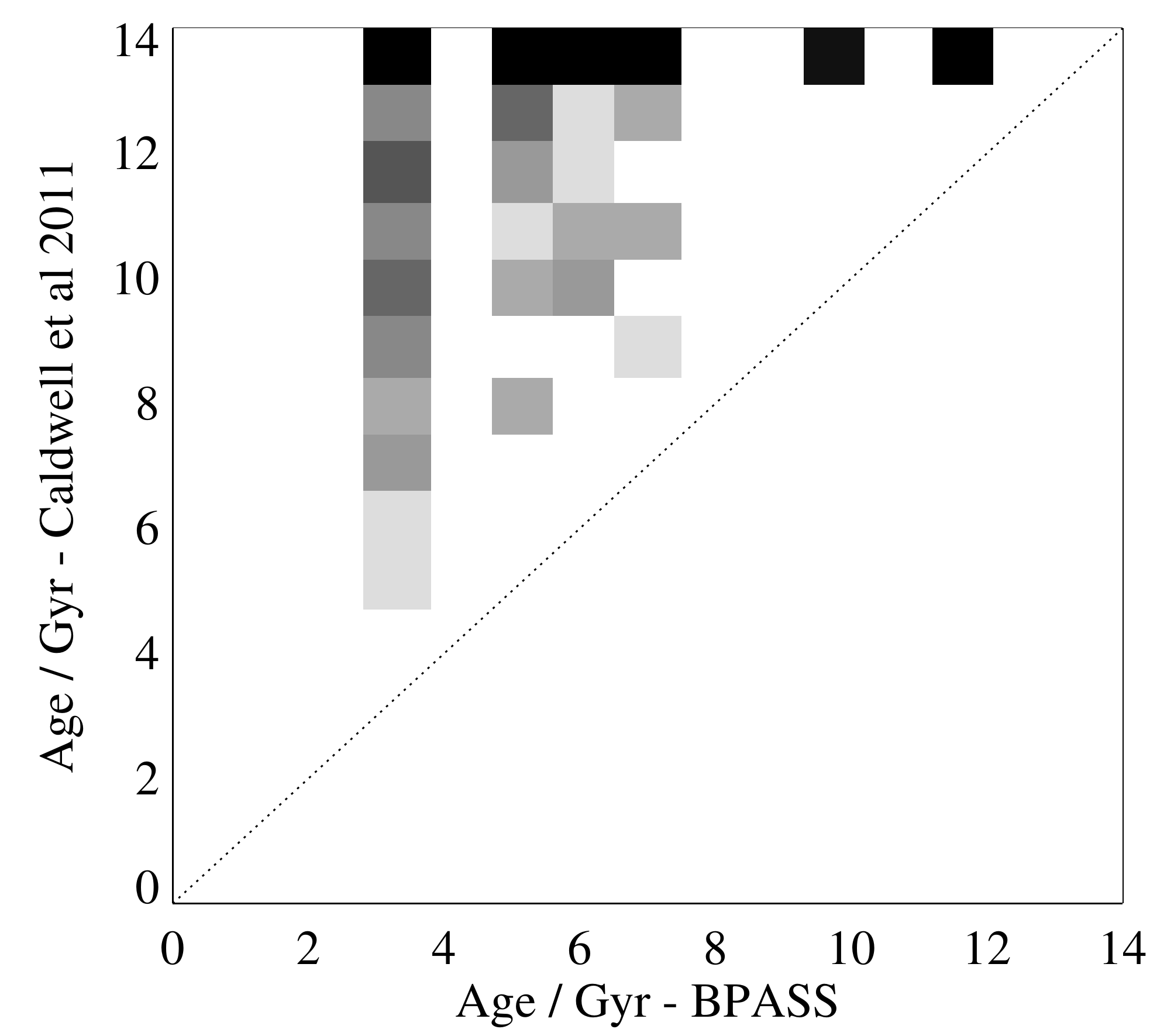}
 \includegraphics[width=0.53\columnwidth]{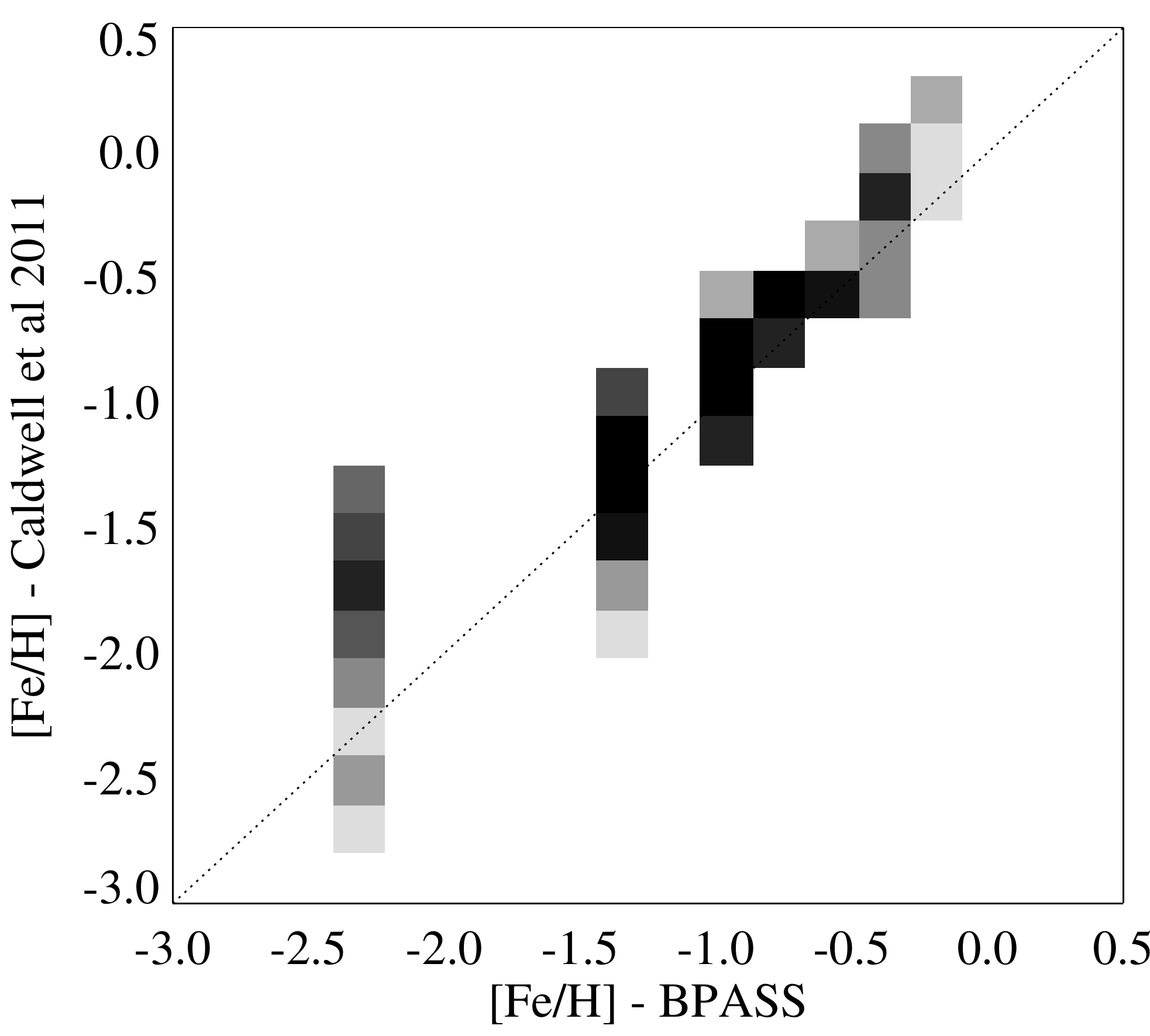}
 \caption{A comparison between the physical parameters inferred from the MgFe50-H$\beta$ plane for BPASS v2.2 models, compared to those derived by \citet{2011AJ....141...61C} for GCs. Bins are shaded by the number of sources in that bin, with black indicating high density regions. \citet{2011AJ....141...61C} values in the uppermost age bin indicate a poor fit in their analysis.\label{fig:gc_comp}}
\end{figure}

A natural next step is to compare the metallicity and ages inferred from our stellar population models to those determined for the same objects from existing model fits \citep{2011AJ....141...61C}. We classify each globular cluster based on its closest match to BPASS v2.2 models in the MgFe50-H$\beta$ index plane. While a more precise fit might be obtained using more indices, this pair provides a good first estimate, with the former index sensitive primarily to iron abundance, while the latter is age-sensitive. Neither index is strongly $\alpha$-dependent. We find that there is a tight correlation between the metallicity inferred from BPASS v2.2 model fitting and earlier calibrations in iron abundance, as Fig. \ref{fig:gc_comp} demonstrates. These tend to be low, with the median and mean (BPASS-inferred) metallicities in the sample given by Z=0.003 ([Fe/H]=-0.82) and 0.004 ([Fe/H]=-0.65) respectively. By contrast, BPASS tends to favour younger ages for the GCs in the  \citet{2012AJ....143...14S} and \citet{2011AJ....141...61C} sample than would be determined by the previous calibration, with some clusters in the sample best fit by ages as young as $\sim$2.5\,Gyr, while the median and mean ages are 7.9 and 7.7\,Gyr respectively.

\subsubsection{SDSS Elliptical galaxies}

\begin{figure*}
\includegraphics[width=0.5\columnwidth]{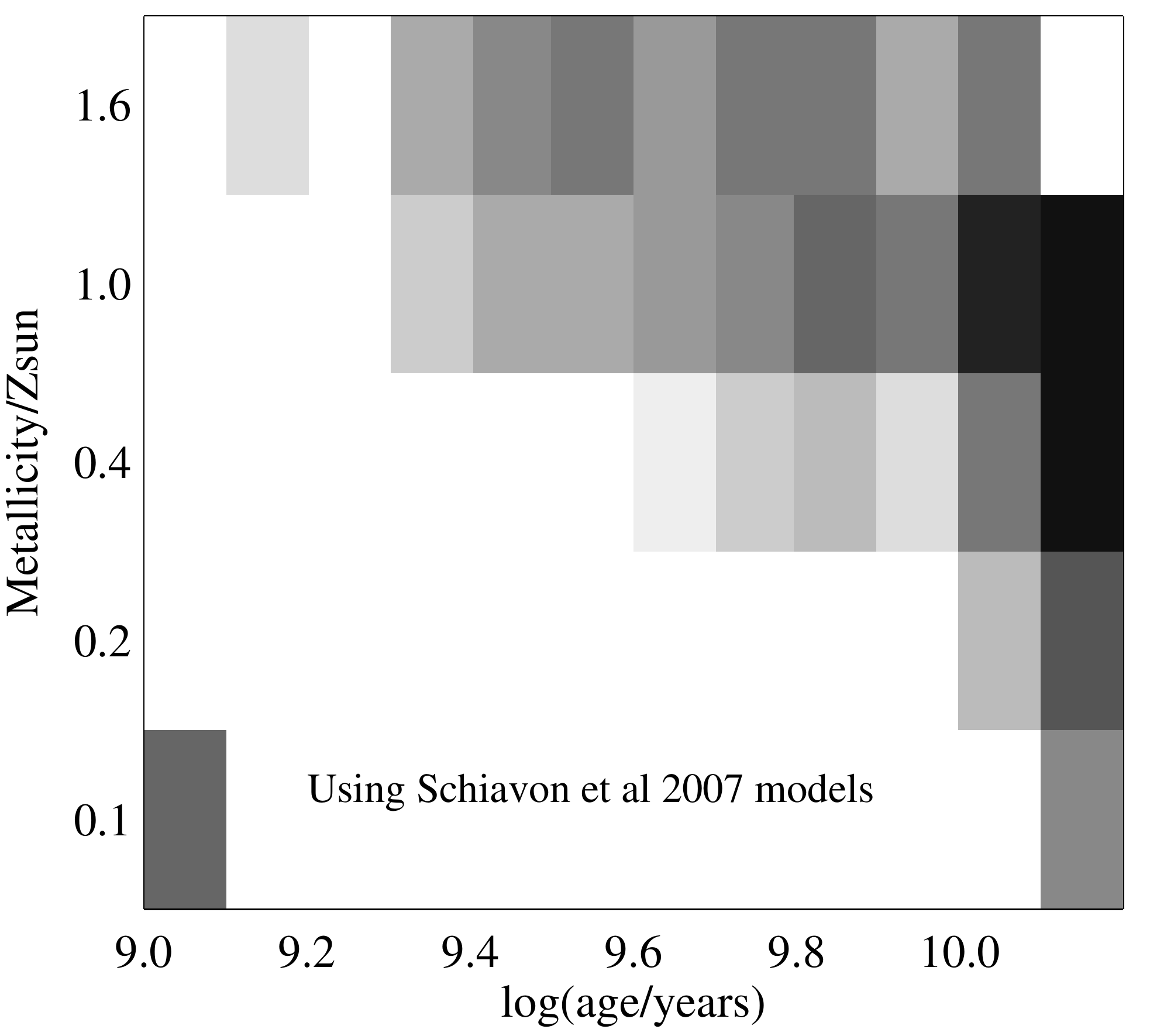}
\includegraphics[width=0.5\columnwidth]{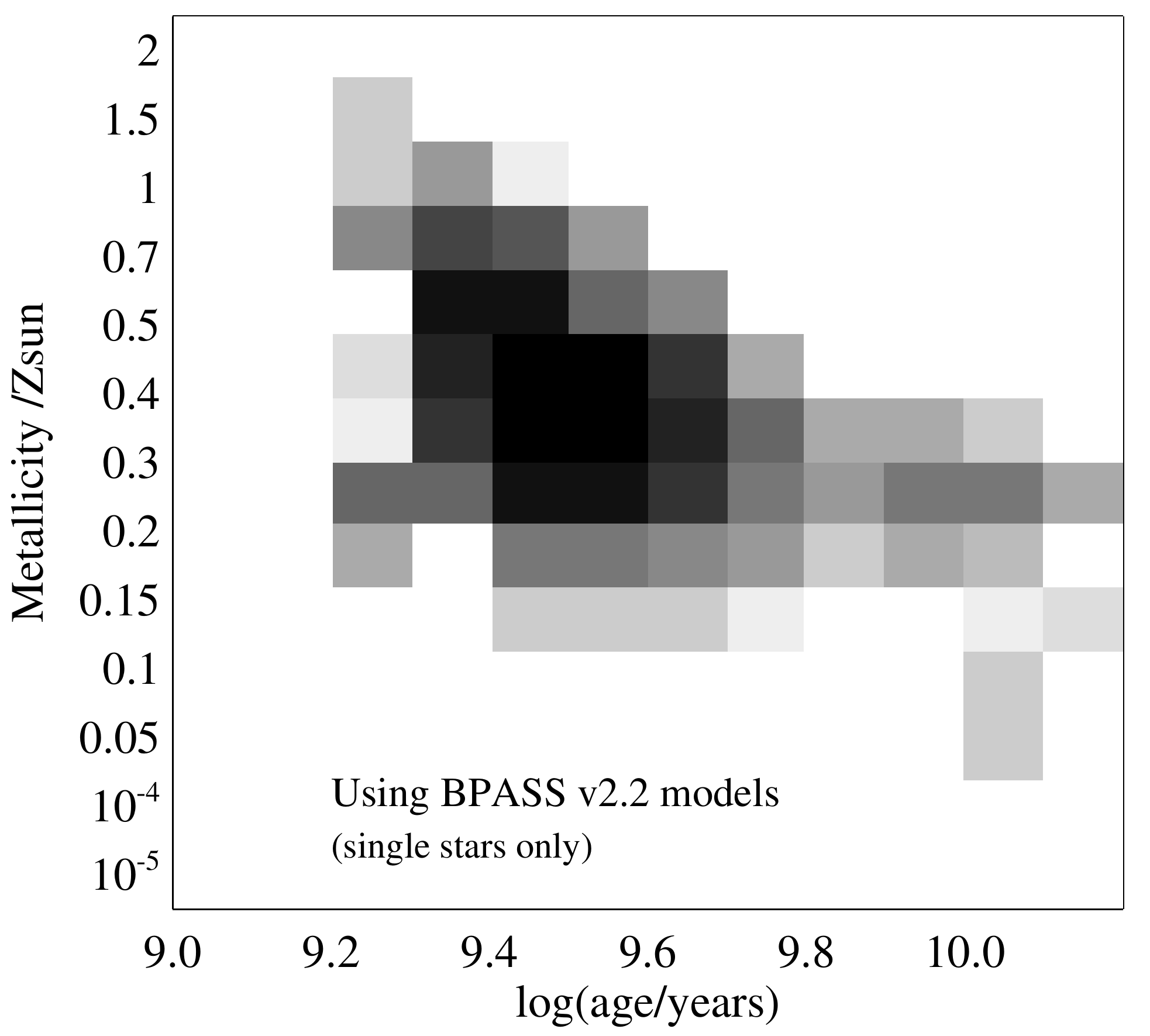}
\includegraphics[width=0.5\columnwidth]{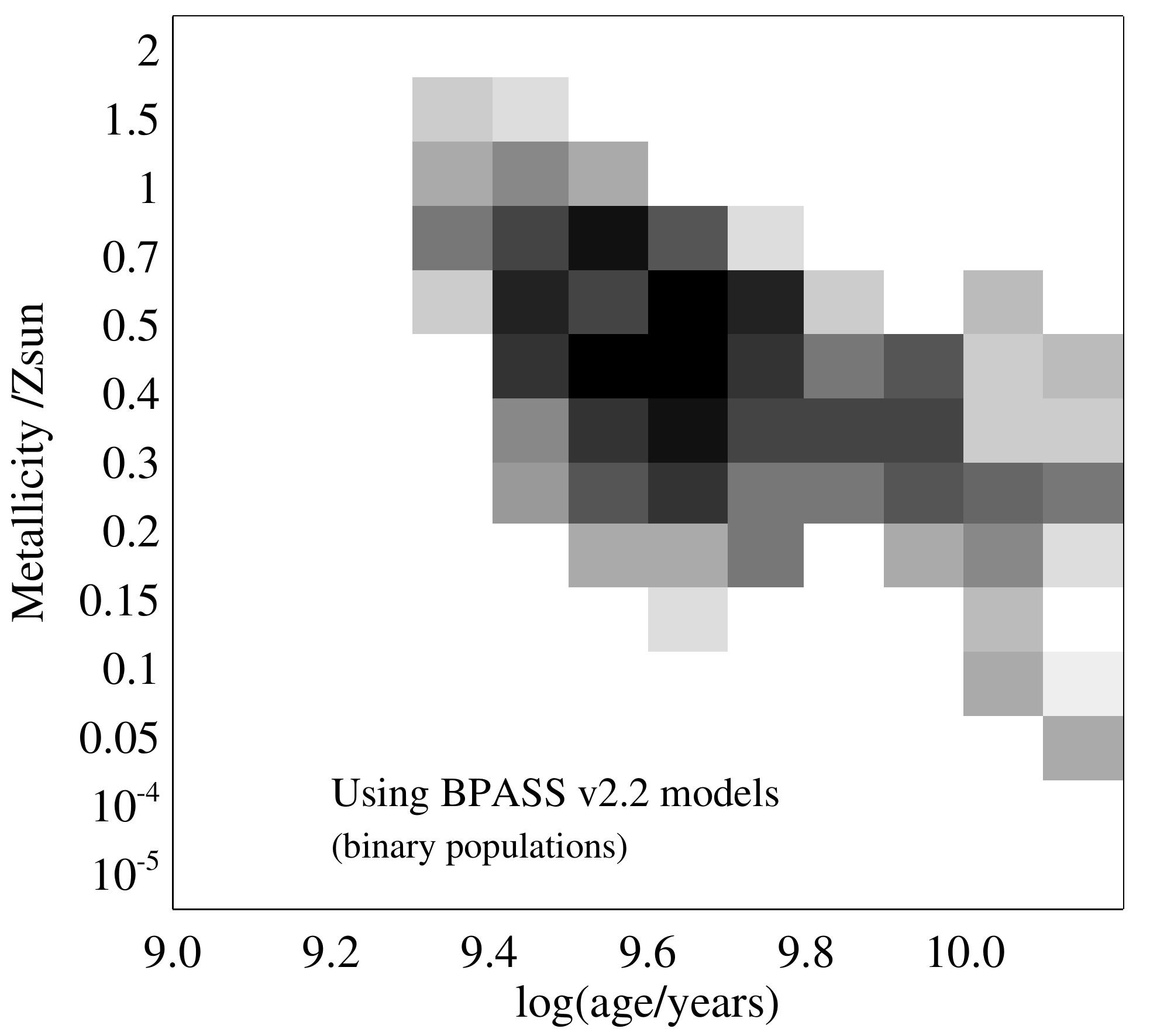}
\includegraphics[width=0.5\columnwidth]{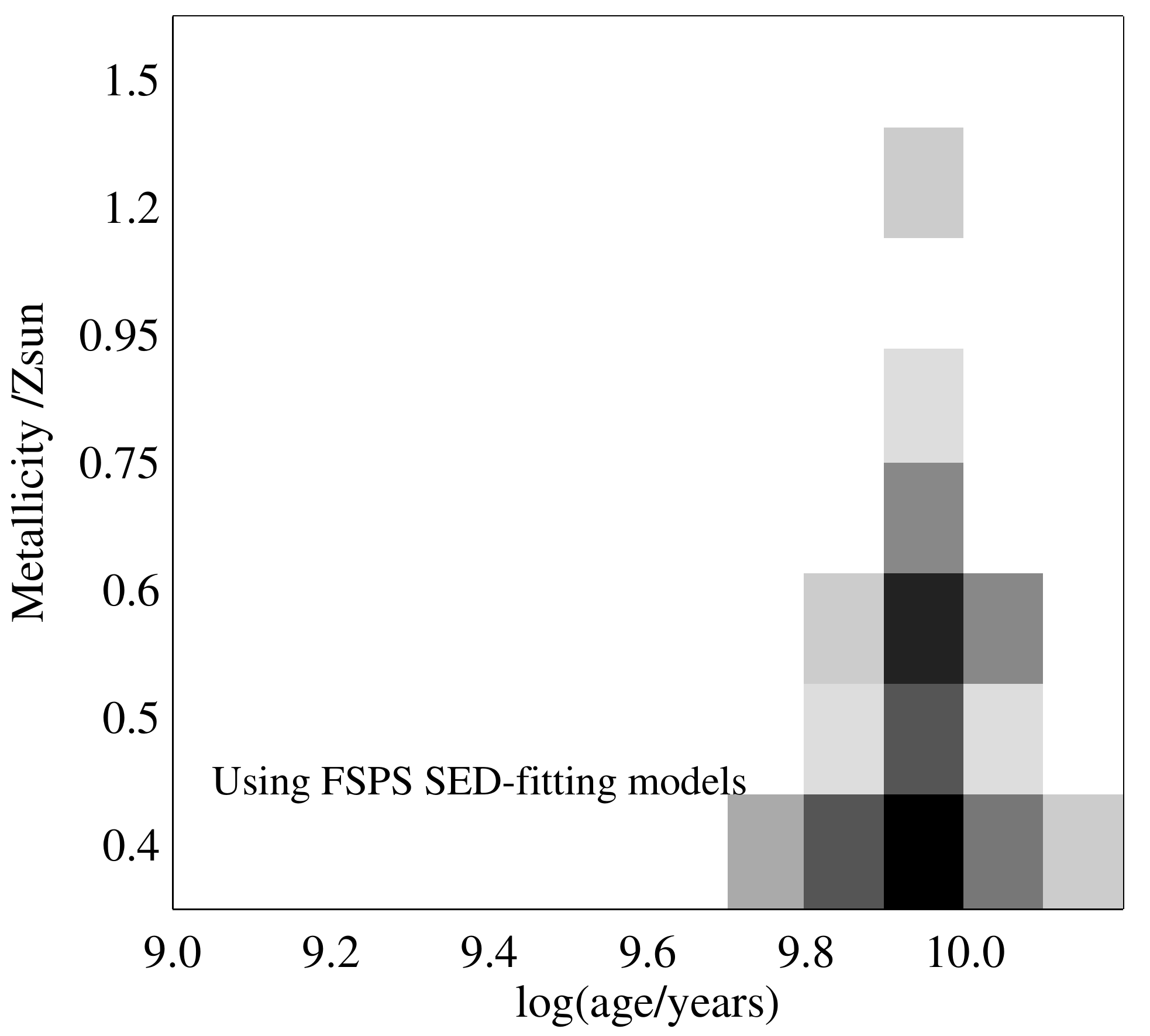}
\caption{Best Fits to SDSS quiescent galaxies based on fitting the MgFe50 vs H$\beta$ line strengths to models using (from left to right) the Schiavon et al 2007 models and BPASS v2.2 single star models and binaries. The right-most  panel shows derived parameters for the same galaxies based on photometric SED fitting using FSPS stellar population models for the same objects.}
\label{fig:best_comp}
\end{figure*}

\begin{figure}
\includegraphics[width=0.9\columnwidth]{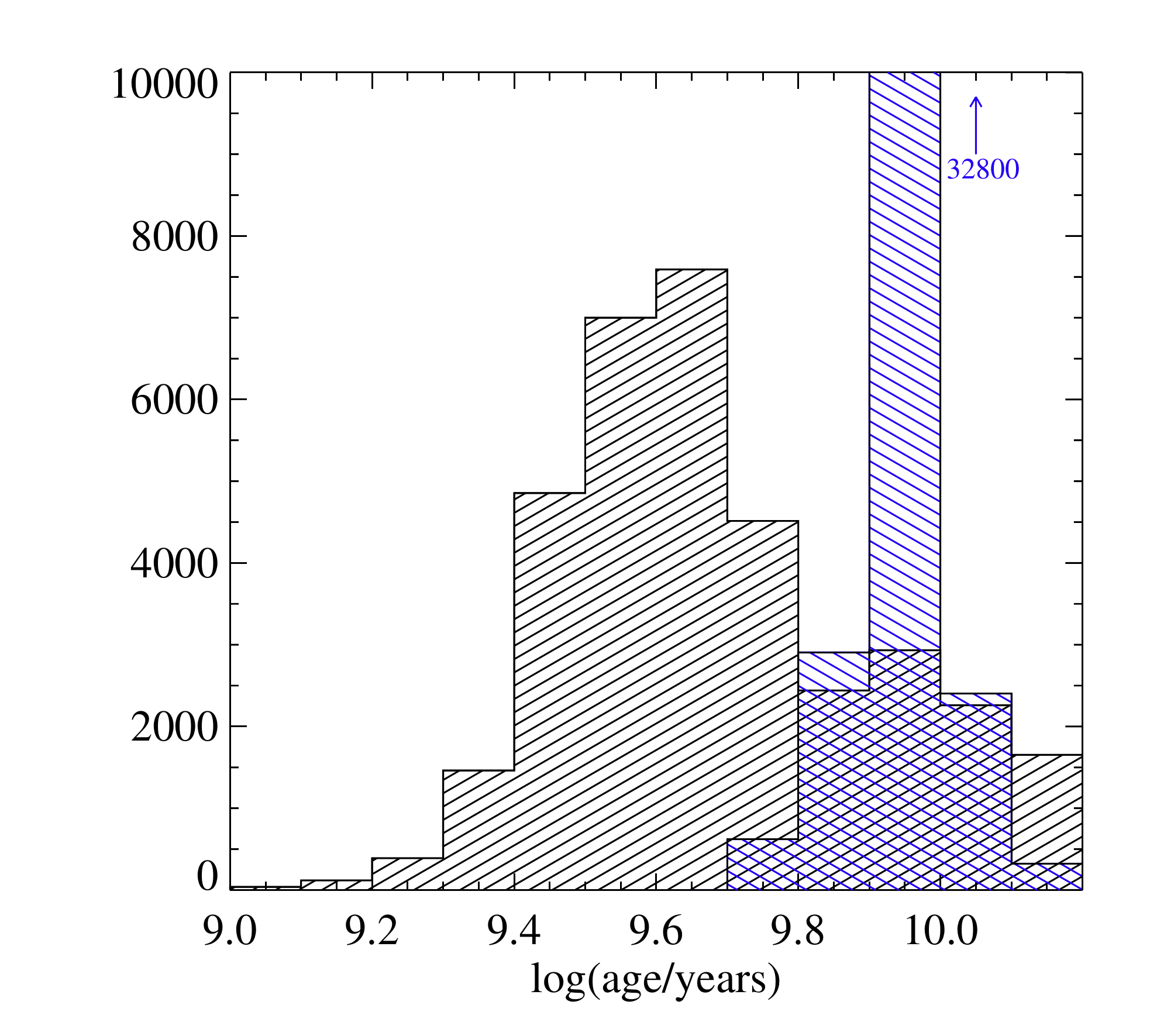}
\includegraphics[width=0.9\columnwidth]{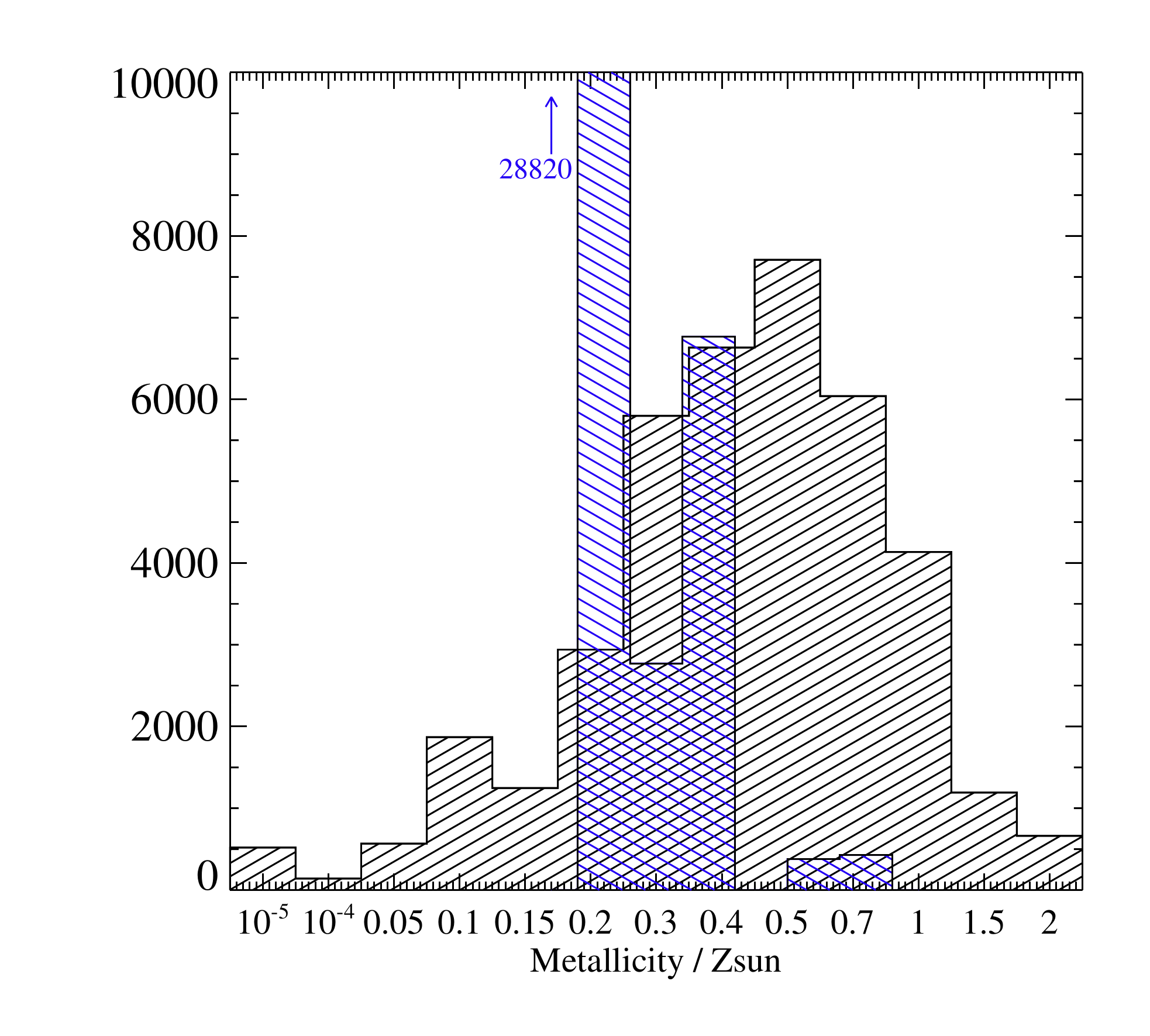}

\caption{Age and metallicity of SDSS quiescent galaxies. The BPASS v2.2 parameters are based on best fitting MgFe50 vs H$\beta$ index (shown in black). These are compared to FSPS stellar population fitting results for the same objects (blue). FSPS histograms extend to number counts well beyond the plotted area, due to the narrow range of fitted age and metallicity values in these models.}
\label{fig:best_ellips}
\end{figure}

The distribution of Lick indices for observed quiescent galaxies, drawn from the same SDSS sample discussed in section \ref{sec:obs_phot}, are near identical to those of GCs, with a slight bias towards higher metallicities (as indicated by stronger Fe-dominated indices). However, while the range of indices is similar, the more complex star formation histories in these systems mean that a simple stellar population may not be the most appropriate model, and there will likely be additional scatter, particularly in age-sensitive indices. 

Again we use the MgFe50 vs H$\beta$ plane to identify the best-fitting simple stellar population template for each galaxy, making use of the GALSPEC database for the observational  measurements. We compare the best BPASS template parameters (for both our single star and binary populations) with the age and metallicities derived using the same data but instead using the model values in table 31 of  \citet{2007ApJS..171..146S} for the line indices. As Fig.~\ref{fig:best_comp} illustrates, the BPASS models place elliptical galaxies in broadly the same parameter space as the older Schiavon model grid, but tends to favour  younger and marginally more metal rich stellar populations. Interestingly our single star models are actually more different from the older models than our binary populations. As discussed in section \ref{sec:comp}, our single star models are redder than others, primarily due to differences in handling of the giant branches. However, since the effect of binary interactions is to strip such stars during their earlier main-sequence phases and so suppress luminosity on the giant branch, this different handling remains completely consistent with our current understanding of stellar populations.

We also compare the best fit age and metallicity with those derived
from spectral energy distribution fitting to the photometry of each galaxy using the FSPS stellar population templates (see section \ref{fig:sdss_cols}). The results of this analysis are shown in the right hand panel of Fig.~\ref{fig:best_comp} and in Fig.~\ref{fig:best_ellips}. BPASS v2.2 Lick index calibrations suggest a broader range of both ages and metallicities than the FSPS SED models. In the latter case, this may be partly an artefact due to an absence of low metallicity FSPS models, since the majority of targets are fit in the lowest metallicity FSPS bin available. However this would not entirely account for the difference in distribution: the BPASS models typically fit to higher metallicities, within the range probe by FSPS. We note that there is a clear degeneracy between age and metallicity in the simple two-index fit adopted for our analysis. However, examination of a range of other indices confirms that younger, higher metallicity BPASS v2.2 models are generally favoured over older, lower metallicity ones.


\section{Discussion}\label{sec:disc}

Analysis and interpretation of old stellar populations is a mature, established field. The determination of age, iron abundance and abundance ratios based on Lick index (or other similar absorption line) calibrations underpins much of the work in this area, while the photometry of galaxies has been used to determine their light-weighted mean characteristics. Increasingly, such analyses are being complemented and superceded by pixel-by-pixel spectrophotometric fitting analyses, which provide increasingly precise matches between observations and templates \citep[e.g.][]{2017MNRAS.466..798C}. However this long history and confidence masks an underlying flaw: that translating the observables to a physical property of the target requires the use of stellar population modelling in one form or another. This may arise from calibrations based on individual stars (in which case stellar evolution and atmosphere modelling is invoked), or from calibrations based on stellar population synthesis (in which case the evolution of not only individual stars but entire populations must be understood). But are the assumptions underlying these calibrations reasonable? As many as forty percent of the low mass stars which dominate old populations are born in binaries. Many more will have interacted with a more massive binary companion before reaching their observed age \citep{2017ApJS..230...15M}. A reasonable fraction, perhaps 10 per cent, occur in tertiary or higher multiples which no current model, including BPASS, currently considers. Given that virtually all of the widely-used models to date have assumed that each star evolves in isolation as a single star, the work undertaken here and similar studies are essential to test the underlying assumptions in the analysis of old stellar populations.

As we have demonstrated, incorporating binary stellar evolution pathways, together with the most up-to-date stellar evolution and atmosphere models for single stars, into stellar population synthesis models can make a significant difference in the interpretation of their integrated light properties. The lifetimes of stars in different mass ranges can be modified, by mass transfer and mixing, as can their temperatures and gravities. Mass transfer onto a secondary can produce more massive, and therefore brighter, stars at late ages, even if their stellar atmospheres are typical of cool red giants. A population can also incorporate stellar types simply not found in the absence of binary evolution. The result is a spectral synthesis model set which  spans a very similar parameter space to existing galaxy and cluster template models, but which nonetheless leads to a significantly different interpretation of the integrated light.  As we demonstrated in Fig. \ref{fig:bc03comp}, BPASS v2.2 models are similar to those of existing spectral synthesis model sets such as the GALAXEV models of \citet{2003MNRAS.344.1000B}, but are both bluer at young ages (as the massive stars interact and dominate the luminosity) and redder at late ages (where the low mass stars are reaching the end stages of their evolution). At its most basic level, this means that we are able to reproduce the photometry of mature, quiescent galaxies and clusters at younger ages than previous model sets (i.e. $\sim5-8$\,Gyr, rather than 10-14\,Gyr).

It is both interesting and reassuring that the same pattern is observed in fitting to those Lick indices that we have considered here. To some extent, this may not be some surprising: if massive red stars are dominating the spectral energy distribution, then the same stars might dominate the absorption features. However it is not clear that the correct interpretation is so clear cut.  A Lick index is fundamentally a measurement of equivalent width -- of line strength relative to the {\it continuum} -- and the continuum light of most stellar populations incorporates contributions from stars which are unlikely to add to line strength.  Similarly, there are stellar types (such as white dwarfs) which can affect the mean absorption line strengths without significantly affecting the continuum colour. Thus, for the line indices and the photometry to match observed data simultaneously suggests that the stellar population synthesis has constrained the relative weighting of different stellar types at a given age correctly.

BPASS v2.2 models do not yield the same calibrations for age and metallicity as some of the models commonly used for fitting to large surveys (e.g. GALAXEV \citep{2003MNRAS.344.1000B}, FSPS \citep{2009ApJ...699..486C} or the \citet{2005MNRAS.362..799M} models). So which is the correct solution?  It is not clear that there is any way to test the age and metallicity of a given stellar population in a fully model-independent way.  One promising future approach, which highlights the advantage of binary evolution modelling, is that the expected rate of astrophysical transients (supernovae, gamma ray bursts and gravitational wave chirps amongst others) can be inferred from a population model and compared to the data \citep[see e.g.][or ES17]{2015MNRAS.452.2597X,2016MNRAS.462.3302E}. In the era of deep, large area transient surveys such as the planned Large Synoptic Survey Telescope (LSST), this may prove to be a very valuable constraint, and should be fit simultaneously with the stellar population of a given galaxy or class of galaxies. Doing so will present a challenge for single star evolution models, in which many such events are not predicted. 

The simple fact that stellar multiple systems are known to exist in such large numbers in the local Universe suggests that they should not be ignored while constructing models. On the other hand, while the binary fraction is now reasonable well known for different mass ranges of the bulk population in the local Universe, it remains unclear whether our decision to use this set of empirical fractions at all metallicities is entirely appropriate. If metal content (in particular the density of CO gas and its resultant opacity) significantly affects the fragmentation of molecular clouds into individual proto-stars, the binary parameters and the underlying initial mass function of the resultant population will need to be reassessed \citep[see e.g.][]{2010A&A...510A.110H}. However the observational constraints needed to impose a metallicity-dependent variation are not currently extant and are unlikely to be available until the coming generation of Extremely Large Telescopes allows resolved stellar population studies in a far larger sample of galaxies.

There also remain significant uncertainties in stellar evolution and atmosphere modelling which lead to consequent uncertainties in all SPS codes, including BPASS and other binary synthesis models. One such is the correct mass loss rates to use for post main-sequence, red giant and AGB stars, and how these scale with metallicity. While progress has been made on constraining these in recent years, particularly through use of direct observations extending to the submillimetre \citep[see e.g.][for a recent review]{2018A&ARv..26....1H}, there is little constraint on how they scale with elemental abundances or whether the inferred prescriptions can be universally applied in any given parameter space (e.g. $M$, $L$,  $T$, $g$ etc). The details of common envelope evolution and its implementation in evolution codes is likely to be critical for the formation of compact binaries, while other post-main sequence effects which remain problematic in stellar models include the modelling of convective dredge-up and thermal pulsations, as well as more dramatic, rapid events such as the helium flash. The prescriptions used for these processes in BPASS have been described here and in Eldridge, Stanway et al (\citeyear{2017PASA...34...58E}), but other possible prescriptions and formalisms exist. It remains to be seen what effect these have on the resultant population, and it is simply infeasible to regenerate the more than quarter of a million individual detailed stellar evolution models on which BPASS is built for each scenario. Here the rapid population synthesis approach used in codes such as {\sc binary\_c} \citep{2009A&A...508.1359I,2015A&A...581A..62A}, in which analytic prescriptions and empirical relations are used to infer the population, or in which interpolation occurs between a much coarser model grid, may have an advantage, although much of the detail in the stellar evolution and binary interaction process is lost.

An additional complication which will need to be considered in future studies is illustrated by Fig. \ref{fig:lick_alpha}. The fixed elemental abundance set used in BPASS models is almost certainly inappropriate for many old stellar populations, in which $\alpha$-element abundance enhancement must be considered. A change in the abundance ratios of elements will affect the relative strength of different core-burning nucleosynthesis chains, while the opacity of heavy elements in a stellar atmosphere is largely responsible for stellar winds. As a result we might expect to see some changes in the results of evolution models when the abundance set is changed, although we note that the [Fe/H] ratio is likely dominant here, and we do not expect such changes to be severe.  We are currently limited in further modelling of $\alpha$-enhancement by the availability to the community of opacity tables and large, uniform, high resolution stellar atmosphere grids with varied composition and iron abundances, not just in the main-sequence but for post-main sequence, stripped and very hot stars. In the short term, it may be possible to make empirical corrections to $\alpha$-element dominated Lick indices as discussed above.

Despite these weaknesses, we highlight again the fact that old stellar populations can be robustly fit by BPASS v2.2 models at younger ages and typically at slightly higher metallicities than considered in many current mainstream model sets. This difference in age behaviour may relieve some of the tension between globular cluster ages and the age of the Universe. While accepted globular ages are currently within the cosmology-derived 13.7\,Gyr age limit imposed by time since the Big Bang, many of them are towards the upper end of this age limit, and imply very large rates of star formation within a few 100\,Myr of the birth of the Universe. This contrasts with observations which suggest that the volume-averaged star formation rate was very low at early times (see e.g. \citet{2014ARA&A..52..415M} although c.f. the tentative inference of early star formation from \citet{edges} based on 21\,cm absorption).

The consequences for interpretation of current giant quiescent galaxies may not be not profound, but they do suggest that caution must be used when using fossil stellar populations or galactic archaeology to reconstruct  the construction and star formation histories of such galaxies, and when relating these to observational cosmology and cosmological simulations. A simulation fine-tuned to yield galaxies of a mass and metallicity inferred from many SPS models at a given lookback time will be producing them earlier and often at a lower metallicity than would be the case had BPASS v2.2 been used for the original galaxy parameter fitting. We strongly encourage those engaged in stellar population fitting to observational data sets to be aware of the differences between models, and to consider carefully the assumptions which underpin the models on which their analysis is based.


\section{Conclusions}\label{sec:conc}

Our main conclusions can be summarised as follows:

\begin{enumerate}

\item The Binary Population and Spectral Synthesis (BPASS) model data release version 2.2 incorporates a larger grid of low mass star models and has modified prescriptions for rejuvenation, dredge up and the AGB phase, which improve its performance over earlier versions at ages $>$1\,Gyr. Changes to the outputs at $<$100\,Myr are negligible.

\item The integrated light of stellar populations at late ages is redder than in older stellar population models. These provide a good match to the optical photometry of globular clusters and quiescent galaxies, although in the later case $A_V\sim0.25$\,mag of dust extinction is also required. Precision fitting  to the near-infrared photometry remains challenging.

\item We calculate Lick indices for BPASS v2.2 and demonstrate that we are able to reproduce the observed values in both globular clusters and galaxy populations, except for indices where adjustment for $\alpha$-element abundance enhancement is required. Well-sampled libraries of $\alpha$-enhanced stellar atmosphere models across a range of luminosities will be required to do much better.

\item Model fits to photometry and spectroscopic indices yield a consistently younger fit, often at slightly higher metallicity, than fits to older calibrations, when new stellar atmosphere models and binary stellar evolution pathways are included.

\item Interpretation of old stellar populations is inevitably model-dependent. Care should be taken to understand and, where possible, test the assumptions built into such models including the effects of stellar multiplicity.

\end{enumerate}

As with previous releases, the BPASS v2.2 spectral energy distribution models, together with other information which now includes tables of Lick index measurements, are publicly available on the project websites at bpass.auckland.ac.nz and warwick.ac.uk/bpass\footnote{If any assistance is required in implementing these in fitting codes or other data analyses, we are always happy to collaborate.}.


\section*{Acknowledgements}

ERS gratefully acknowledges travel funding from the Physics Department of the University of Auckland, and also thanks the Australian Astronomical Observatory (AAO) for hospitality and travel support through their Shaw Distinguished Visitor scheme. JJE acknowledges travel funding and support from the University of Auckland. We thank Charlie Conroy for making his CKC14 atmosphere models available to the authors.

This publication makes use of data products from the Two Micron All Sky Survey, which is a joint project of the University of Massachusetts and the Infrared Processing and Analysis Center/California Institute of Technology, funded by the National Aeronautics and Space Administration and the National Science Foundation. 
It also uses results and data from SDSS-III. Funding for SDSS-III has been provided by the Alfred P. Sloan Foundation, the Participating Institutions, the National Science Foundation, and the U.S. Department of Energy Office of Science. The SDSS-III web site is http://www.sdss3.org/.
This research has made use of the SVO Filter Profile Service (http://svo2.cab.inta-csic.es/theory/fps/) supported from the Spanish MINECO through grant AyA2014-55216, and of NASA's Astrophysics Data System Bibliographic Services. This research has made use of the VizieR catalogue access tool, CDS, Strasbourg, France. The original description of the VizieR service was published in \citet{2000A&AS..143...23O}.

The authors wish to acknowledge the use of NeSI high-performance computing facilities for this research. NZ's national facilities are provided by the NZ eScience Infrastructure and funded jointly by NeSI's collaborator institutions and through the Ministry of Business, Innovation \& Employment's Research Infrastructure programme. https://www.nesi.org.nz. We also made use of computing resources provided by Warwick's Scientific Computing Research Technology Platform (SCRTP).




%
\appendix

\section{Quantitative Results}

We provide here quantitative results from the BPASS v2.2 models for future reference.  In tables \ref{tab:a1} and \ref{tab:a2} we give the rest-frame $K$-band  and $V$-band mass to light ratios derived from our models as a function of metallicity at stellar population ages log(age)=9.0-10.2.  In table \ref{tab:a3} we present a table of selected Lick indices in the same age range, at a tenth Solar, half Solar, Solar and twice Solar metallicities. Note that the values given here have been smoothed over two adjacent time bins to remove jumpiness due to residual sampling issues in our low mass binary models as discussed in section \ref{sec:comp}. Full tables of these quantities at all metallicities and ages sampled by BPASS and for nine different IMFs are available as part of our v2.2 data release.

\begin{table*}
\caption{$K$-band Mass to Light ratios}\label{tab:a1}
\begin{tabular}{ccrcccccccccccc}
\multicolumn{2}{c}{Metallicity} &  \multicolumn{13}{c}{M/L in units of $M_\odot/L_\odot$ at log (Stellar Population Age / years)}\\
 $Z$ & [Fe/H] & log(Age) = 9.0 & 9.1 & 9.2 & 9.3 & 9.4 & 9.5 & 9.6 & 9.7 & 9.8 & 9.9 & 10.0 & 10.1 & 10.2\\
 \hline\hline
$10^{-5}$ &    -3.30 &     0.24 &     0.33 &     0.41 &     0.45 &     0.48 &     0.53 &     0.69 &     0.77 &     0.85 &     1.04 &     1.02 &     1.42 &     1.62 \\ 
$10^{-4}$ &    -2.30 &     0.22 &     0.25 &     0.31 &     0.33 &     0.38 &     0.44 &     0.60 &     0.69 &     0.78 &     0.96 &     0.95 &     1.22 &     1.35 \\ 
0.001 &    -1.30 &     0.13 &     0.13 &     0.17 &     0.24 &     0.30 &     0.36 &     0.52 &     0.61 &     0.69 &     0.87 &     0.87 &     1.24 &     1.33 \\ 
0.002 &    -1.00 &     0.12 &     0.12 &     0.14 &     0.20 &     0.27 &     0.37 &     0.45 &     0.53 &     0.69 &     0.80 &     1.00 &     0.95 &     1.19 \\ 
0.003 &    -0.82 &     0.10 &     0.11 &     0.14 &     0.20 &     0.26 &     0.32 &     0.39 &     0.48 &     0.65 &     0.77 &     0.99 &     0.92 &     1.28 \\ 
0.004 &    -0.70 &     0.10 &     0.10 &     0.14 &     0.18 &     0.24 &     0.30 &     0.36 &     0.51 &     0.60 &     0.69 &     0.93 &     0.94 &     1.30 \\ 
0.006 &    -0.52 &     0.10 &     0.10 &     0.13 &     0.17 &     0.22 &     0.28 &     0.40 &     0.48 &     0.58 &     0.76 &     0.86 &     1.06 &     0.99 \\ 
0.008 &    -0.40 &     0.13 &     0.17 &     0.19 &     0.20 &     0.23 &     0.29 &     0.35 &     0.42 &     0.61 &     0.73 &     0.82 &     0.95 &     0.98 \\ 
0.010 &    -0.30 &     0.18 &     0.22 &     0.24 &     0.20 &     0.23 &     0.28 &     0.33 &     0.41 &     0.59 &     0.70 &     0.94 &     0.95 &     0.95 \\ 
0.014 &    -0.15 &     0.10 &     0.14 &     0.17 &     0.20 &     0.22 &     0.29 &     0.37 &     0.46 &     0.51 &     0.70 &     0.80 &     0.86 &     0.97 \\ 
0.020 &     0.00 &     0.11 &     0.16 &     0.19 &     0.22 &     0.24 &     0.29 &     0.35 &     0.44 &     0.53 &     0.76 &     0.87 &     1.15 &     1.14 \\ 
0.030 &     0.18 &     0.09 &     0.19 &     0.22 &     0.26 &     0.23 &     0.28 &     0.35 &     0.50 &     0.63 &     0.75 &     0.77 &     1.17 &     1.31 \\ 
0.040 &     0.30 &     0.10 &     0.16 &     0.21 &     0.25 &     0.26 &     0.29 &     0.39 &     0.54 &     0.69 &     0.81 &     1.14 &     1.31 &     1.47 \\ 
 \hline
\end{tabular}
\end{table*}

\begin{table*}
\caption{$V$-band Mass to Light ratios}\label{tab:a2}
\begin{tabular}{ccrcccccccccccc}
\multicolumn{2}{c}{Metallicity} &  \multicolumn{13}{c}{M/L in units of $M_\odot/L_\odot$ at log (Stellar Population Age / years)}\\
 $Z$ & [Fe/H] & log(Age) = 9.0 & 9.1 & 9.2 & 9.3 & 9.4 & 9.5 & 9.6 & 9.7 & 9.8 & 9.9 & 10.0 & 10.1 & 10.2\\
 \hline\hline
 $10^{-5}$ &    -3.30 &     0.32 &     0.41 &     0.49 &     0.55 &     0.62 &     0.72 &     0.92 &     1.12 &     1.30 &     1.65 &     1.72 &     2.59 &     3.29 \\ 
$10^{-4}$ &    -2.30 &     0.30 &     0.36 &     0.45 &     0.51 &     0.61 &     0.71 &     0.93 &     1.13 &     1.32 &     1.64 &     1.71 &     2.35 &     2.91 \\ 
0.001 &    -1.30 &     0.25 &     0.28 &     0.36 &     0.49 &     0.62 &     0.76 &     1.06 &     1.28 &     1.49 &     1.88 &     1.96 &     2.81 &     3.23 \\ 
0.002 &    -1.00 &     0.27 &     0.28 &     0.35 &     0.47 &     0.64 &     0.86 &     1.08 &     1.29 &     1.64 &     1.95 &     2.40 &     2.50 &     3.22 \\ 
0.003 &    -0.82 &     0.27 &     0.30 &     0.38 &     0.51 &     0.68 &     0.85 &     1.05 &     1.28 &     1.71 &     2.04 &     2.62 &     2.61 &     3.76 \\ 
0.004 &    -0.70 &     0.28 &     0.31 &     0.40 &     0.51 &     0.67 &     0.85 &     1.05 &     1.41 &     1.71 &     2.00 &     2.62 &     2.75 &     3.94 \\ 
0.006 &    -0.52 &     0.30 &     0.34 &     0.43 &     0.55 &     0.71 &     0.90 &     1.22 &     1.51 &     1.84 &     2.36 &     2.75 &     3.36 &     3.52 \\ 
0.008 &    -0.40 &     0.35 &     0.43 &     0.52 &     0.63 &     0.77 &     0.97 &     1.23 &     1.49 &     1.98 &     2.43 &     2.84 &     3.45 &     3.71 \\ 
0.010 &    -0.30 &     0.38 &     0.47 &     0.56 &     0.65 &     0.81 &     1.01 &     1.26 &     1.54 &     2.10 &     2.55 &     3.28 &     3.64 &     3.83 \\ 
0.014 &    -0.15 &     0.38 &     0.49 &     0.60 &     0.72 &     0.87 &     1.11 &     1.45 &     1.81 &     2.11 &     2.63 &     3.14 &     3.58 &     4.34 \\ 
0.020 &     0.00 &     0.44 &     0.55 &     0.68 &     0.83 &     1.02 &     1.27 &     1.60 &     2.02 &     2.44 &     3.23 &     3.87 &     4.98 &     5.57 \\ 
0.030 &     0.18 &     0.52 &     0.65 &     0.79 &     0.94 &     1.13 &     1.39 &     1.75 &     2.34 &     2.99 &     3.66 &     4.14 &     5.73 &     6.93 \\ 
0.040 &     0.30 &     0.59 &     0.72 &     0.88 &     1.05 &     1.24 &     1.52 &     2.00 &     2.73 &     3.46 &     4.39 &     5.63 &     7.19 &     8.41 \\ 
\hline
\end{tabular}
\end{table*}

\begin{table*}
\caption{Numerical values of key Lick indices at selected metallicities. The MgFe50 and Ca\,Triplet indices are additions to the original index set of \citet{1994ApJS...94..687W} and \citet{1997ApJS..111..377W}, as discussed in the text.}\label{tab:a3}
\begin{tabular}{cccrccccccccccccc}
\multicolumn{2}{c}{Metallicity}  & Lick &   \multicolumn{13}{c}{Index Value in Angstroms at log (Stellar Population Age / years)}\\
 $Z$ & [Fe/H] & Index &  log(Age) = 9.0 & 9.1 & 9.2 & 9.3 & 9.4 & 9.5 & 9.6 & 9.7 & 9.8 & 9.9 & 10.0 & 10.1 & 10.2\\
\hline\hline
   0.002 &    -1.00 &      H\,$\beta$ &     4.89 &     4.32 &     3.86 &     3.46 &     3.13 &     2.77 &     2.39 &     2.09 &     1.87 &     1.65 &     1.59 &     1.23 &     1.24 \\ 
         &          &       Fe5015 &     2.49 &     2.81 &     2.98 &     3.06 &     3.04 &     3.09 &     3.24 &     3.38 &     3.43 &     3.53 &     3.44 &     3.84 &     3.83 \\ 
         &          &        Mg\,b &     1.12 &     1.22 &     1.27 &     1.29 &     1.30 &     1.35 &     1.43 &     1.51 &     1.57 &     1.65 &     1.69 &     1.83 &     1.91 \\ 
         &          &        G4300 &     0.51 &     1.23 &     1.89 &     2.50 &     3.02 &     3.43 &     4.00 &     4.51 &     4.92 &     5.41 &     5.54 &     6.58 &     6.58 \\ 
         &          &       Ca4227 &     0.22 &     0.25 &     0.26 &     0.25 &     0.24 &     0.25 &     0.27 &     0.29 &     0.31 &     0.35 &     0.37 &     0.45 &     0.51 \\ 
         &          &       Ca4455 &     0.35 &     0.42 &     0.48 &     0.52 &     0.55 &     0.58 &     0.63 &     0.67 &     0.70 &     0.74 &     0.74 &     0.85 &     0.87 \\ 
         &          &       Fe5270 &     1.36 &     1.52 &     1.60 &     1.64 &     1.62 &     1.65 &     1.72 &     1.79 &     1.83 &     1.90 &     1.88 &     2.09 &     2.14 \\ 
         &          &        Na\,D &     0.56 &     0.60 &     0.61 &     0.62 &     0.60 &     0.65 &     0.70 &     0.74 &     0.80 &     0.88 &     0.95 &     1.03 &     1.19 \\ 
         &          &  H\,$\delta$\,F &     6.33 &     5.56 &     4.72 &     3.85 &     3.03 &     2.33 &     1.63 &     1.08 &     0.66 &     0.24 &     0.11 &    -0.52 &    -0.35 \\ 
         &          &       MgFe50 &     1.63 &     1.82 &     1.93 &     1.97 &     1.97 &     2.01 &     2.11 &     2.21 &     2.26 &     2.33 &     2.31 &     2.55 &     2.57 \\ 
         &          &        Ca\,T &     6.57 &     6.57 &     6.43 &     6.20 &     5.92 &     5.78 &     5.72 &     5.70 &     5.64 &     5.64 &     5.57 &     5.74 &     5.81 \\ 
\hline
   0.010 &    -0.30 &      H\,$\beta$ &     4.53 &     4.09 &     3.52 &     2.83 &     2.37 &     1.99 &     1.69 &     1.47 &     1.41 &     1.27 &     1.21 &     0.95 &     0.81 \\ 
         &          &       Fe5015 &     3.53 &     3.70 &     4.13 &     4.86 &     5.27 &     5.62 &     5.90 &     6.08 &     6.01 &     6.12 &     6.07 &     6.43 &     6.61 \\ 
         &          &        Mg\,b &     1.78 &     1.80 &     1.99 &     2.36 &     2.59 &     2.78 &     2.96 &     3.09 &     3.14 &     3.25 &     3.31 &     3.49 &     3.61 \\ 
         &          &        G4300 &     1.27 &     2.12 &     2.97 &     3.96 &     4.83 &     5.69 &     6.40 &     6.93 &     7.18 &     7.46 &     7.65 &     8.03 &     8.24 \\ 
         &          &       Ca4227 &     0.26 &     0.21 &     0.24 &     0.34 &     0.42 &     0.51 &     0.60 &     0.68 &     0.69 &     0.78 &     0.82 &     1.00 &     1.13 \\ 
         &          &       Ca4455 &     0.64 &     0.71 &     0.82 &     0.98 &     1.10 &     1.21 &     1.31 &     1.38 &     1.39 &     1.46 &     1.48 &     1.61 &     1.69 \\ 
         &          &       Fe5270 &     1.97 &     1.95 &     2.08 &     2.40 &     2.57 &     2.72 &     2.85 &     2.94 &     2.94 &     3.02 &     3.04 &     3.22 &     3.32 \\ 
         &          &        Na\,D &     1.46 &     1.17 &     1.19 &     1.51 &     1.64 &     1.71 &     1.79 &     1.86 &     1.89 &     2.01 &     2.10 &     2.33 &     2.46 \\ 
         &          &  H\,$\delta$\,F &     4.73 &     3.60 &     2.60 &     1.62 &     0.82 &     0.07 &    -0.56 &    -1.04 &    -1.23 &    -1.49 &    -1.63 &    -2.03 &    -2.28 \\ 
         &          &       MgFe50 &     2.38 &     2.47 &     2.75 &     3.24 &     3.53 &     3.77 &     3.97 &     4.11 &     4.09 &     4.18 &     4.18 &     4.42 &     4.55 \\ 
         &          &        Ca\,T &     7.91 &     7.52 &     7.41 &     7.72 &     7.72 &     7.63 &     7.55 &     7.46 &     7.29 &     7.22 &     7.10 &     7.16 &     7.16 \\ 
\hline
   0.020 &     0.00 &      H\,$\beta$ &     4.15 &     3.62 &     3.12 &     2.72 &     2.37 &     2.00 &     1.67 &     1.41 &     1.22 &     1.15 &     0.99 &     0.90 &     0.67 \\ 
         &          &       Fe5015 &     4.53 &     4.79 &     5.18 &     5.59 &     6.03 &     6.42 &     6.81 &     7.06 &     7.25 &     7.23 &     7.40 &     7.41 &     7.76 \\ 
         &          &        Mg\,b &     2.36 &     2.45 &     2.66 &     2.91 &     3.19 &     3.43 &     3.68 &     3.88 &     4.05 &     4.12 &     4.28 &     4.35 &     4.55 \\ 
         &          &        G4300 &     2.38 &     3.37 &     4.29 &     5.03 &     5.70 &     6.36 &     6.95 &     7.40 &     7.73 &     7.89 &     8.08 &     8.20 &     8.39 \\ 
         &          &       Ca4227 &     0.25 &     0.24 &     0.30 &     0.40 &     0.54 &     0.68 &     0.85 &     1.00 &     1.14 &     1.18 &     1.35 &     1.42 &     1.68 \\ 
         &          &       Ca4455 &     0.88 &     0.97 &     1.08 &     1.19 &     1.31 &     1.44 &     1.58 &     1.69 &     1.79 &     1.82 &     1.92 &     1.97 &     2.11 \\ 
         &          &       Fe5270 &     2.33 &     2.36 &     2.48 &     2.65 &     2.84 &     3.01 &     3.19 &     3.32 &     3.43 &     3.46 &     3.57 &     3.62 &     3.79 \\ 
         &          &        Na\,D &     1.65 &     1.44 &     1.43 &     1.58 &     1.79 &     1.95 &     2.15 &     2.32 &     2.47 &     2.56 &     2.75 &     2.89 &     3.17 \\ 
         &          &  H\,$\delta$\,F &     3.30 &     2.22 &     1.34 &     0.70 &     0.09 &    -0.57 &    -1.22 &    -1.72 &    -2.11 &    -2.28 &    -2.56 &    -2.74 &    -3.10 \\ 
         &          &       MgFe50 &     3.08 &     3.24 &     3.51 &     3.80 &     4.12 &     4.39 &     4.67 &     4.87 &     5.02 &     5.03 &     5.17 &     5.21 &     5.45 \\ 
         &          &        Ca\,T &     8.45 &     8.09 &     7.87 &     7.77 &     7.75 &     7.66 &     7.61 &     7.55 &     7.48 &     7.37 &     7.32 &     7.20 &     7.26 \\ 
\hline
   0.040 &     0.30 &      H\,$\beta$ &     3.96 &     3.47 &     3.04 &     2.69 &     2.32 &     1.93 &     1.62 &     1.41 &     1.35 &     1.18 &     1.07 &     0.79 &     0.61 \\ 
         &          &       Fe5015 &     5.55 &     5.97 &     6.39 &     6.83 &     7.41 &     8.00 &     8.40 &     8.63 &     8.72 &     8.86 &     8.82 &     9.07 &     9.27 \\ 
         &          &        Mg\,b &     3.04 &     3.26 &     3.47 &     3.74 &     4.12 &     4.49 &     4.79 &     5.01 &     5.11 &     5.30 &     5.40 &     5.68 &     5.87 \\ 
         &          &        G4300 &     3.86 &     4.82 &     5.58 &     6.12 &     6.53 &     6.99 &     7.40 &     7.69 &     7.61 &     7.61 &     7.60 &     7.80 &     7.87 \\ 
         &          &       Ca4227 &     0.29 &     0.32 &     0.40 &     0.55 &     0.80 &     1.08 &     1.31 &     1.49 &     1.60 &     1.82 &     1.95 &     2.33 &     2.69 \\ 
         &          &       Ca4455 &     1.16 &     1.29 &     1.41 &     1.54 &     1.71 &     1.89 &     2.05 &     2.18 &     2.23 &     2.35 &     2.43 &     2.62 &     2.76 \\ 
         &          &       Fe5270 &     2.62 &     2.75 &     2.90 &     3.08 &     3.33 &     3.57 &     3.74 &     3.86 &     3.92 &     4.03 &     4.08 &     4.24 &     4.37 \\ 
         &          &        Na\,D &     1.40 &     1.47 &     1.64 &     1.92 &     2.33 &     2.70 &     2.97 &     3.16 &     3.31 &     3.59 &     3.75 &     4.14 &     4.44 \\ 
         &          &  H\,$\delta$\,F &     1.83 &     0.88 &     0.15 &    -0.41 &    -0.95 &    -1.62 &    -2.26 &    -2.75 &    -2.71 &    -2.88 &    -3.00 &    -3.50 &    -3.74 \\ 
         &          &       MgFe50 &     3.83 &     4.11 &     4.39 &     4.70 &     5.13 &     5.55 &     5.85 &     6.05 &     6.12 &     6.26 &     6.28 &     6.50 &     6.66 \\ 
         &          &        Ca\,T &     7.96 &     7.97 &     8.03 &     8.10 &     8.16 &     8.21 &     8.17 &     8.06 &     8.00 &     7.85 &     7.60 &     7.44 &     7.37 \\ 
\hline
 \hline
\end{tabular}
\end{table*}

\begin{figure}
\includegraphics[width=0.9\columnwidth]{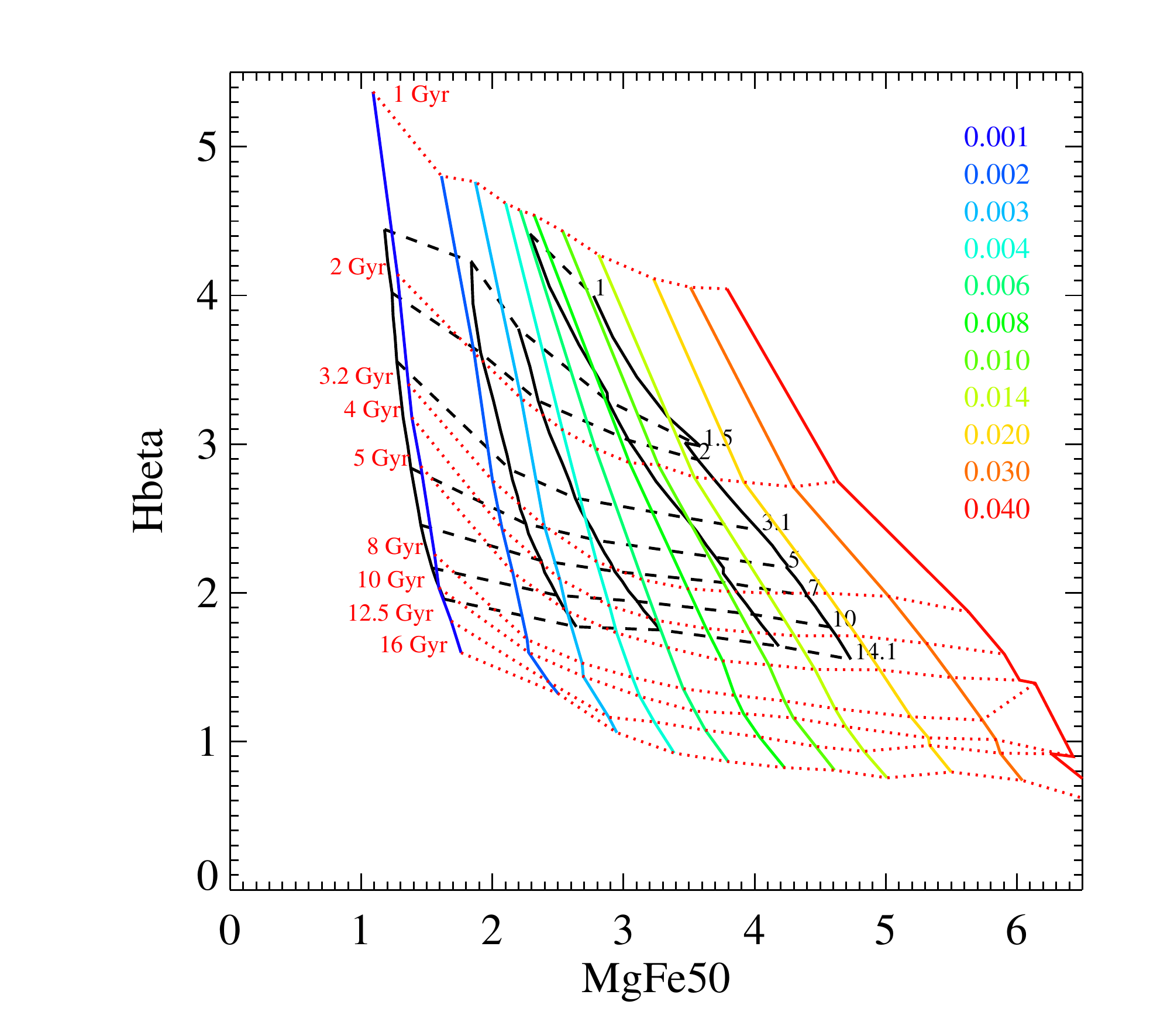}
\includegraphics[width=0.9\columnwidth]{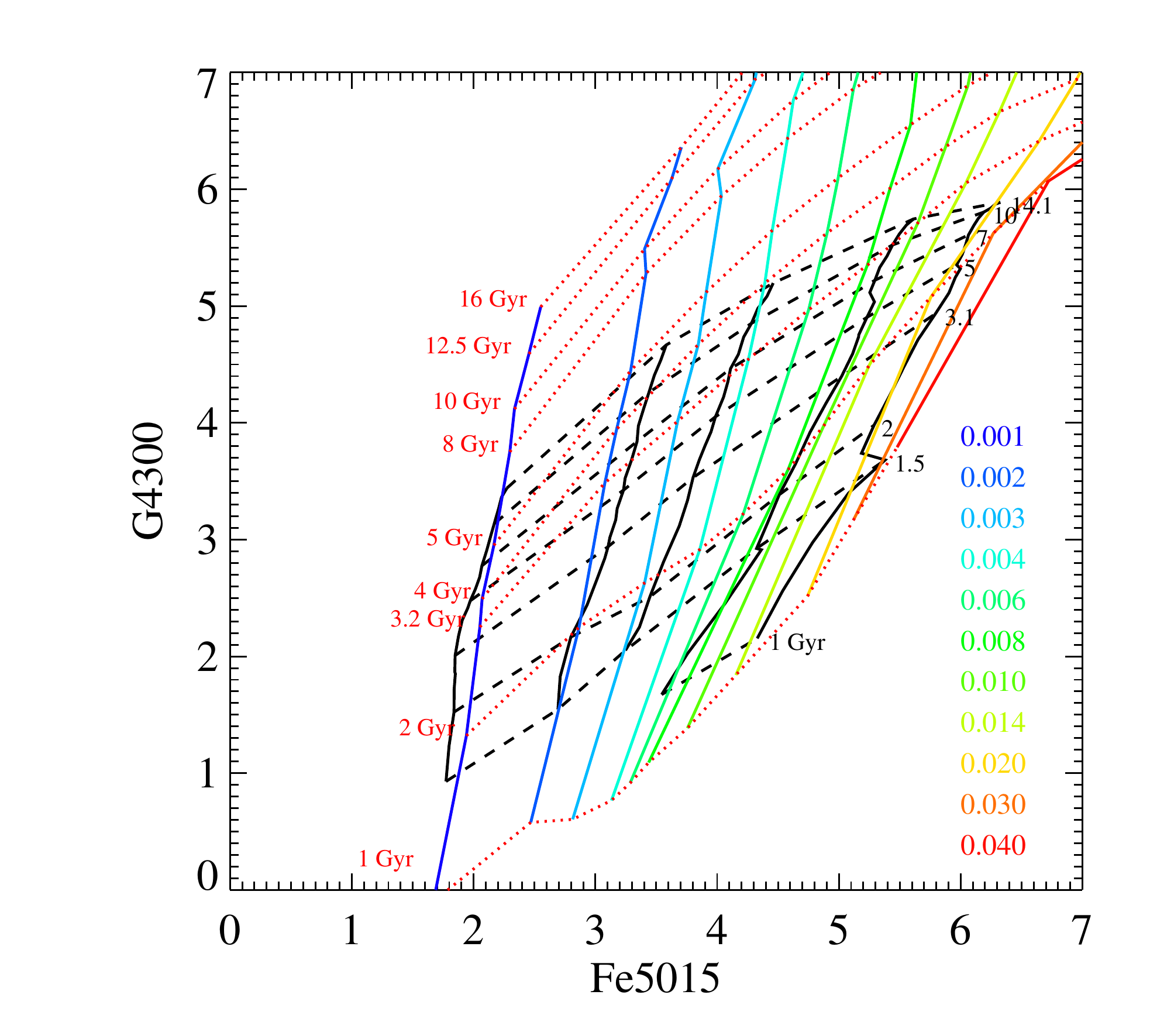}

\caption{The (MgFe50 versus H$\beta$) and (Fe5015 versus G4300) planes as a diagnostic of age and metallicity. BPASS v2.2 model indices are shown as lines coloured by metallicity as shown in the legend, with selected ages marked in red. The older \citet{2007ApJS..171..146S} models are shown in black, again with ages indicated, and increase monotonically in metallicity with increasing MgFe50 or Fe5015 index. Model values (indicated by solid lines) are [Fe/H]=-1.3, -0.7, -0.4, and 0.0 (equivalent to $Z=$0.001, 0.003, 0.004, 0.008 and 0.020).}
\label{fig:grid}
\end{figure}

We also provide a comparison of diagnostic index-index plots between BPASS v2.2 and the models of \citet{2007ApJS..171..146S} in Fig. \ref{fig:grid}. As discussed in the main text, BPASS model grids tend to span a wider range in index value than the older calibrations. In particularly, the same age range is spanned by a larger range of H$\beta$ or G4300 index, allowing a younger ($\sim8$\,Gyr) model to match an index which would yield an older ($\sim14$\,Gyr)  \citet{2007ApJS..171..146S} model.

\begin{figure}
\includegraphics[width=0.9\columnwidth]{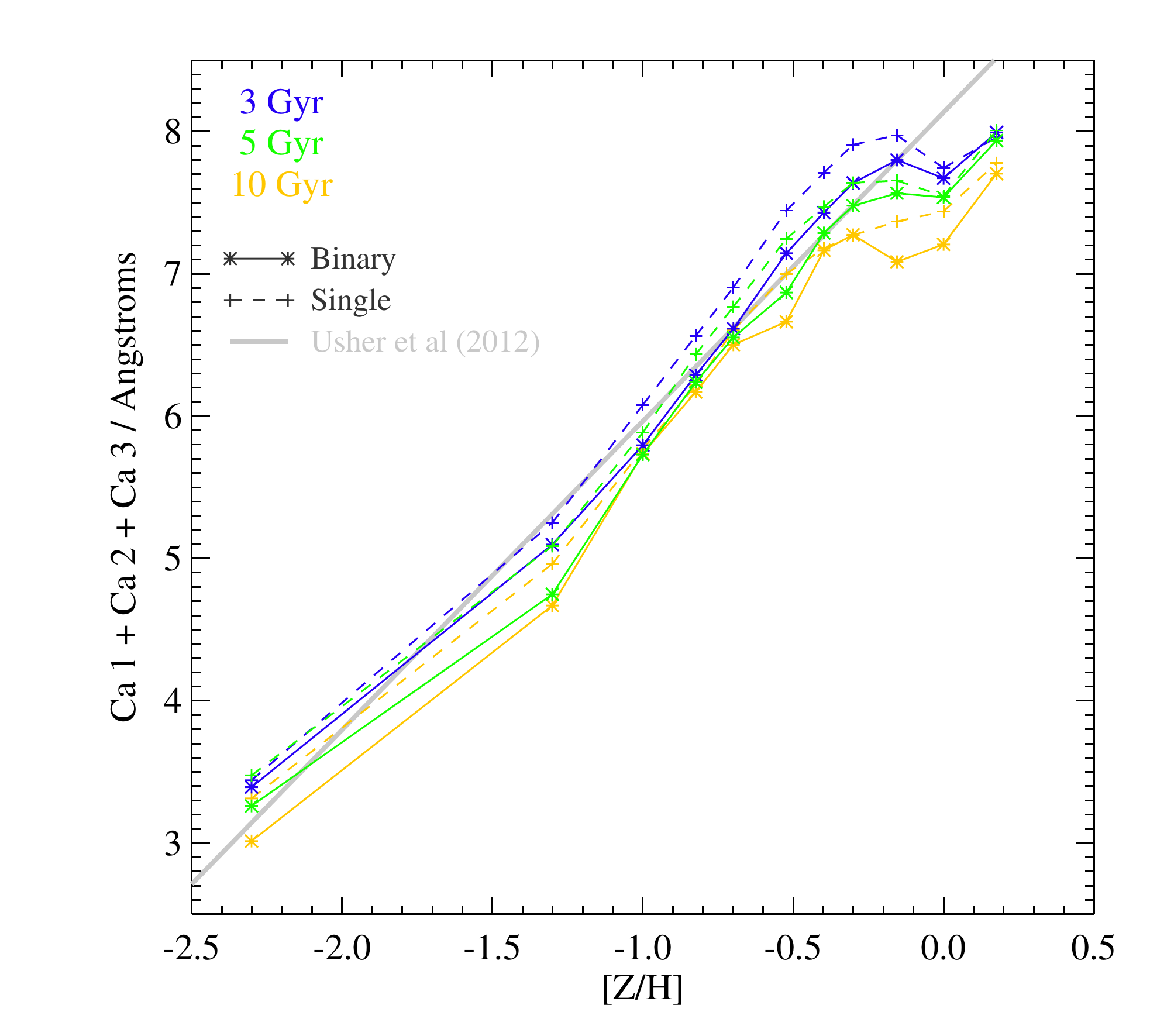}
\caption{Variation in the strength of the near-infrared calcium triplet (Ca\,T = Ca\,1 + Ca\,2 + Ca\,3) with age and metallicity. We compare BPASS v2.2 predictions for both single and binary star populations at three ages. We also show the linear relation derived by \citet{2012MNRAS.426.1475U}, based on the models of \citet{2010MNRAS.404.1639V}.}
\label{fig:catriplet}
\end{figure}

The strength of the near-infrared calcium triplet ($\lambda$=8498, 8542, 8662\,\AA) has been proposed as a reliable metallicity indicator, independent of stellar population age and $\alpha$-element enhancement.  In Fig. \ref{fig:catriplet} we compare predictions from BPASS v2.2 \citep[using the line index definitions of][]{2012AJ....143...44D} with the calibration derived by \citet{2012MNRAS.426.1475U}, based on the models of \citet{2010MNRAS.404.1639V}. Given the slight differences in line index and continuum definition, the agreement is rather remarkably good. We note that our binary models produce marginally weaker line indices than our single star models, and that we also see a weak dependence on stellar population age at fixed metallicity. The turn-over in the models at high metallicity is largely due to changes in the line blanketing of the pseudo-continuum bands, rather than the strength of the Ca triplet absorption.

\begin{figure*}
\includegraphics[width=0.9\columnwidth]{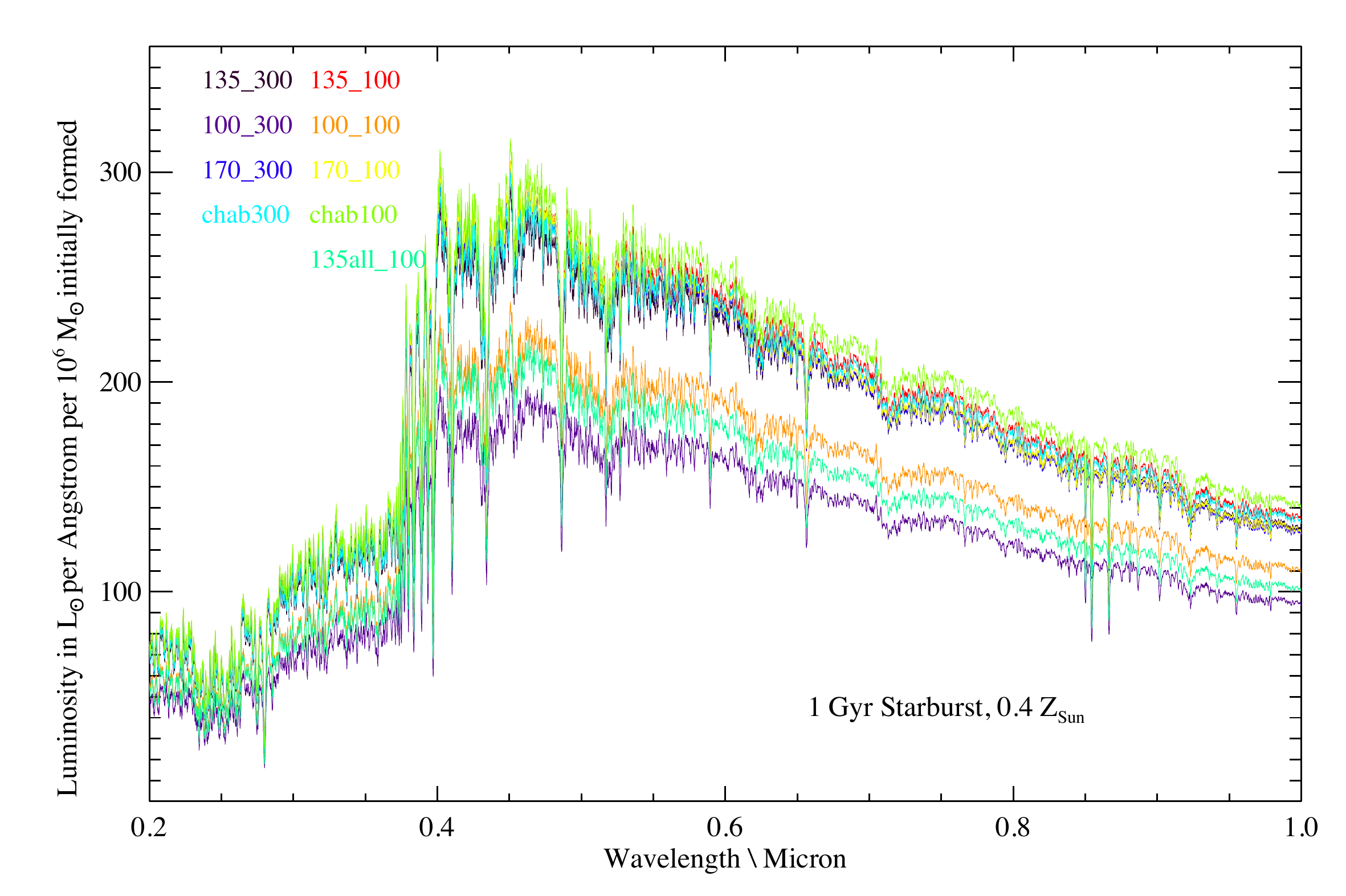}
\includegraphics[width=0.9\columnwidth]{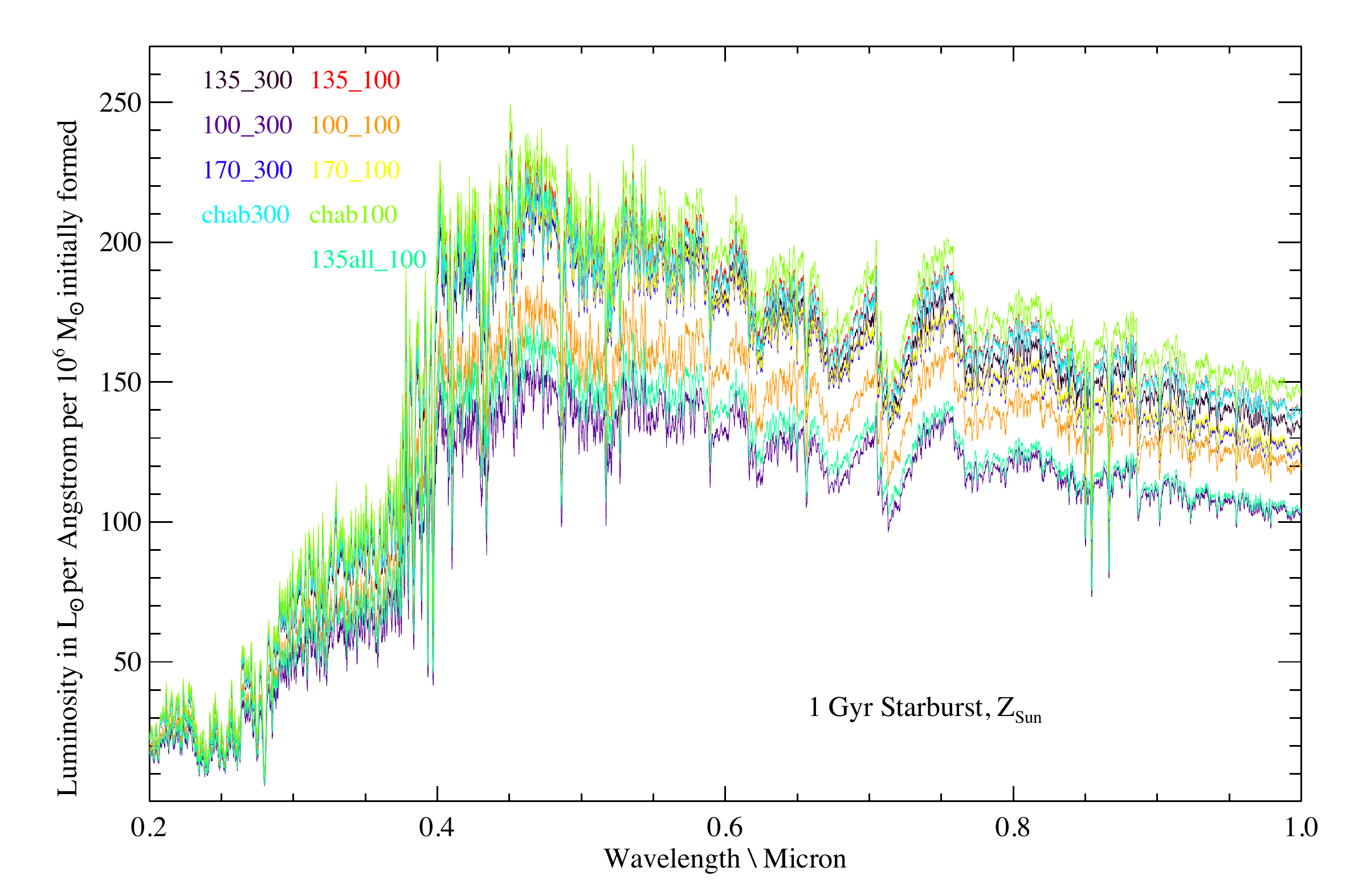}
\includegraphics[width=0.9\columnwidth]{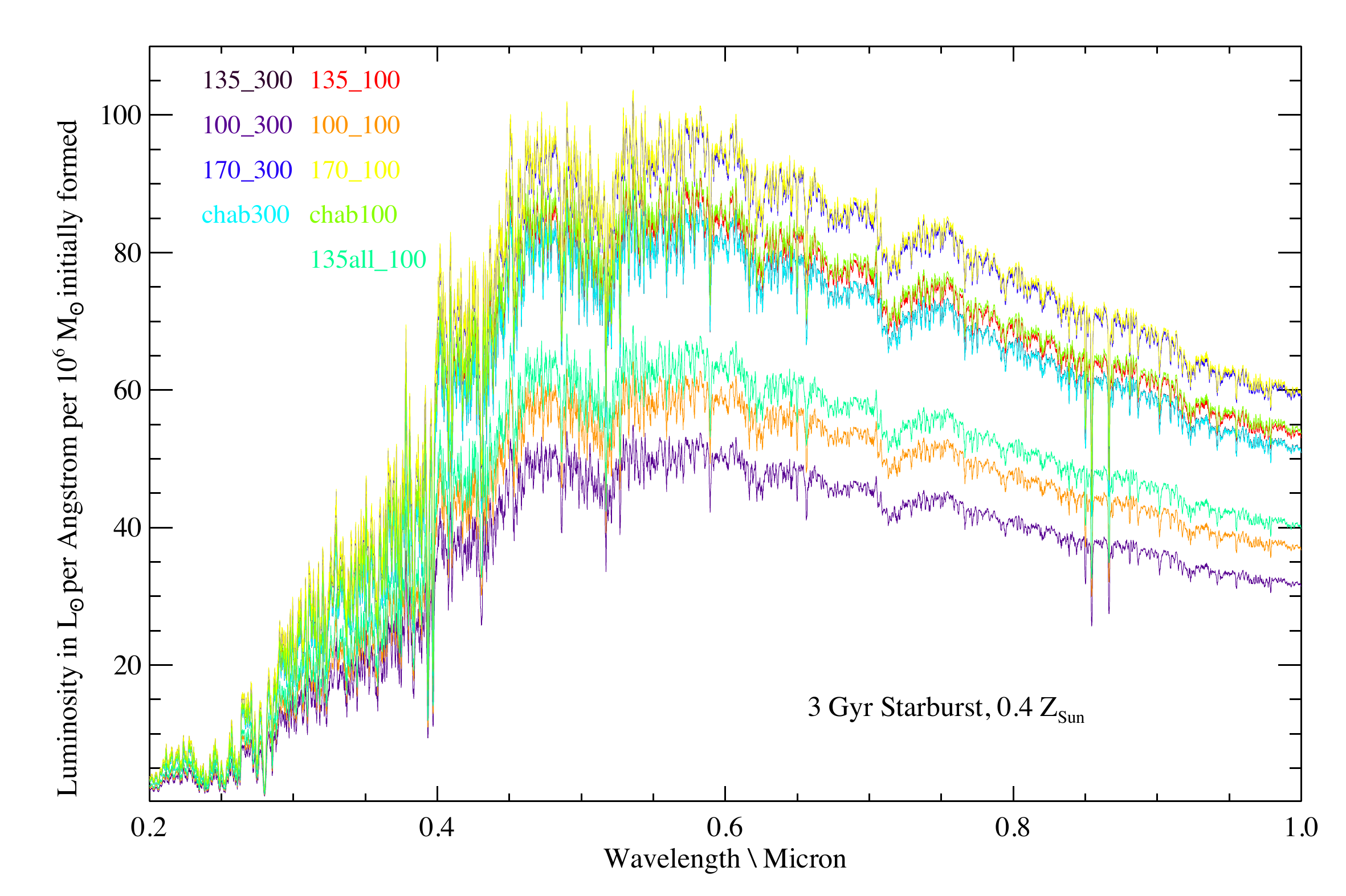}
\includegraphics[width=0.9\columnwidth]{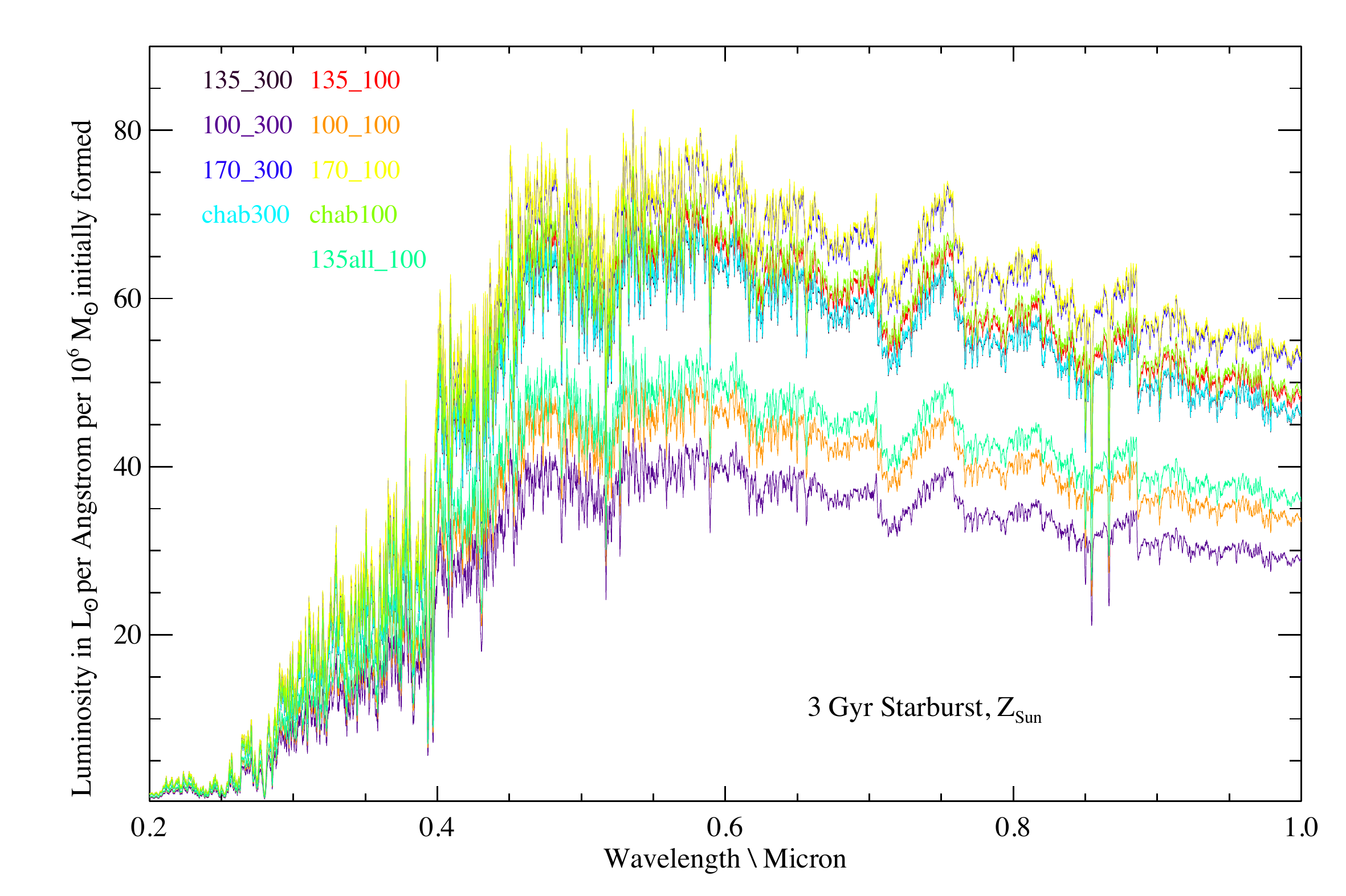}
\includegraphics[width=0.9\columnwidth]{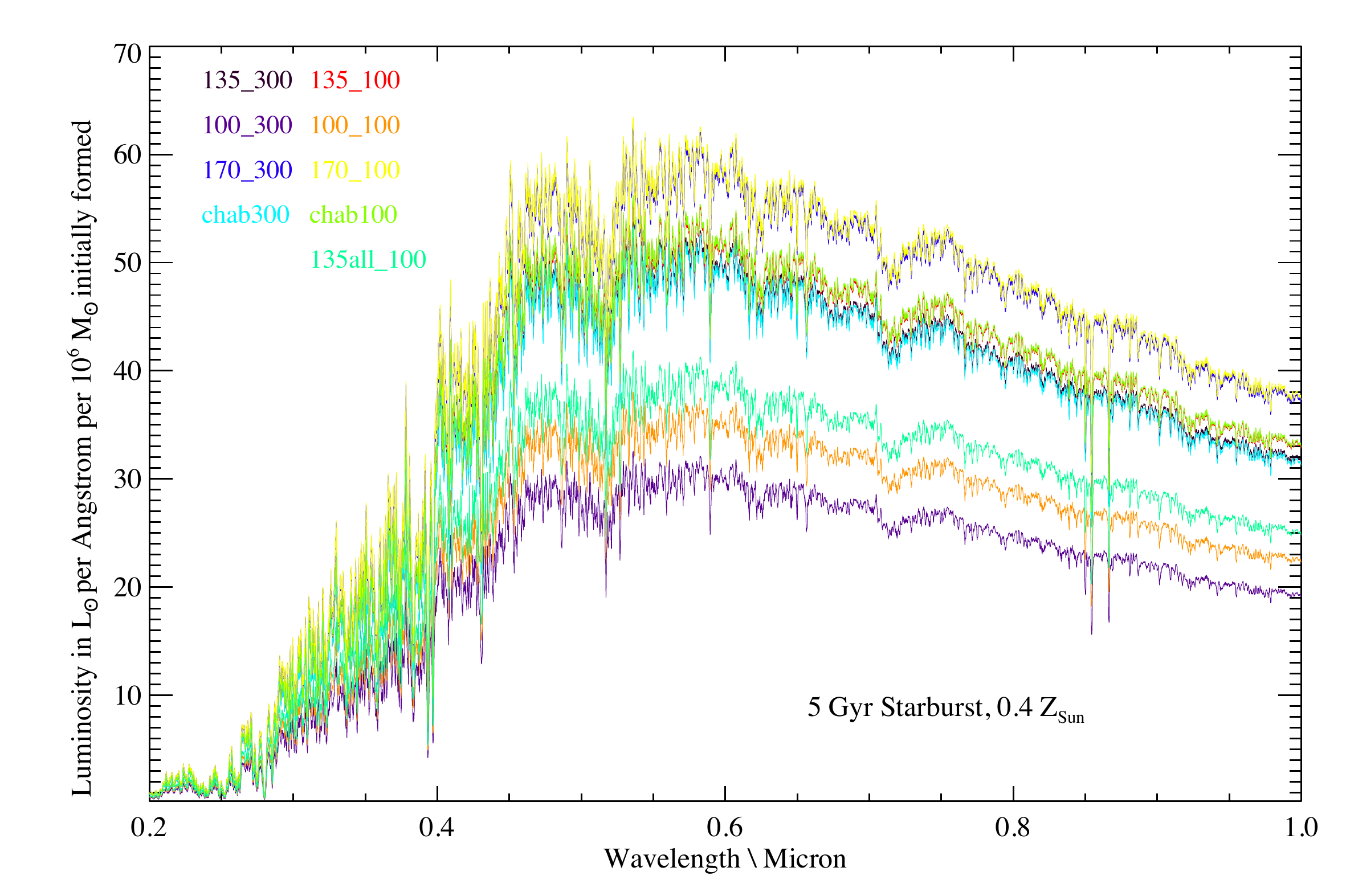}
\includegraphics[width=0.9\columnwidth]{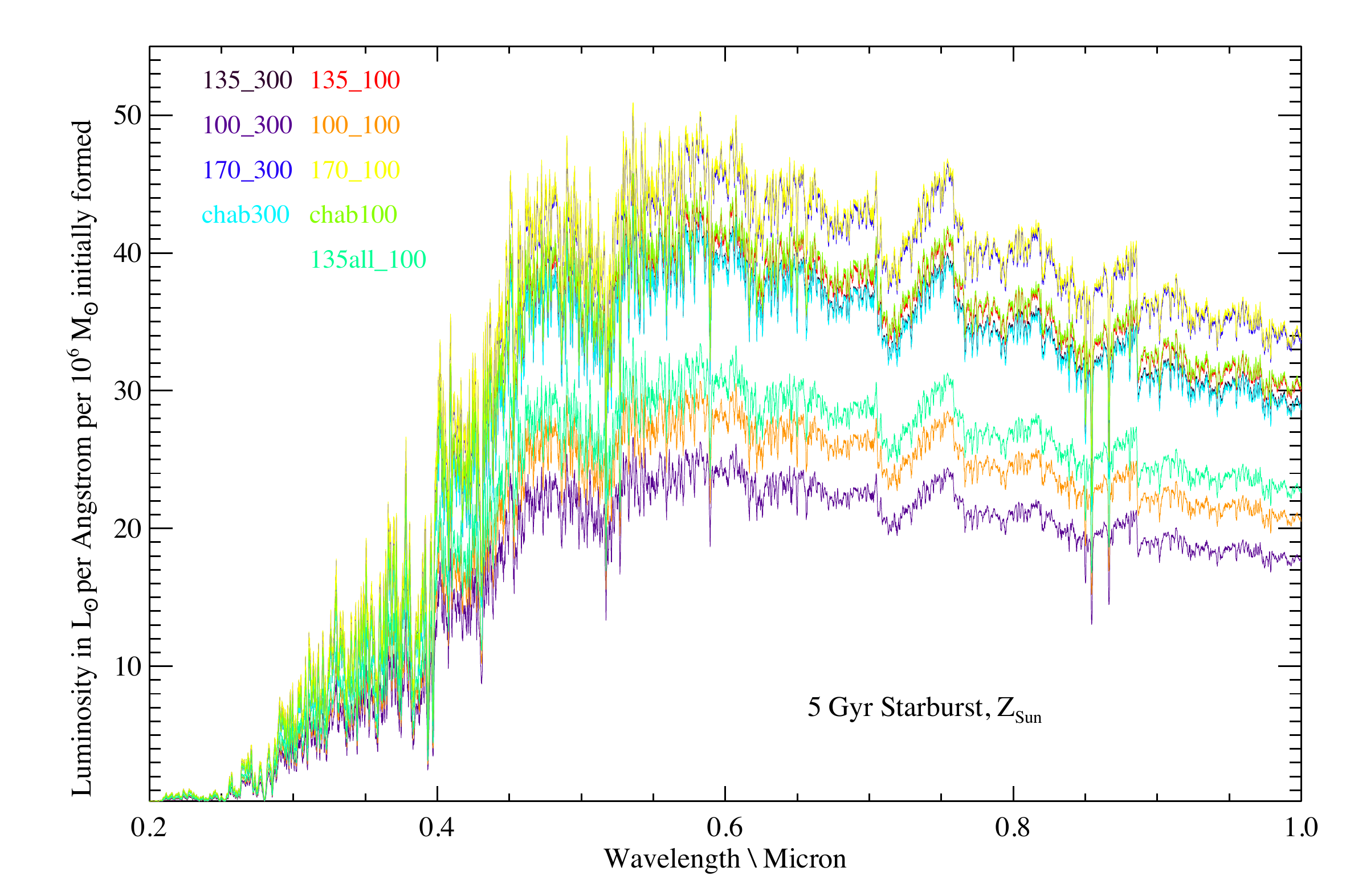}
\includegraphics[width=0.9\columnwidth]{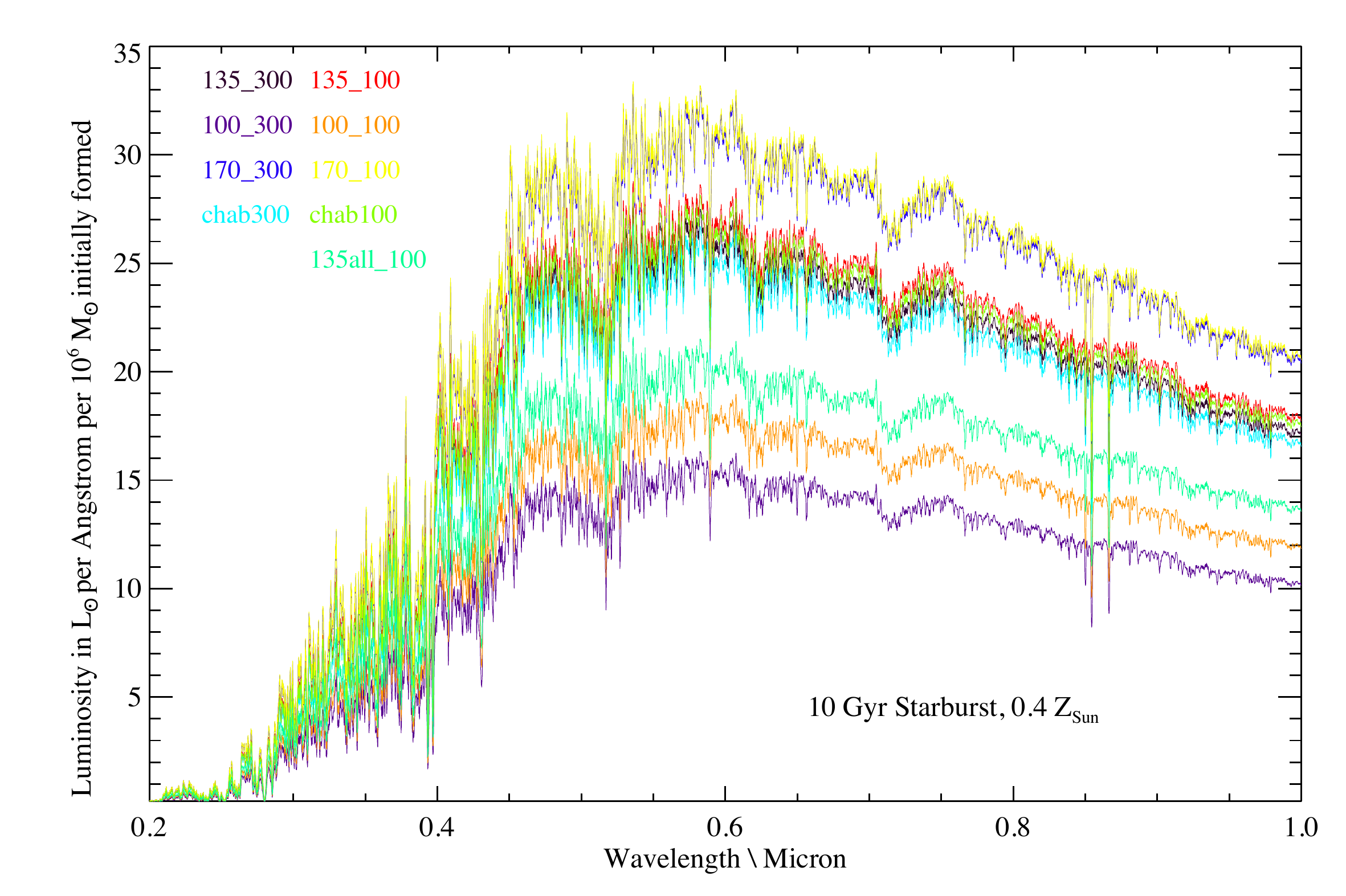}
\includegraphics[width=0.9\columnwidth]{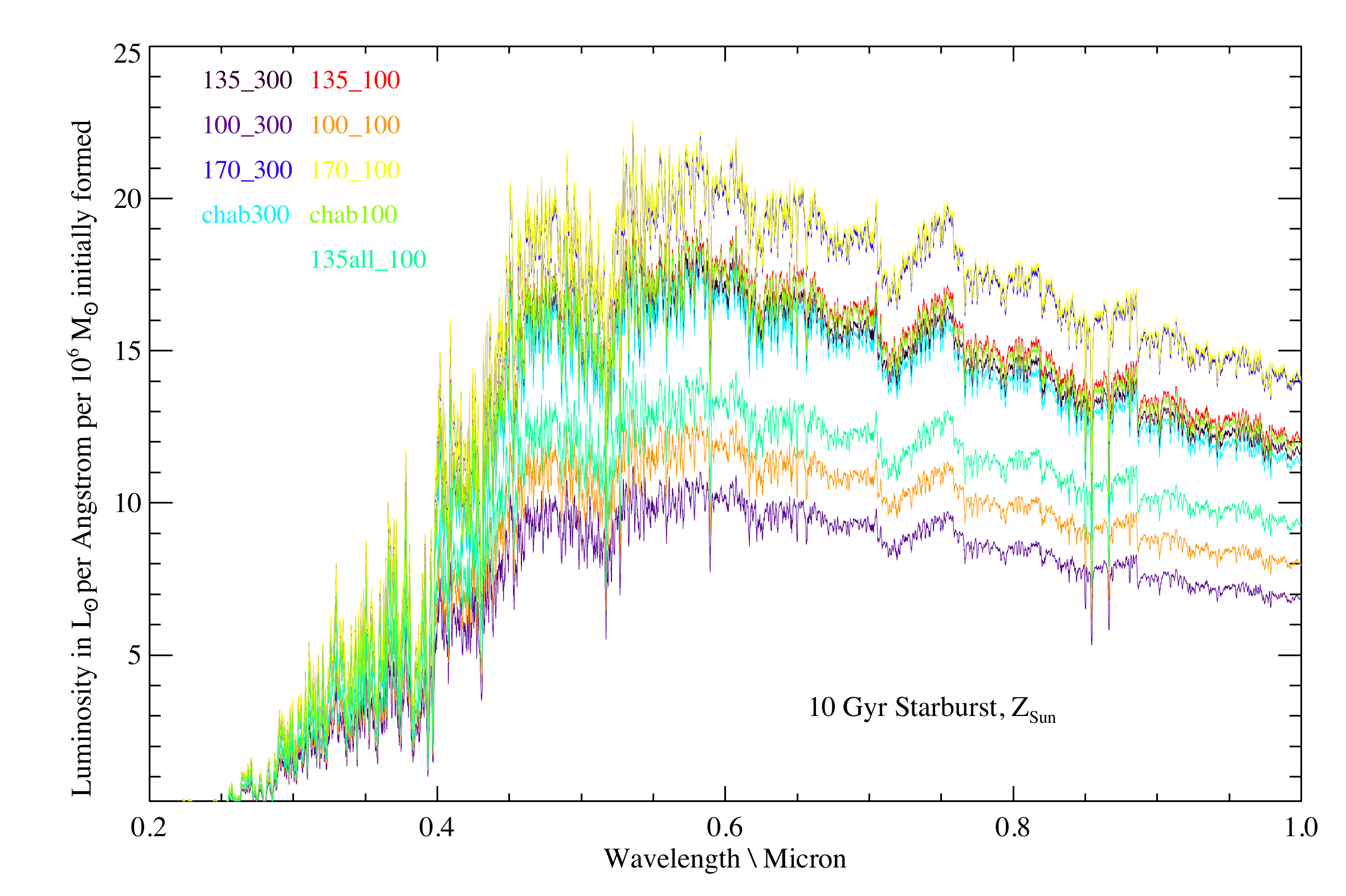}
\caption{The effect of IMF on the output spectra of old stellar populations in BPASS v2.2. Results are shown for our nine input IMFs (defined in table \ref{tab:imfs}) at four ages and two metallicities.}\label{fig:seds_imfs}
\end{figure*}

Finally, in Fig. \ref{fig:seds_imfs}, we demonstrate the effect of the adopted IMF on the evolution of the spectral shape and integrated luminosity of a stellar population between 1\,Gyr and 10\,Gyr in age, at two metallicities.

%

\bsp	
\label{lastpage}
\end{document}